\documentclass[fleqn,usenatbib]{mnras}

\usepackage{newtxtext} 

\usepackage[T1]{fontenc}

\DeclareRobustCommand{\VAN}[3]{#2}
\let\VANthebibliography\thebibliography
\def\thebibliography{\DeclareRobustCommand{\VAN}[3]{##3}\VANthebibliography}

\usepackage{graphicx}	
\usepackage{amsmath}	
\usepackage{amssymb}	

\newcommand{\GGbig}{{\bf G}}      
\newcommand{\GGbigcomp}{G}        
\newcommand{\Ggvert}{{\bf\bar g}} 
\newcommand{\Ggvertcomp}{\bar g}  
\newcommand{\ggloc}{{\bf g}}      
\newcommand{\ggloccomp}{g}        
\newcommand{\ggloctens}{g}        
\newcommand{\llunit}{{\bf l}}     
\newcommand{\LLbig}{{\bf L}}      
\newcommand{\LLbigcomp}{{L}}      
\newcommand{\TTbig}{{\bf T}}      
\newcommand{\tvisc}{t^{(v)}}       
\newcommand{\ttot}{t}             
\newcommand{\shear}{s}            
\newcommand{\shbig}{S}            
\newcommand{\whp}{W}              
\newcommand{\Xii}{g_0}            
\newcommand{\psidir}{{\bf e}_\psi} 
\newcommand{\warpvector}{{\boldsymbol\psi}} 
\newcommand{\vx}{v_x'}
\newcommand{\vy}{v_y'}
\newcommand{\vz}{v_z'}
\newcommand{\vi}{v_i'}
\newcommand{\vxyz}{v_{xyz}'}
\newcommand{\VVx}{V_x}
\newcommand{\VVy}{V_y}
\newcommand{\VVz}{V_z}

\title[Warped disc equations]{On the equations of warped disc dynamics}

\author[C. P. Dullemond et al.]{
C. P. Dullemond,$^{1}$\thanks{E-mail: dullemond@uni-heidelberg.de}
C. N. Kimmig$^{1},$
and J. J. Zanazzi$^{2}$
\\
$^{1}$Institute for Theoretical Astrophysics, Zentrum f\"ur Astronomie, Heidelberg University, Albert Ueberle Str. 2, 69122 Heidelberg, Germany\\
$^{2}$Canadian Institute for Theoretical Astrophysics, University of Toronto, 60 St. George Street, Toronto, Ontario, M5S 3H8, Canada
}

\date{Accepted XXX. Received YYY; in original form ZZZ}

\pubyear{2021}

\begin{document}
\label{firstpage}
\pagerange{\pageref{firstpage}--\pageref{lastpage}}
\maketitle

\begin{abstract}
  The 1-D evolution equations for warped discs come in two flavors: For very
  viscous discs the internal torque vector $\mathbf{G}$ is uniquely determined
  by the local conditions in the disc, and warps tend to damp out rapidly if
  they are not continuously driven. For very inviscid discs, on the other hand,
  $\mathbf{G}$ becomes a dynamic quantity, and a warp will propagate through the
  disc as a wave. The equations governing both regimes are usually treated
  separately. A unified set of equations was postulated recently by Martin et
  al. (2019), but not yet derived from the underlying physics. The standard
  method for deriving these equations is based on a perturbation series
  expansion, which is a powerful, but somewhat abstract technique. A more
  straightforward method is to employ the warped shearing box framework of
  Ogilvie \& Latter (2013), which so far has not yet been used to derive the
  equations for the wavelike regime. The goal of this paper is to analyze the
  warped disc equations in both regimes using the warped shearing box framework,
  to derive a unified set of equations, {valid for
  small warps}, and to discuss how our results can be
  interpreted in terms of the affine tilted-slab approach of Ogilvie (2018).
\end{abstract}




\section{Introduction}

In the last few years, numerous examples of non-planar protoplanetary discs have
been observed. The first direct observational indication of such non-standard
geometries came with the interpretation of two mysterious shadows on the disc
around HD 142527 as being cast by an inner disc that is inclined 70$^\circ$ with
respect to the outer disc \citep{2015ApJ...798L..44M}. Since then numerous
additional examples have been found \citep[e.g.][]{2017A&A...597A..42B,
  2017ApJ...849..143S, 2018A&A...619A.171B, 2020A&A...639A..62K}. Lately, even
more complex warped, twisted and broken disc geometries have been discovered,
for instance in the disc around HD 139614 \citep{2020A&A...635A.121M} and GW Ori
\citep{2020Sci...369.1233K, 2020ApJ...895L..18B}. Clearly, the topic of warped and twisted discs has
been cast back into the limelight by these discoveries, even though the theory
goes back several decades \citep[e.g.][]{1983MNRAS.202.1181P,
  1992MNRAS.258..811P, 1993ApJ...409..360L}.

{Also in other areas of astrophysics warped disk geometries are common. For
instance, some X-ray binaries are thought to host warped and tilted accretion
disks.  The occultation of an accreting neutron star by a precessing, tilted
accretion disk is thought to explain the superorbital modulation in the
light-curves of LMC X-4, SMC X-1, and Her X-1
\citep[e.g.][]{Charles+(2008),Brumback+(2020),Brumback+(2021)}.  In addition,
the narrow Fe K emission line in the X-ray binary and black hole candidate MAXI
J1535-571 is ascribed to a warp which locally alters the profile of the
accretion disk \citep{Miller+(2018)}. Active Galactic Nuclei (AGN) disks around
supermassive black holes have significant evidence for warps as well.  The maser
emission from NGC 4258 \citep{Herrnstein+(2005)}, Circinus
\citep{Greenhill+(2003)}, and four of the seven megamaser disks in the Megamaser
Cosmology Project \citep{Kuo+(2011)} are best fit by warped AGN disk models.
Also, the jets of multiple AGN are not perpendicular to the galactic plane,
implying misalignment of an inner AGN accretion disk with the galactic disk
\citep{Kinney+(2000)}.}

While it has become increasingly clear that a full understanding of warped discs
requires 3-D numerical simulations \citep[e.g.][]{2007MNRAS.381.1287L,
  2013MNRAS.433.2142F, 2013MNRAS.434.1946N, 2013ApJ...768..133S,
  2016MNRAS.455L..62N, 2020ApJ...898L..26M}, these simulations are extremely
costly and therefore cannot be propagated in time over millions of years.
Furthermore, the complexity of these 3-D models can make it difficult to gain
physical and mathematical insight into the mechanisms responsible for the
observed dynamics. Simple 1-D models of interacting concentric rings remain
therefore an important tool for the study of warped discs.

The equations for warped discs in the interacting concentric rings approach have
been formulated in several papers including e.g. \citet{1999MNRAS.304..557O} and
\citet{2000ApJ...538..326L}. In these papers the equations were derived using a
higher-order perturbation theory approach, leading to equations that showed the
dynamic nature of the internal torque vector $\GGbig{}$ and the wavelike nature
of the propagation of a warp (bending waves). For very viscous discs, however,
$\GGbig{}$ loses its dynamic nature, and will instead be purely a function of the
local conditions of the disc. The warp then propagates as a diffusive mode, with
the torque vector $\GGbig{}$ acting to damp out the warp and viscously transport
mass.  The expressions for $\GGbig{}$ as a function of the local conditions in
the disc were derived by \citet{2013MNRAS.433.2403O} by introducing a local
shearing box formulation of the disc hydrodynamics. In contrast to the higher
order perturbation analysis method, this approach does not yield the global disc
equations: Only the $\GGbig{}$ vector as a function of the local conditions is
obtained. But the advantage is that it is much more straightforward to extend
the local warped shearing box formulation into the highly non-linear regime.
Moreover, it is more intuitive than the perturbation analysis approach, since
it directly solves for the motions of the local fluid variables.
Although the two methods
yield mutually compatible results in the relevant limits, the relation between
the two is not fully clarified. As a consequence, the time-dependent evolution
equations for the interacting concentric rings model for the two regimes
(low-viscosity and high-viscosity regime) are somewhat disjunct.

\citet{2019ApJ...875....5M} introduced a generalized set of equations for the
interacting concentric rings model, which bridges the gap between the two
regimes. In their set of equations the dynamical nature of $\GGbig{}$
automatically appears for low viscosity, and automatically vanishes for high
viscosity. The equations for both limiting cases are reproduced. In addition,
they add two damping terms proportional to a parameter they call $\beta$,
which are necessary to eliminate an unphysical and spurious behavior of
the viscous evolution of the surface density $\Sigma(r,t)$ of the disc.

It is the purpose of this paper to derive a unified set of equations directly
from the warped shearing box model of \citet{2013MNRAS.433.2403O}, and through
this, obtain a clearer picture of how the wavelike and diffusive regimes are
related. We show that they are in agreement with the limiting cases of
\citet{1999MNRAS.304..557O}, \citet{2000ApJ...538..326L} and
\citet{2013MNRAS.433.2403O}, and that the general case agrees with
\citet{2019ApJ...875....5M}, with the exception of Martin's $\beta$ terms.
We elucidate the role of Martin's $\beta$ terms,
and introduce an alternative way to eliminate the unphysical behavior of the
unmodified equations.

In addition, an analytical theory of the nature of the gas motions
in a warped disc can be used as a starting point for further investigations
of physical processes occurring inside of warped discs, such as hydrodynamic
instabilities, the physics of dust in these discs, and the interaction of
the warped disc with planetesimal or planetary objects that have formed
in them.

\section{Preview}
Since the derivations to come are somewhat lengthy, we start with a preview of
our approach. Consider two neighboring disc annuli, A and B, which are slightly
inclined with respect to each other.  Annulus A is the inner one, B the outer
one. As a convention we define the unit vector perpendicular to annulus A to be
along the $Z$-axis: $\llunit{}_A=(0,0,1)$. That of anulus $B$ is
$\llunit{}_B=(\epsilon,0,1-\tfrac{1}{2}\epsilon^2)$ to second order in
$\epsilon$, for a small positive value of $\epsilon$. Annulus B is therefore
tilted in positive $X$-direction with respect to A. We assume the orbital motion
to be counter-clockwise, when viewed in the $(X,Y)$-plane, and we define the
azimuthal angle $\phi$ such that $\phi=0$ lies on the $Y=0$, $X>0$ plane and
increases in the direction of the orbital motion.

The question now is: How do these two annuli affect each other's orbital
orientation ($\llunit{}_A$ and $\llunit{}_B$)? 

At first glance one may be tempted to compute the out-of-plane component of the
pressure force between the two annuli. This force is maximal at $\phi=0$ and
$\phi=\pi$ as these are the locations where the two annuli are maximally
vertically offset from each other. This pressure force would lead to a torque
that annulus A exerts on annulus B that lies in the $Y$-direction,
i.e.\ perpendicular to both $\llunit{}_A$ and $\llunit{}_B$. The opposite torque
acts on annulus A. As a consequence, both annuli would start precessing around
their mean angular momentum axis. However, a more detailed calculation would show
that this precession only happens under special circumstances, while in most
circumstances it is only a very minor effect.

On ``second glance'' one may be tempted to compute the viscous friction force
between the two annuli as they switch sides (near $\phi=\pi/2$ and
$\phi=3\pi/2$). During the passage near $\pi/2$ the gas of annulus B moves
upward with respect to the gas of annulus A. Shear viscosity thus exerts a
torque on annulus B that lies along the $X$-axis and points in negative $X$
direction, while the opposite torque acts on annulus A. As a result, the
orientation vectors of two annuli $\llunit{}_A$ and $\llunit{}_B$ approach each
other: They align to each other and their mutual inclination damps
out. While this picture is correct, it turns out that this torque plays only a
minor role compared to the torque originating from oscillatory motions in the
gas \citep{1983MNRAS.202.1181P}.

\begin{figure}
  \centerline{\includegraphics[width=0.35\textwidth]{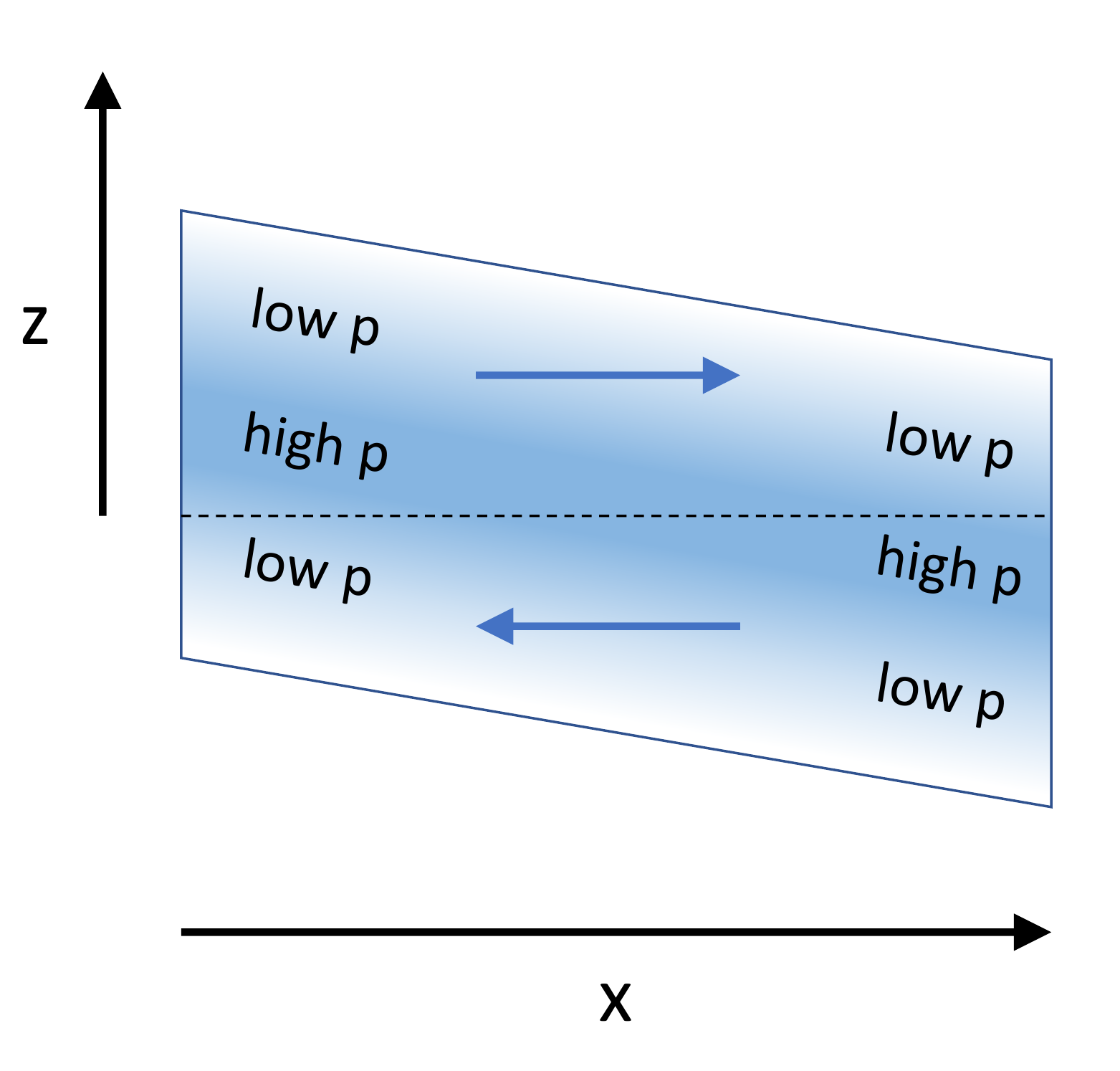}}
  \caption{\label{fig-cartoon-pressure-gradient} A cartoon of how a warp leads
    to horizontal pressure gradients, which lead to the ``sloshing motion'' of
    gas in the disc (blue horizontal arrows).  At the inclined midplane, the gas
    pressure is the highest ('high p' in the figure), while in the disc
    atmosphere it is lower ('low p' in the figure). The coordinates $x$ and $z$
    are the local coordinates used for the shearing box in this paper.  This
    cartoon was inspired by the cartoon in Figure 5 of
    \citet{2013MNRAS.433.2403O}.  }
\end{figure}

It was shown by \citet{1983MNRAS.202.1181P} that the oscillating horizontal pressure
gradients produced by the oscillating vertical offset between adjacent
annuli (see Fig.~\ref{fig-cartoon-pressure-gradient}) leads to strong horizontal
epicyclic motions in the disc with an amplitude proportional to the distance
$z'$ from the disc's midplane. These oscillations, in turn, produce an
orbit-averaged torque that completely dominates any of the viscous
torques. These motions are similar to the ``sloshing motion'' of a layer of
water on a tray that undergoes an oscillating tilt. {Although these
  are usually referred to as ``resonant motions'' in the literature, we will call them}
``sloshing motions'' from here on. The key to understanding the internal torque
in the disc is therefore to understand the behavior of these sloshing motions.

\citet{1983MNRAS.202.1181P} and \citet{1995ApJ...438..841P} showed that for
viscosities $\alpha_t>h_p/r$ (where $h_p$ is the pressure scale height of the
disc and $\alpha_t$ the usual viscosity parameter) the global behavior of a disc
warp is to damp out in a diffusive manner, while for lower viscosity the warp
propagates as a wave. Using an asymptotic expansion method,
\citet{1999MNRAS.304..557O} and \citet{2000ApJ...538..326L} derived, from first
principles, self-consistent equations for these two regimes. In the diffusive
regime ($\alpha_t>h_p/r$), the sloshing motion is, at all times, in a local steady state of
oscillation that depends only on the local conditions and the local warp
amplitude. The resulting internal torque vector $\GGbig{}$ can therefore be
computed uniquely from these local conditions. In the wave-like regime ($\alpha_t<h_p/r$), the
sloshing motion never finds the time to reach a local oscillatory steady-state,
because the disc geometry changes faster than this steady state can be
reached. The resulting internal torque vector $\GGbig{}$ therefore becomes a
dynamic quantity. The local conditions do not determine the torque, but only
determine its time derivative. As a result, a warp propagates as a wave.

Finding a set of equations that is valid in both regimes, and also self-consistently
includes the radial viscous mass transport in the disk, requires an
understanding of how the sloshing motions and the resulting internal torque
behave when the time scale for the sloshing motion to reach a local steady-state
oscillation is similar to the time scale by which the disc geometry itself
changes. \citet{2019ApJ...875....5M} have done this empirically by starting from
the time-dependent equation for the internal torque from the wavelike regime,
and adding terms such that, for sufficiently large $\alpha_t$, the resulting
asymptotic torque becomes equal to the one derived for the diffusive regime.

The aim of our paper is to derive the time-dependent equations for $\GGbig{}$,
valid in both the diffusive and the wavelike regime, from first principles, by
studying the sloshing motion itself. To this end we will zoom in to a local
annulus of the disc, and employ the warped shearing box framework of
\citet{2013MNRAS.433.2403O} to derive these equations. The resulting equations
{(Eqs.~\ref{eq-eom-slosh-cmplx-vx}, \ref{eq-eom-slosh-cmplx-vy})}
are those of a driven and damped harmonic oscillator. When initiated with a given initial
condition, the oscillation evolves, and eventually reaches a steady-state
oscillation {(Eqs.~\ref{eq-part-sol-vx}, \ref{eq-part-sol-vy},}
provided the local warp does not change). The steady-state
oscillation solutions were described in \citet{2013MNRAS.433.2403O}. The key to
finding the link between the diffusive and wavelike regime lies in including the
dynamics that can occur before this steady state solution is reached
{(Eqs.~\ref{eq-hom-sol-simple}, \ref{eq-freq-hom-sol})}. For large
amplitudes of the sloshing motion the equations become non-linear, which would
require a numerical treatment. For sufficiently small amplitudes, however, the
linear set of equations allow the solution to be written as a steady-state
particular solution plus a transient homogeneous solution {(Eqs.~\ref{eq-sum-xp-xh},
\ref{eq-sum-yp-yh})}. The steady-state
particular solution is identical to the solution described in
\citet{2013MNRAS.433.2403O}. The transient homogeneous solution describes how
the sloshing motion approaches this particular solution for any given initial
condition. {After transforming these solutions to the lab frame
  (Eqs.~\ref{eq-Vx-linear-full-tau-phi}, \ref{eq-Vy-linear-full-tau-phi}),
  we derive the resulting internal torque vector components
  (Eqs.~\ref{eq-ggbig-re-x}, \ref{eq-ggbig-re-y}).}

Given that the decay of the homogeneous solution takes often much more time than
the change in the disc geometry, we cannot just use this description of the
sloshing motion and the resulting internal torque vector. Instead, we cast this
behavior of the internal torque vector into a local ordinary differential
equation {(Eq.~\ref{eq-vect-form-of-G-hom-to-part})}. This leads to
the {first version of the} generalized warped disc equations we propose in this
paper {(Eq.~\ref{eq-vect-form-of-G-hom-to-part-alt-Qtilde}, together
  with the equations in Section \ref{sec-global-equations}, valid for
  small warps)}.
We will compare our equations to those in the literature in the
appropriate limits (diffusion limit and wavelike limit) and find general
agreement.

However, in agreement with \citet{2019ApJ...875....5M}, we find that the full
set of equations display a spurious behavior in the viscous evolution of the
surface density $\Sigma(r,t)$ of the disc, even in regions of the disc where the
warp wave has already passed. The cause of this behavior lies in the fact that
our equations, and those in the literature, do not account for what happens when
the orbital plane of the disc annulus changes its orientation (which will
doubtlessly happen as a result of the torques themselves, and possibly due to an
external torque as well). The corresponding correction terms to the equations
cannot be readily derived from the shearing box analysis, since in that analysis
the box is kept at a fixed orientation. Instead, we argue that the internal
torque vector co-rotates along with any rotation of the orientation vector,
because otherwise the torque vector will become unphysical. We propose a
rotation-inducing correction term to our equations and demonstrate that this
yields a physically correct behavior of the combined evolution of the warp and
the surface density. {This leads to the second and final version of
  our proposed generalized equation for the internal torque, valid for
  small warps (Eq.~\ref{eq-vect-form-of-Gslosh-with-rot}).}

\citet{2019ApJ...875....5M} follow a different approach: they use damping terms
to damp away the unphysical parts of the internal torque vector. We compare our
approach to theirs and show that both approaches lead to compatible results,
though our approach results in a less stiff set of equations and does not
require a free tuning parameter such as the $\beta$ parameter of
\citet{2019ApJ...875....5M}.

Finally, we will show how our set of equations can be intuitively interpreted
using the affine tilted slab picture of \citet{2018MNRAS.477.1744O}, thereby
resolving the apparent ``first glance misconception'' mentioned above.

\section{Setting the scene: Global conservation equations}
\label{sec-global-equations}
{
Before zooming in onto the shearing box, it is useful to recall the global
conservation laws that govern the evolution of the disc. Mass conservation is
given by the following partial differential equation:
\begin{equation}\label{eq-mass-cons}
\frac{\partial\Sigma}{\partial t} + \frac{1}{r}\frac{\partial}{\partial r}\big(r\Sigma v_r\big) = 0
\end{equation}
where $v_r$ is the radial velocity of the gas and $\Sigma$ is the surface
density. Angular momentum conservation is a vector-valued partial differential
equation. Define the angular momentum per unit
surface area
\begin{equation}\label{eq-def-llbig}
\LLbig(r,t) \equiv \Sigma(r,t)\, \Omega(r)\, r^2 \llunit{}(r,t)
\end{equation}
where $\Omega(r)$ is the orbital angular frequency at radius $r$, and
$\llunit{}(r,t)$ is the unit vector perpendicular to the disk annulus of radius
$r$. The vector $\llunit{}(r,t)$ is a dynamic quantity describing the warp geometry
and its evolution.
Angular momentum conservation is now given by
\begin{equation}\label{eq-angmom-cons}
  \frac{\partial\LLbig}{\partial t} + \frac{1}{r}\frac{\partial}{\partial r}\big(
  r\LLbig{}v_r+r\GGbig{}\big) = \TTbig{}
\end{equation}
where $\GGbig{}$ is the internal torque vector and $\TTbig{}$ is a possible external
torque. Although Eq.~(\ref{eq-angmom-cons}) is formulated in terms of $\LLbig$, 
it is, actually, the equation of motion for $\llunit{}(r,t)$. 
The two conservation equations can be combined to find an expression for
the radial velocity $v_r$ \citep{2019ApJ...875....5M}:
\begin{equation}\label{eq-vr-formula}
v_r = -\frac{\partial(r\GGbig)/\partial r\cdot\llunit}{r\Sigma\,\partial(\Omega r^2)/\partial r}
\end{equation}
The computation of the internal torque vector $\GGbig{}$ is the subject of this
paper, which we will do by studying the disk with a shearing box analysis. We
will show that $\GGbig{}$ is governed by Eqs.~(\ref{eq-split-G-into-s-and-v},
\ref{eq-G-visc-coordfree}, \ref{eq-vect-form-of-Gslosh-with-rot-limit}). Some
readers may be more familiar with another form of the warped disk equations. We
will discuss the relation between these two forms in Appendix
\ref{app-global-warp-eqs-alternative}.  }

\section{Local warped shearing box equations}
\subsection{Basics}
\label{sec-local-warped-shearing-box}
\citet{2013MNRAS.433.2403O} presented the ``warped shearing box'' approach
for studying the local internal dynamics of the gas in a warped disc. We will
largely follow their path, with only minor modifications of notation. We refer
to that paper for an introduction to the concepts of this approach. 

In classical viscous disc theory, the disc is described as a continuous set of
concentric annuli as a function of radius $r$. The mass distribution is
described by the surface density $\Sigma(r)$, and viscous disc theory describes
how this function changes with time: $\Sigma(r,t)$. In a warped disc, also the
inclination is dependent on radial coordinate. Let $\llunit{}(r)$ be the unit
vector perpendicular to the disc annulus at radius $r$, then a non-zero
$d\llunit{}/dr$ is what is called a warp. Let us define the warp vector
as
\begin{equation}\label{eq-def-psivec-warp}
\warpvector(r) = \frac{d\llunit(r)}{d\ln r}
\end{equation}
and the warp amplitude \cite[][]{1999MNRAS.304..557O} as
\begin{equation}\label{eq-def-psi-warp}
\psi(r) = \left|\warpvector(r)\right|
\end{equation}

\begin{figure}
  \centerline{\includegraphics[width=0.45\textwidth]{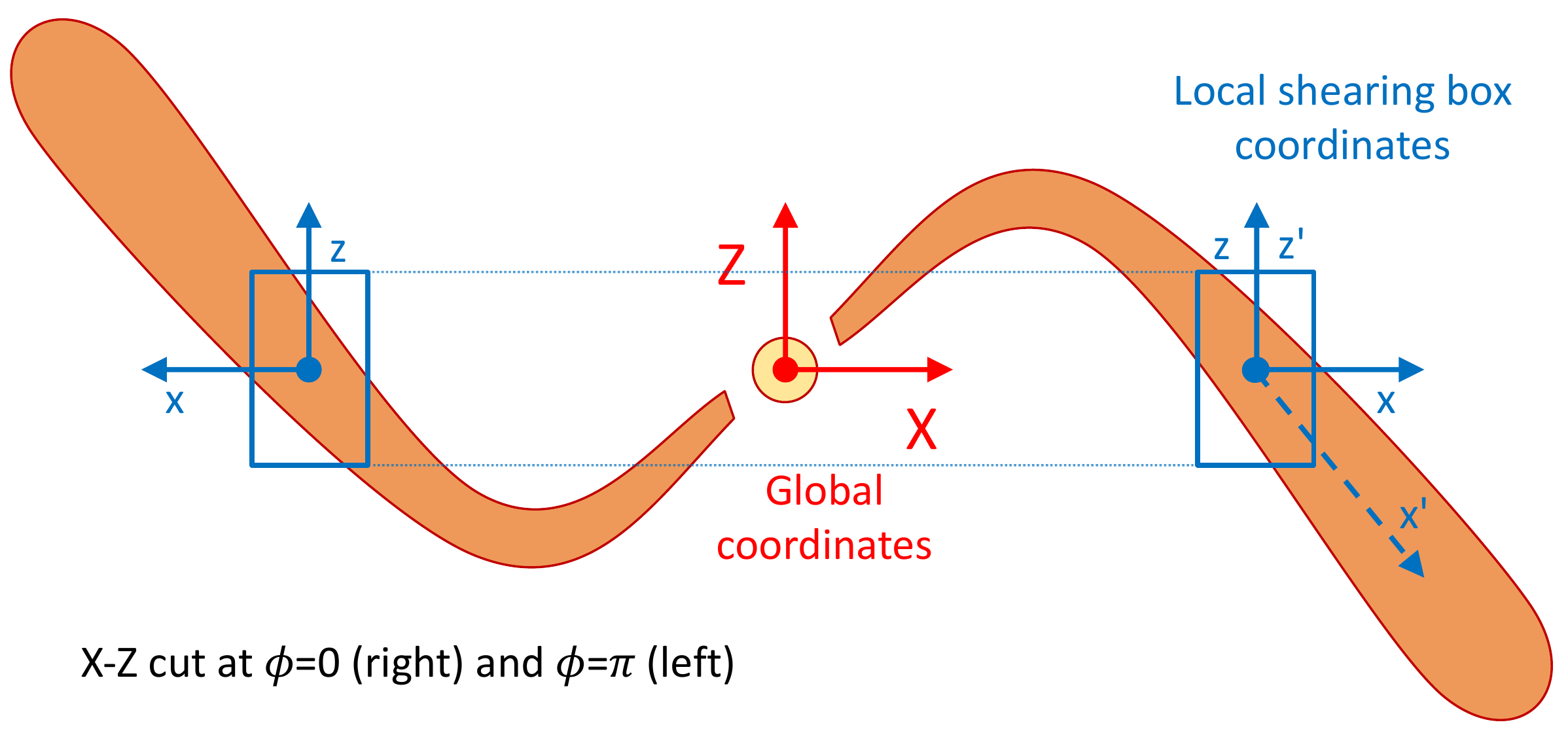}}
  \caption{\label{fig-cartoon-warp-coordinates} {A cartoon of the geometry of the
    warped disk and coordinate systems used, shown as a vertical cut through the
    disk. Here only the $\phi=0$, $\phi=\pi$ (i.e.~the $X-Z$) plane is shown. The
    global coordinates $(X,Y,Z)$ are, for the derivation of the equations, chosen
    such that $d\llunit/d\ln r$ points in $X$-direction.}}
\end{figure}

The viscous evolution of a warped disc describes the time-dependence of
$\Sigma(r,t)$ and $\llunit{}(r,t)$ for a given initial condition. Since this
evolution is driven by the conservation and transport of angular momentum
through the disc (see Section \ref{sec-global-equations}), we need to derive
equations for the internal torque $\GGbig{}(r,t)$. This is where the shearing
box model comes in. To apply the shearing box framework to a warped disc we
choose a radius $r_0$ and define the global laboratory frame $(X,Y,Z)$
coordinate system such that $\llunit{}(r_0)$ points in the $Z$-direction, and
that the warp vector $d\llunit{}/d\ln r$ points into positive $X$ direction {(see
Fig.~\ref{fig-cartoon-warp-coordinates}). Note that 
Fig.~1 of \citet{2013MNRAS.433.2403O} gives a 3-D illustration of this geometry},
where their ${\bf m}$ vector is the unit vector in $X$-direction, ${\bf e}_X$,
and their ${\bf n}$ vector is the unit vector in $Y$-direction, ${\bf
  e}_Y$. Along the circular orbit at $r=r_0$ we define the azumuthal coordinate
$\phi$ counter-clockwise, with $\phi=0$ at the positive $X-$axis,
i.e.~$X(r=r_0,\phi)=r_0\cos(\phi)$ and $Y(r=r_0,\phi)=r_0\sin(\phi)$. The gas
rotates in the direction of increasing $\phi$ with an orbital angular frequency
$\Omega_0=\Omega(r=r_0)$. As \citet{2013MNRAS.433.2403O} we define $q$ as
\begin{equation}\label{eq-definition-q}
q=-\frac{d\ln\Omega}{d\ln r}
\end{equation}
which, for perfectly Keplerian orbits, is $q=3/2$. However, in realistic discs,
$q$ can deviate slightly from $3/2$. In protoplanetary discs this is due to the
radial pressure gradient, or external influences, such as a binary companion
\citep[e.g.][]{2018MNRAS.477.5207Z}, or magnetic
torques \citep[e.g.][]{1999ApJ...524.1030L, 2003ApJ...591L.119L}.

\subsection{Unwarped shearing box coordinates}\label{sec-flat-shearing-box}
In the classical shearing box framework we follow the orbital motion of the gas
near radius $r_0$ and define local coordinates $(x,y,z)$ such that the origin
comoves along the orbit at $r=r_0$ at orbital angular frequency
$\Omega_0=\Omega(r=r_0)$, and rotates such that $x$ always points outwards, $y$
always stays tangent to the orbit, and $z$ always points upward (parallel to
$\llunit{}(r=r_0)$). The velocity components $(u_x,u_y,u_z)$ are defined as the
comoving time derivatives of the location of a test particle or fluid parcel in
this local coordinate system: $u_x=D_tx(t)$, $u_y=D_ty(t)$, $u_z=D_tz(t)$.

In this local coordinate system the equations of motion of a test particle
or a fluid parcel are:
\begin{eqnarray}
  D_t x &=& u_x\label{eq-eom-standard-box-dt-x}\\
  D_t y &=& u_y\label{eq-eom-standard-box-dt-y}\\
  D_t z &=& u_z\label{eq-eom-standard-box-dt-z}\\
  D_t u_x - 2\Omega_0 u_y &=& f_{x} + 2q\Omega_0^2x \label{eq-eom-standard-box-dt-ux}\\
  D_t u_y + 2\Omega_0 u_x &=& f_{y} \label{eq-eom-standard-box-dt-uy}\\
  D_t u_z &=& f_{z}-\Omega_0^2z\label{eq-eom-standard-box-dt-uz}
\end{eqnarray}
where $f_i$ (with $i=x,y,z$) are the forces per unit mass acting on the
test particle or fluid parcel. In case of a fluid parcel, these forces include
the pressure gradient force and the viscous forces, which we will discuss later,
plus external forces, if present. The $2q\Omega_0^2x$ term in
Eq.~(\ref{eq-eom-standard-box-dt-ux}) is the sum of the outward-pointing
centrifugal force and the inward-pointing gravitational force, which, at $x=0$,
are in perfect balance.  The second terms on the left-hand-side of
Eqs.~(\ref{eq-eom-standard-box-dt-ux}, \ref{eq-eom-standard-box-dt-uy}) are the
Coriolis forces. The $-\Omega_0^2z$ term in
Eq.~(\ref{eq-eom-standard-box-dt-uz}) is the vertical component of the
gravitational force.

In the simple case of zero viscosity and no external forcing, only the
gradient of the gas pressure $p$ enters in the $f_i$ terms (let us call
these $f^p_i$):
\begin{equation}
  f_x^p = \rho^{-1}\partial_xp, \qquad
  f_y^p = \rho^{-1}\partial_yp, \qquad
  f_z^p = \rho^{-1}\partial_zp\label{eq-pressure-force}
\end{equation}
where $\rho$ is the gas density.
For a non-warped laminar disc one can set $\partial_yp=0$ and $\partial_xp=0$ (a
global radial pressure gradient cannot be consistently included in a local shearing box
model, except as a pseudo-force). A simple solution is then $u_x=0$,
$u_y=-q\Omega_0x$ and $u_z=0$. What remains is to solve for the vertical
density structure from Eq.~(\ref{eq-eom-standard-box-dt-uz}):
\begin{equation}
\frac{1}{\rho(z)}\frac{\partial p(z)}{\partial z} = -\Omega_0^2z
\end{equation}
For the simple isothermal case we set $p=\rho c_s^2$ with the isothermal
sound speed $c_s$ set to a constant. The solution is then
\begin{equation}\label{eq-simple-gaussian-vertical-structure}
\rho(z) = \frac{\Sigma}{\sqrt{2\pi}h_p}\exp\left(-\frac{z^2}{2h_p^2}\right)
\end{equation}
with $\Sigma$ the surface density, and $h_p=c_s/\Omega_0$ the pressure
scale height.

\subsection{Warped shearing box coordinates}\label{sec-warped-shearing-box}
{For a warped disk the geometry is pictographically shown in Fig.~\ref{fig-cartoon-warp-coordinates},
where the amplitude of the warp has been exaggerated for clarity.}
The warped version of the shearing box framework of \citet{2013MNRAS.433.2403O} introduces
``warped local coordinates'' $(x',y',z')$ which adjust themselves to the
warped geometry:
\begin{eqnarray}
  x   &=& x' \label{eq-def-xprime}\\
  y   &=& y' \label{eq-def-yprime}\\
  z   &=& z' - \psi x' \cos(\phi) \label{eq-def-zprime}
\end{eqnarray}
The $z'$ coordinate thus follows the up-and-down oscillation of the gas at
$x'\neq 0$ as a result of the disc warp. The $(x',z')$ coordinate geometry is
shown {in Fig.~\ref{fig-cartoon-warp-coordinates} on the right side of
the figure. It is also nicely visualized}
in the right panel of figure 4 of \citet{2013MNRAS.433.2403O}. However, as
one can see from Eq.~(\ref{eq-def-yprime}), in constrast to
\citet{2013MNRAS.433.2403O}, we do not modify $y'$ to follow the azimuthal
shearing motion of the gas in the disc because it would lead to an incessant
``winding up'' of the $y'$ coordinate. Therefore, the left panel of figure 4 of
\citet{2013MNRAS.433.2403O} does not apply to our $(x',y')$ geometry.

The new vertical coordinate $z'$ is the vertical coordinate with respect to the
midplane of the warped disc. Therefore, in first approximation, the vertical
density structure for the warped disc can thus be described by
Eq.~(\ref{eq-simple-gaussian-vertical-structure}) with $z$ replaced by $z'$.

As the fluid parcel orbits around the star, the azimuth will change according to
$\phi=\Omega_0t$, where the zero time $t=0$ is chosen to be when the parcel was
at $\phi=0$. 

In the warped shearing box coordinates one can define the velocity components in
a similar way as in the unwarped case: $u_x'=D_tx'(t)$, $u_y'=D_ty'(t)$,
$u_z'=D_tz'(t)$. However, it is useful to define new velocities $(\vx,\vy,\vz)$
such that
\begin{equation}
  u_x'=\vx, \quad u_y'=\vy-q\Omega_0x',\quad
  u_z'=\vz+\psi\cos(\phi)\vx
\end{equation}
with $\phi=\Omega_0t$ for the fluid parcel we follow. In relation to the
unwarped coordinate velocities we then have:
\begin{eqnarray}
  u_x &=& \vx\label{eq-def-vxprime}\\
  u_y &=& \vy - q\Omega_0 x'\label{eq-def-vyprime}\\
  u_z &=& \vz + \psi \Omega_0 x' \sin(\phi)\label{eq-def-vzprime}
\end{eqnarray}
where we used Eq.~(\ref{eq-rel-u-uprime-z}) of Appendix
\ref{app-derivations-unwarped-warped}.  The advantage of the velocities
$(\vx,\vy,\vz)$ compared to $(u_x',u_y',u_z')$ is that they are orthogonal
velocities, in spite of the skewed coordinate system $(x',y',z')$. We call this
a ``half-mixed frame'' because it is mixed-frame in $y$-direction but fully
comoving frame in $z$-direction (and in $x$-direction it remains fixed to the
shearing box).

In terms of these ``half-mixed frame'' warped coordinates and velocities,
the equations of motion of a test particle or a fluid parcel are (see Appendix
\ref{app-derivations-unwarped-warped}):
\begin{eqnarray}
  D_t x' &=& \vx\label{eq-eom-warped-box-dt-x}\\
  D_t y' &=& \vy - q\Omega_0 x'\label{eq-eom-warped-box-dt-y}\\
  D_t z' &=& \vz + \psi \vx\cos(\phi)\label{eq-eom-warped-box-dt-z}\\
  D_t \vx - 2\Omega_0 \vy &=& f_{x} \label{eq-eom-warped-box-dt-vx}\\
  D_t \vy + (2-q)\Omega_0 \vx &=&  f_{y} \label{eq-eom-warped-box-dt-vy}\\
  D_t \vz + \psi\Omega_0\sin(\phi) \vx &=& f_{z} -\Omega_0^2z' \label{eq-eom-warped-box-dt-vz}
\end{eqnarray}
For the fluid dynamics we also need the continuity equation:
\begin{equation}
D_t\rho = -\rho \nabla\cdot {\bf u}
\end{equation}
which can, with the help of Eq.~(\ref{eq-div-u-warped}), be written in warped
coordinates as:
\begin{equation}\label{eq-eom-warped-box-dt-rho}
D_t\ln\rho = -(\partial_{x'} + \psi\cos(\phi)\partial_{z'})\vx - \partial_{y'} \vy - \partial_{z'} \vz
\end{equation}

Next we turn to the forces $f_i$ (with $i=x,y,z$), which consist of the pressure
gradient force $f_i^p$, the shear viscosity force $f_i^v$, and possibly an
external force $f_i^e$:
\begin{equation}\label{eq-all-body-forces}
f_i = f_i^p + f_i^v + f_i^e
\end{equation}
The pressure gradient force components in warped coordinates become (see
Eq.~\ref{eq-partial-x-in-prime}):
\begin{eqnarray}
f_x^p &=& -(1/\rho)(\partial_{x'}+\psi\cos(\phi)\partial_{z'}) p \label{eq-fp-x-prime}\\
f_y^p &=& -(1/\rho)\partial_{y'} p  \label{eq-fp-y-prime}\\
f_z^p &=& -(1/\rho)\partial_{z'} p  \label{eq-fp-z-prime}
\end{eqnarray}

The expressions for the shear viscosity forces $f^v_i$ in
warped coordinates are complex and their derivation cumbersome, so we defer this
to Appendix \ref{app-f-shearvisc}.

The equations of this section are the fluid equations in Lagrange form. The
comoving time derivative $D_t$ in all the above equations can be written as
$\partial_{t'}$ plus partial derivatives in $x'$, $y'$ and $z'$ using
Eq.~(\ref{eq-warped-comoving-derivative}), yielding the equations in the
comoving laboratory frame (defined as the $x'$, $y'$ and $z'$ warped coordinate
system).  This forms a set of coupled partial differential equations for the
motion of the fluid (gas) in the warped shearing box.

\subsection{Dimensionless time}
For the following analysis it will be convenient to scale all the equations
to a dimensionless time defined by
\begin{equation}\label{eq-tau-t}
\tau=\Omega_0t
\end{equation}
It is no coincidence that for a particle or fluid parcel $\tau=\phi$, as
the dimensionless time corresponds to the location along the orbit. In the
following we will use $\tau$ when we follow the fluid parcel along its orbit,
and $\phi$ when we put emphasis on the geometric location along the orbit.
We can then write
\begin{equation}\label{eq-Dtau-in-Dt}
D_t = \Omega_0D_\tau
\end{equation}
Equations (\ref{eq-eom-warped-box-dt-rho}, \ref{eq-eom-warped-box-dt-vx},
\ref{eq-eom-warped-box-dt-vy}, \ref{eq-eom-warped-box-dt-vz}) then become:
\begin{eqnarray}
  \Omega_0D_\tau\ln\rho &=& -(\partial_{x'} + \psi\cos(\phi)\partial_{z'})\vx\nonumber\\ & & - \partial_{y'} \vy - \partial_{z'} \vz\label{eq-eom-warped-box-dtau-rho}\\
  D_\tau \vx - 2 \vy &=& \Omega_0^{-1}f_{x} \label{eq-eom-warped-box-dtau-vx}\\
  D_\tau \vy + (2-q) \vx &=&  \Omega_0^{-1}f_{y} \label{eq-eom-warped-box-dtau-vy}\\
  D_\tau \vz + \psi\sin(\phi) \vx &=& \Omega_0^{-1}f_{z} -\Omega_0z' \label{eq-eom-warped-box-dtau-vz}
\end{eqnarray}

\subsection{Equations for laminar solutions}\label{sec-eqs-laminar}
Although the equations derived so far are valid for general flows in the warped
shearing box framework, in the remainder of this paper we are concerned with
laminar solutions that are locally translationally symmetric in $x'$ and
$y'$. This allows an analytic treatment.  It should, however, be kept in mind
that in making this assumption (and in fact, by using the shearing box approach
in the first place) we are rejecting potentially important physics, which may
affect the outcome (see further discussion in Section
\ref{sec-caveats-shearing-box}).

The assumption of translational symmetry in $x'$ and $y'$ has the simplifying
consequence that $\partial_{x'}=0$ and $\partial_{y'}=0$. Furthermore we can
safely set $x'=0$ and $y'=0$. This already removes a number of terms in the
equations.

Next we make the simplifying assumption of a vertically isothermal equation of
state, as we did at the end of Subsection \ref{sec-flat-shearing-box}. The
vertical density structure is then identical to the Gaussian solution of
Eq.~(\ref{eq-simple-gaussian-vertical-structure}), but with $z$ replaced by
$z'$:
\begin{equation}\label{eq-simple-gaussian-vertical-structure-warped}
\rho(z',\tau) = \frac{\Sigma}{\sqrt{2\pi}h_p(\tau)}\exp\left(-\frac{(z')^2}{2h_p(\tau)^2}\right)
\end{equation}
In fact, if we make the additional assumption that all velocities $\vx$, $\vy$
and $\vz$ are zero at $z'=0$ and linearly proportional to $z'$, then the
Gaussian structure of Eq.~(\ref{eq-simple-gaussian-vertical-structure-warped})
remains valid even while $\partial_{z'}\vz\neq 0$, i.e.~during vertical
compression or expansion. The time-dependence of the entire vertical
density profile can then be described by the time-dependence of a single
parameter: the pressure scale height $h_p(\tau)$.

Following \citet{2013MNRAS.433.2403O} we therefore look for solutions
to the velocity variables of the form\footnote{For the symbol $V$ we omit the prime
to reduce notational cluttering.}:
\begin{eqnarray}
\vx(z',\tau) &=& \VVx(\tau)\Omega_0 z' \label{eq-vx-in-VOmz}\\
\vy(z',\tau) &=& \VVy(\tau)\Omega_0 z' \label{eq-vy-in-VOmz}\\
\vz(z',\tau) &=& \VVz(\tau)\Omega_0 z' \label{eq-vz-in-VOmz}
\end{eqnarray}
Inserting these into Eqs.~(\ref{eq-eom-warped-box-dtau-rho}-\ref{eq-eom-warped-box-dtau-vz})
yields:
\begin{eqnarray}
  D_\tau\ln\rho &=& -\psi\cos(\phi)\VVx - \VVz\label{eq-eom-warped-box-dtau-rho-laminar}\\
  \bar D_\tau \VVx - 2 \VVy &=& (\Omega_0^2z')^{-1}f_{x} \label{eq-eom-warped-box-dtau-vx-laminar}\\
  \bar D_\tau \VVy + (2-q) \VVx &=&  (\Omega_0^2z')^{-1}f_{y} \label{eq-eom-warped-box-dtau-vy-laminar}\\
  \bar D_\tau \VVz + \psi\sin(\phi) \VVx &=& (\Omega_0^2z')^{-1}f_{z} - 1 \label{eq-eom-warped-box-dtau-vz-laminar}
\end{eqnarray}
where $\bar D_\tau V_i$ (with $i=x,y,z$) is defined as
\begin{equation}\label{eq-def-bardtau}
\begin{split}
  \bar D_\tau V_i &\equiv D_\tau V_i + V_i\frac{1}{z'}D_\tau z'\\
  &=  \partial_\tau V_i + V_i\big(\VVz+\psi\cos(\phi)\VVx\big)
\end{split}
\end{equation}
where in the second term we used Eqs.~(\ref{eq-Dtau-in-Dt}, \ref{eq-warped-comoving-derivative}).

Clearly only forces are allowed that vanish at $z'=0$, for otherwise these
equations become singular. In fact, for solutions to exist, the $f_i$ forces
also have to be linear in $z'$. Indeed, this happens to be true automatically
for the pressure gradient force $f^p_i$ and the shear viscosity force $f^v_i$,
if one applies the Gaussian vertical structure of
Eq.~(\ref{eq-simple-gaussian-vertical-structure-warped}).
For the pressure gradient force $f^p_i$ we start with Eqs.~(\ref{eq-fp-x-prime}-\ref{eq-fp-z-prime})
and set $\partial_{x'}=0$ and $\partial_{y'}=0$ to obtain:
\begin{eqnarray}
f_x^p &=& -(1/\rho)\psi\cos(\phi)\partial_{z'} p \label{eq-fp-lam-x-prime}\\
f_y^p &=& 0 \label{eq-fp-lam-y-prime}\\
f_z^p &=& -(1/\rho)\partial_{z'} p  \label{eq-fp-lam-z-prime}
\end{eqnarray}
Now use Eq.~(\ref{eq-simple-gaussian-vertical-structure-warped}) for computing
the $\partial_{z'}p$ by setting $p=\rho c_s^2$ and keeping $c_s^2$ constant.
This yields
\begin{equation}
\partial_{z'}p = c_s^2 \partial_{z'}\rho = -\rho c_s^2 \frac{z'}{h_p^2}
\end{equation}
where $h_p(\tau)$ can be time-dependent. From the non-warped disc geometry
we know that the equilibrium value of $h_p$ is $c_s/\Omega_0$ (see
Section \ref{sec-flat-shearing-box}). So let us define the dimensionless
pressure scale height $H$ as
\begin{equation}\label{eq-def-dimless-H}
h_p(\tau) = H(\tau) \frac{c_s}{\Omega_0}
\end{equation}
This then leads to
\begin{equation}
\frac{1}{\rho}\partial_{z'}p = -\frac{\Omega_0^2}{H^2}\; z'
\end{equation}
which leads to the following expressions for the pressure gradient forces
\begin{eqnarray}
(\Omega_0^2z')^{-1}f_x^p &=& \psi\cos(\phi)H^{-2} \label{eq-fp-dimless-x-prime}\\
(\Omega_0^2z')^{-1}f_y^p &=& 0 \label{eq-fp-dimless-y-prime}\\
(\Omega_0^2z')^{-1}f_z^p &=& H^{-2} \label{eq-fp-dimless-z-prime}
\end{eqnarray}
Before we devote our attention to the viscosity forces, let us rewrite the
$D_\tau\ln\rho$ term in Eq.~(\ref{eq-eom-warped-box-dtau-rho-laminar}). Using
again the Gaussian vertical density structure of
Eq.~(\ref{eq-simple-gaussian-vertical-structure-warped}) and realizing that in
the comoving derivative the $(z')^2/h_p(\tau)^2$ inside the exponent stays
constant (the vertical structure shrinks or expands vertically in a self-similar
way), we find
\begin{equation}
D_\tau\ln\rho = -D_\tau\ln H = -\partial_\tau\ln H
\end{equation}
This allows us to write the equations for $\VVx(\tau)$, $\VVy(\tau)$,
$\VVz(\tau)$ and $H(\tau)$ (Eqs.~\ref{eq-eom-warped-box-dtau-rho-laminar}-\ref{eq-eom-warped-box-dtau-vz-laminar})
as
\begin{eqnarray}
  \partial_\tau\ln H &=& \psi\cos(\phi)\VVx + \VVz\label{eq-eom-warped-box-dtau-rho-laminar-withpres}\\
  \bar D_\tau \VVx - 2 \VVy &=& \psi\cos(\phi)H^{-2} + F^{\mathrm{ve}}_{x} \label{eq-eom-warped-box-dtau-vx-laminar-withpres}\\
  \bar D_\tau \VVy + (2-q) \VVx &=&  F^{\mathrm{ve}}_{y} \label{eq-eom-warped-box-dtau-vy-laminar-withpres}\\
  \bar D_\tau \VVz + \psi\sin(\phi) \VVx &=& H^{-2} - 1 + F^{\mathrm{ve}}_{z} \label{eq-eom-warped-box-dtau-vz-laminar-withpres}
\end{eqnarray}
where $F^{\mathrm{ve}}_i$ are the viscous and external forces in the form:
\begin{equation}\label{eq-bigf-ve}
F^{\mathrm{ve}}_i \equiv F^{\mathrm{v}}_i + F^{\mathrm{e}}_i \equiv (\Omega_0^2z')^{-1}(f^v_{i}+f^e_{i})
\end{equation}
Bulk viscosity can be included as a hysteresis factor in the pressure, dependent
on $\nabla\cdot {\bf u}$, but we will not include this in this analysis.

Note, incidently, that the right-hand-side of Eq.~(\ref{eq-eom-warped-box-dtau-rho-laminar-withpres})
happens to be the same as the term in brackets in Eq.~(\ref{eq-def-bardtau}). And so one can
write Eq.~(\ref{eq-def-bardtau}) as
\begin{equation}\label{eq-def-bardtau-alt}
  \bar D_\tau V_i 
  =  \partial_\tau V_i + V_i\,\partial_\tau\ln H
\end{equation}

Together with the expressions for the viscous forces $F^{\mathrm{v}}_i$ from
appendix \ref{app-f-shearvisc} (Eqs.~\ref{eq-Fvisc-x}-\ref{eq-Fvisc-z}), and any
possible external $F^{\mathrm{v}}_e$, the set of equations
Eqs.~(\ref{eq-eom-warped-box-dtau-rho-laminar-withpres}-\ref{eq-eom-warped-box-dtau-vz-laminar-withpres})
with $\phi=\tau$ is complete, and can be integrated in time $\tau$ for
any initial condition of $H$, $\VVx$, $\VVy$ and $\VVz$. A {recommended} way to
integrate these is by using a numerical integrator such as the {\sf
  solve\_ivp()} method from the {\sf scipy.integrate} library of {\sf Python},
which is a higher-order integration scheme that automatically adjusts step size
to control the error, and is easy to use \citep{2020SciPy-NMeth}.

\section{Solutions for the sloshing motion}

\subsection{Vertical and horizontal oscillations}
As pointed out by \citet{2013MNRAS.433.2403O}, for the non-warped case
($\psi=0$), and for zero viscosity, the equation set
Eqs.~(\ref{eq-eom-warped-box-dtau-rho-laminar-withpres}-\ref{eq-eom-warped-box-dtau-vz-laminar-withpres})
has two oscillating modes: A vertical oscillation (called
the ``breathing mode'' by \citet{2013MNRAS.433.2403O}) coupling
Eqs.~(\ref{eq-eom-warped-box-dtau-rho-laminar-withpres} and
\ref{eq-eom-warped-box-dtau-vz-laminar-withpres}), and a horizontal epicyclic
oscillation mode (which we call the ``sloshing motion'')
coupling Eqs.~(\ref{eq-eom-warped-box-dtau-vx-laminar-withpres}
and \ref{eq-eom-warped-box-dtau-vy-laminar-withpres}).

This can be seen a bit clearer if we linearize
Eqs.~(\ref{eq-eom-warped-box-dtau-rho-laminar-withpres}-\ref{eq-eom-warped-box-dtau-vz-laminar-withpres}).
Let us define an alternative variable to $H$:
\begin{equation}
H=e^\whp{}=1+\whp{}+{\cal O}(\whp{}^2)
\end{equation}
and assume $|\whp{}|\ll 1$. Now we remove all terms that are of second or higher
order in $(\whp{},\VVx,\VVy,\VVz)$. We arrive at:
\begin{eqnarray}
  \partial_\tau \whp{} &=& \psi\cos(\phi)\VVx + \VVz\label{eq-eom-warped-box-dtau-rho-laminar-withpres-lin}\\
  \partial_\tau \VVx - 2 \VVy &=& \psi\cos(\phi)(1-2\whp{}) + F^{\mathrm{ve}}_{x} \label{eq-eom-warped-box-dtau-vx-laminar-withpres-lin}\\
  \partial_\tau \VVy + (2-q) \VVx &=&  F^{\mathrm{ve}}_{y} \label{eq-eom-warped-box-dtau-vy-laminar-withpres-lin}\\
  \partial_\tau \VVz + \psi\sin(\phi) \VVx &=& -2\whp{} + F^{\mathrm{ve}}_{z} \label{eq-eom-warped-box-dtau-vz-laminar-withpres-lin}
\end{eqnarray}
For $\psi=0$ and $F^{\mathrm{ve}}_i=0$, the two modes decouple. The breathing
mode has a frequency $\Omega_b=\sqrt{2}\Omega_0$, while the sloshing mode
oscillates at the epicyclic frequency
\begin{equation}
\Omega_e = \sqrt{2(2-q)}\,\Omega_0
\end{equation}
It will be convenient for the remainder of this paper to define these (and
other) frequencies in units of the dimensionless time $\tau$. From here
onward we define the dimensionless breathing mode frequency
$\omega_b=\Omega_b/\Omega_0=\sqrt{2}$, and the dimensionless epicyclic frequency
$\kappa$ as
\begin{equation}\label{eq-epicyclic-freq-dimless}
\kappa = \Omega_e/\Omega_0 = \sqrt{2(2-q)}
\end{equation}
For an exactly keplerian disc, $q=3/2$, and therefore $\kappa=1$. 

If no bulk viscosity is included, the breathing mode remains undamped. In
practice it is likely that such modes will propagate as waves through the disc
in radial direction, but that cannot be described within the framework used
here. If no shear viscosity is included, the epicylic oscillation will also be
undamped. Also in this case it may be that the epicyclic oscillations of neighboring
annuli interact, but, again, this is outside of the scope of the present
framework.

For a warped disc, $\psi\neq 0$, the two modes couple, albeit only weakly if
$\psi\ll 1$. 


\subsection{Solutions to the linearized equations}\label{sec-sol-V-tau-linear}
The linearized equations
Eqs.~(\ref{eq-eom-warped-box-dtau-rho-laminar-withpres-lin}-\ref{eq-eom-warped-box-dtau-vz-laminar-withpres-lin})
allow simple analytic solutions if all terms proportional to the product of
$\psi$ with one of the variables $(\whp{},\VVx,\VVy,\VVz)$ are considered small
and are ignored. The equations then reduce to:
\begin{eqnarray}
  \partial_\tau \whp{} &=& \VVz\label{eq-eom-warped-box-dtau-rho-laminar-withpres-linn}\\
  \partial_\tau \VVx - 2 \VVy &=& \psi\cos(\phi) + F^{\mathrm{ve}}_{x} \label{eq-eom-warped-box-dtau-vx-laminar-withpres-linn}\\
  \partial_\tau \VVy + (2-q) \VVx &=&  F^{\mathrm{ve}}_{y} \label{eq-eom-warped-box-dtau-vy-laminar-withpres-linn}\\
  \partial_\tau \VVz  &=& -2\whp{} + F^{\mathrm{ve}}_{z} \label{eq-eom-warped-box-dtau-vz-laminar-withpres-linn}
\end{eqnarray}
The viscous forces (Eqs.~\ref{eq-Fvisc-x}-\ref{eq-Fvisc-z}) then reduce to:
\begin{eqnarray}
F_x^v &=& - \alpha_t \big(\VVx + \psi\sin(\phi) \big)\label{eq-Fvisc-x-linn} \\
F_y^v &=& - \alpha_t \big(\VVy  - q\psi\cos(\phi) \big) \label{eq-Fvisc-y-linn} \\
F_z^v &=& - \alpha_t \big(\tfrac{4}{3}\VVz + \psi^2\sin(\phi)\cos(\phi)\big) 
\end{eqnarray}
The removal of these $\psi\whp{}$ and $\psi
V_i$ terms has the consequence that the vertical and horizontal oscillations
decouple completely. Of relevance to the internal torque is only the oscillation
in $\VVx$ and $\VVy$: the sloshing motion. Let us set the external force
$F^{e}_i=0$, and insert $F^{v}_x$ and $F^{v}_y$ into
Eqs.~(\ref{eq-eom-warped-box-dtau-vx-laminar-withpres-linn},\ref{eq-eom-warped-box-dtau-vy-laminar-withpres-linn}):
\begin{eqnarray}
  \partial_\tau \VVx - 2 \VVy     \kern-0.5em &=& \kern-0.5em \psi(\cos(\phi)-\alpha_t\sin(\phi)) -\alpha_t \VVx \label{eq-eom-slosh-vx}\\
  \partial_\tau \VVy + (2-q) \VVx \kern-0.5em &=& \kern-0.5em \alpha_t q\psi\cos(\phi) -\alpha_t \VVy \label{eq-eom-slosh-vy}
\end{eqnarray}
where, again, we use\footnote{The reason why we do not immediately replace
  $\phi$ by $\tau$ in these equations will become clear in Subsection
  \ref{sec-phi-tau}.} $\phi=\tau$.
And so, after a long journey, we have arrived at two coupled linear ordinary
differential equations for the sloshing motion that can be solved analytically
for $\VVx(\tau)$ and $\VVy(\tau)$.

For this analytical treatment it is convenient to replace $\cos(\phi)$
with $e^{i\phi}$ and $\sin(\phi)$ with $-ie^{i\phi}$, solve for the complex
versions of $\VVx(\tau)$ and $\VVy(\tau)$, and then take the real part of
these. Eqs.~(\ref{eq-eom-slosh-vx},\ref{eq-eom-slosh-vy}) become:
\begin{eqnarray}
  \partial_\tau \VVx - 2 \VVy     \kern-0.5em &=& \kern-0.5em \psi(1+i\alpha_t)e^{i\phi} -\alpha_t \VVx \label{eq-eom-slosh-cmplx-vx}\\
  \partial_\tau \VVy + (2-q) \VVx \kern-0.5em &=& \kern-0.5em \alpha_t q\psi e^{i\phi} -\alpha_t \VVy \label{eq-eom-slosh-cmplx-vy}
\end{eqnarray}
We now seek solutions of the form
\begin{eqnarray}
\VVx(\tau) &=& V_{xp}(\tau) + V_{xh}(\tau)\label{eq-sum-xp-xh}\\
\VVy(\tau) &=& V_{yp}(\tau) + V_{yh}(\tau)\label{eq-sum-yp-yh}
\end{eqnarray}
where $V_{ip}(\tau)$ are the harmonic particular solution and $V_{ih}(\tau)$ are the
homogeneous solution. The particular solution can be written as
\begin{equation}
V_{xp}(\tau) = V_{xp0} e^{i\tau}, \qquad V_{yp}(\tau) = V_{yp0} e^{i\tau}
\end{equation}
with
\begin{eqnarray}
  V_{xp0} &=& \frac{\alpha_t(4-\kappa^2)+i(1+\alpha_t^2)}{\kappa^2+(i+\alpha_t)^2}\;\psi\label{eq-part-sol-vx}\\
  V_{yp0} &=& \frac{2\alpha_t(i+\alpha_t)-\tfrac{1}{2}\kappa^2(\alpha_t^2+2i\alpha_t+1)}{\kappa^2+(i+\alpha_t)^2}\;\psi \label{eq-part-sol-vy}
\end{eqnarray}
where $\kappa=\sqrt{2(2-q)}$. The homogeneous solution can be written
as
\begin{equation}
V_{xh}(\tau) = V_{xh0} e^{i\omega\tau}, \qquad V_{yh}(\tau) = V_{yh0} e^{i\omega\tau}\label{eq-hom-sol-simple}
\end{equation}
with
\begin{equation}\label{eq-freq-hom-sol}
\omega=\kappa+i\alpha_t
\end{equation}
The values of $V_{xh0}$ and $V_{yh0}$ are related to the initial conditions
$\VVx(\tau=0)$ and $\VVy(\tau=0)$ through
\begin{equation}
V_{ih0} = V_i(\tau=0) - V_{ip0}
\end{equation}

With this, we now have the complete family of solutions for the sloshing
oscillation in the linear regime for sufficiently small $\psi$ that the
$\psi\whp{}$ and $\psi V_i$ terms can be neglected. For non-small $\psi$, the
inclusion of these terms still keeps the problem linear in
$(\whp{},\VVx,\VVy,\VVz)$, but the solution will acquire higher-order modes, and
the problem will become substantially more difficult. In that case, as well as
in the case that the linear approximation becomes invalid, a numerical treatment
is preferable.

\subsection{The $(\tau,\phi)$ picture and the real time-dependence}
\label{sec-phi-tau}
So far we have looked at a parcel of gas in the azimuthal comoving frame, where
$\tau$ can be regarded as equivalent to azimuth $\phi$. However, if the solution
at $\tau=2\pi$ is not the same as the starting point at $\tau=0$, then this
equivalence of $\tau$ and $\phi$ is invalid. Furthermore, while we see
time-dependent behavior when moving along with a fluid parcel as it orbits
around the star, the behavior of the entire annulus, seen in the lab frame, may
be stationary or only slowly varying in time.

It is therefore better to look at the dynamics as a function of time $\tau$ {\em
  and} azimuthal angle $\phi$, i.e.\ $\VVx(\tau,\phi)$ and likewise for the other
quantities. We will now assume that the $\phi$-dependence is $e^{im\phi}$
at all times, and choose $m=1$, because a warp is by definition an $m=1$
mode. So we have
\begin{equation}\label{eq-V-tau-phi-with-expiphi}
\VVx(\tau,\phi) = \VVx(\tau)e^{i\phi}
\end{equation}
and likewise for the other quantities. This assumption is valid for the
linearized equations in which all terms proportional to $\psi\whp{}$ and $\psi
V_i$ are neglected
(i.e.~Eqs.~\ref{eq-eom-warped-box-dtau-rho-laminar-withpres-linn}-\ref{eq-eom-warped-box-dtau-vz-laminar-withpres-linn}).
And since in this case the vertical ``breathing'' and horizontal ``sloshing''
motions decouple, we will from here on focus only on the ``sloshing'' motion,
the solution of which was presented in Section \ref{sec-sol-V-tau-linear}.

In the $(\tau,\phi)$-picture, $\VVx(\tau)$ is a property of the entire $2\pi$
circumference of the annulus instead of a single fluid parcel. By definition
$\VVx(\tau)$ is the value of $\VVx(\tau,\phi)$ at dimensionless time $\tau$ and
azimuth $\phi=0$. The value at any other azimuth is then a rotation $e^{i\phi}$
in the complex plane of this value. If in the previous comoving picture the mode
under consideration had angular frequency $\omega=1$ (i.e.\ in dimensional
units: angular frequency $\Omega_0$), then in the present $m=1$ mode picture
$\VVx(\tau)$ does not vary with $\tau$. If, on the other hand, in the previous
comoving picture the mode under consideration had angular frequency $\omega\neq
1$, then in the $m=1$ mode picture, $\VVx(\tau)$ varies with time as
$\VVx(\tau)\propto e^{i(\omega-1)\tau}$.

The comoving time derivative $D_\tau$ of the comoving picture now gets replaced by
\begin{equation}\label{eq-ddtau-comoving-partial}
  D_\tau \rightarrow \partial_\tau + \partial_\phi
  = \partial_\tau + i
\end{equation}
The $i$ arises due to $\partial_\phi e^{i\phi} = ie^{i\phi}$.
The $\partial_\tau$ now stands for the non-comoving (lab frame) time
derivative at $\phi=0$.

The new form of the dynamic equations for $\VVx$ and $\VVy$
(Eqs.~\ref{eq-eom-slosh-cmplx-vx}, \ref{eq-eom-slosh-cmplx-vy}) now becomes:
\begin{eqnarray}
\partial_\tau \VVx - 2 \VVy     &=& \psi(1+i\alpha_t) -(i+\alpha_t)\VVx\label{eq-slosh-Vx-fullring}\\
\partial_\tau \VVy + (2-q) \VVx &=& \alpha_t q\psi - (i+\alpha_t)\VVy\label{eq-slosh-Vy-fullring}
\end{eqnarray}
We look for solutions of the form:
\begin{eqnarray}
\VVx(\tau) &=& V_{xp0} + V_{xh0} e^{i\omega_0\tau}\label{eq-Vx-fullring-fullsol}\\
\VVy(\tau) &=& V_{yp0} + V_{yh0} e^{i\omega_0\tau}\label{eq-Vy-fullring-fullsol}
\end{eqnarray}
where the first terms are the particular solution given by
Eqs.~(\ref{eq-part-sol-vx}, \ref{eq-part-sol-vy}), and the second terms are the
homogeneous solution, where $\omega_0=\omega-1$ is the lab-frame frequency of
the homogeneous solution, where $\omega$ is given by Eq.~(\ref{eq-freq-hom-sol}),
hence
\begin{equation}\label{eq-omega0-dispersion-definition}
\omega_0 = \kappa - 1 + i\alpha_t
\end{equation}
What this says is that for slightly non-Keplerian discs ($0<|\kappa-1|\ll 1$)
the homogeneous part of the sloshing motion of $\VVx(\tau,\phi)$ slowly
phase-shifts in time, and at the same time (for $\alpha_t>0$) decays, leaving
eventually only the steady-state sloshing motion $V_{xp0}e^{i\phi}$, which does
not phase-shift in time. The slow phase-shift of the homogeneous solution is
simply the apsidal precession of the epicyclic motion for $\kappa\neq 1$. 

Note that the $V_{xp0}$ and $V_{yp0}$ are {\em nearly} perpendicular in the
complex plane ($V_{yp0}$ being a phase shift $\sim\pi/2$ ahead of $V_{xp0}$), but
not exactly.

For the homogeneous solution we can choose $V_{xh0}$ (a complex number) at will:
it determines the initial condition of the homogeneous solution. For a chosen
$V_{xh0}$, the $V_{yh0}$ follows as:
\begin{equation}
V_{yh0} = \frac{1}{2}i\kappa V_{xh0}
\end{equation}
So, for the homogeneous solution, $V_{yh}(\tau)$ has exactly a phase shift $\pi/2$
ahead of $V_{xh}(\tau)$.

If we wish to express the initial conditions explicitly, we can write the
solutions as:
\begin{eqnarray}
\VVx(\tau) &=& V_{xp0} + (V_{x0}-V_{xp0}) e^{i\omega_0\tau}\\
\VVy(\tau) &=& V_{yp0} + (V_{y0}-V_{yp0}) e^{i\omega_0\tau}
\end{eqnarray}
where $V_{x0}=\VVx(\tau=0)$ and $V_{y0}=\VVy(\tau=0)$ are the initial
conditions. In other words: $V_{ih0}=V_{i0}-V_{ip0}$. Clearly, for $\alpha_t>0$
the solution converges to the steady-state particular solution on a
dimensionless time scale $1/\alpha_t$. And for $\kappa\neq 1$ the solution also
rotates (in the complex plane) around the steady-state particular solution on a
time scale $1/|\kappa-1|$.

For completeness, let us write the full solution of the shoshing motion,
including the $\phi$-dependence, as well:
\begin{eqnarray}
\VVx(\tau,\phi) &=& V_{xp0}e^{i\phi} + (V_{x0}-V_{xp0}) e^{i\omega_0\tau+i\phi}\label{eq-Vx-linear-full-tau-phi}\\
\VVy(\tau,\phi) &=& V_{yp0}e^{i\phi} + (V_{y0}-V_{yp0}) e^{i\omega_0\tau+i\phi}\label{eq-Vy-linear-full-tau-phi}
\end{eqnarray}
This gives, in the linear regime for sufficiently small $\psi$, a complete
description of the sloshing motion in an annulus of the disc. It is
time-dependent as long as the steady-state particular solution is not reached. But this
time-dependence vanishes as the solution approaches the steady-state particular solution, in
which case only the $m=1$ dependence on azimuth $\phi$ remains:
\begin{eqnarray}
\VVx(\tau\rightarrow\infty,\phi) &=& V_{xp0}e^{i\phi}\label{eq-Vx-linear-full-tau-phi-stst}\\
\VVy(\tau\rightarrow\infty,\phi) &=& V_{yp0}e^{i\phi}\label{eq-Vy-linear-full-tau-phi-stst}
\end{eqnarray}
Note that this assumes that $\psi$ stays constant, or more precisely: that
$\psi$ changes slower than the convergence of the solution to the steady-state
particular solution.

\begin{figure*}
  \centerline{\includegraphics[width=0.97\textwidth]{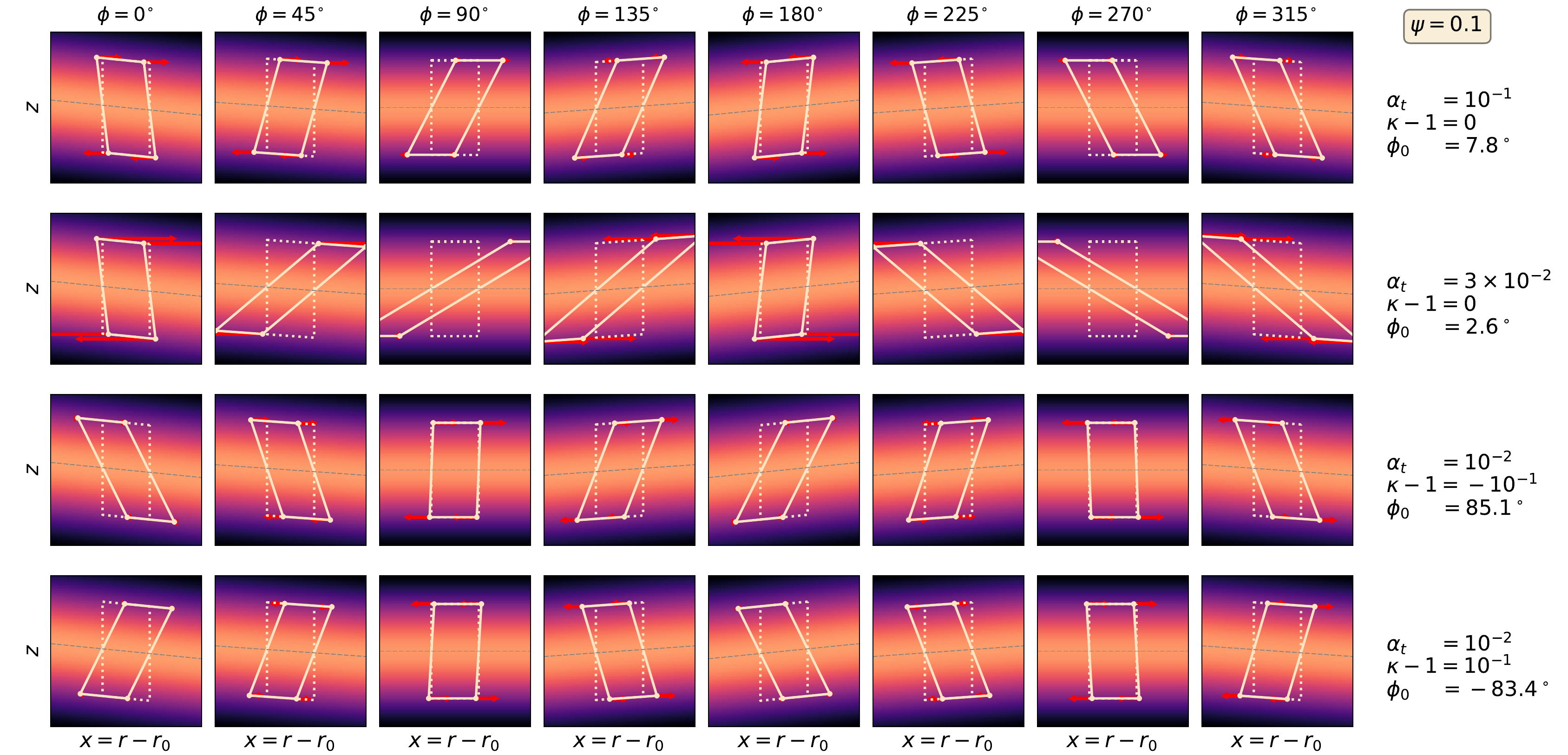}}
  \caption{\label{fig-sloshing-panels}The local $(x,z)$ geometry of the sloshing
    motion of the steady-state particular solution
    Eq.~(\ref{eq-Vx-linear-full-tau-phi-stst}) for a warp amplitude of
    $\psi=0.1$. The panels are arranged horizontally according to azimuth $\phi$
    along the annulus of the disc, and vertically according to four pairs of
    model parameters $(\alpha_t,\kappa-1)$. In each of the panels, the
    grey line marks the rocking equatorial plane, and the color scale represents
    the gas density in arbitrary units. The dashed rectangle is the
    $(x,z)$ cross section of an annulus of the disc, if it would be unperturbed
    by the sloshing motion. The solid rectangle is the skewed shape this annulus
    acquires as a result of the horizontal sloshing motion. The red arrows are
    the horizontal velocities. For each of these four models, the
    value of $\phi_0$ is also given, which is the azimuth angle of minus the
    resulting internal torque vector, defined in Section
    \ref{sec-internal-torque-lam-sols} in Eq.~(\ref{eq-phi0-from-q3-q2}). This
    angle is closely related to the phase of the sloshing motion.
  }
\end{figure*}

Fig.~\ref{fig-sloshing-panels} is a geometric representation of the sloshing
motion, seen in an $(x,z)$-cut through the local disc. The sloshing motion is
shown by the red velocity arrows. The skewed box shows the effect this motion
has on a rectangular slab of disc material. The original slab is shown with
dotted lines and it is, in actuality, an annulus of disc material. This figure
shows, for four different pairs of $(\alpha_t,\kappa-1)$, the skewing of the
slab as a function of the azimuth $\phi$ along the annulus. Two things are
particularly noteworthy of the results shown in this figure: Firstly, it shows
that for $\kappa-1=0$ the amplitude of the sloshing, and the resulting degree of
skewing of the box, becomes very large for $\alpha_t$ lower than $10^{-1}$. This
is the result of the resonance between the orbital and epicyclic frequencies
(though note that these high amplitudes may not be reached before $\psi$
changes). For the two $|\kappa-1|=0.1$ models (bottom two panels) this
divergence does not occur. Secondly, the skewing in the bottom three panels are
strongly phase-shifted with respect to each other. This plays a fundamental role
in the nature of the internal torque arising from this sloshing motion, which is
the topic of Section \ref{sec-from-sloshing-to-torque}.

\subsection{A note on non-linear solutions}\label{sec-non-lin-sol-tau-phi}
In principle the above procedure can also be applied to numerical solutions of
Eqs.~(\ref{eq-eom-warped-box-dtau-rho-laminar-withpres}-\ref{eq-eom-warped-box-dtau-vz-laminar-withpres}),
which are more general since they can also include solutions in the non-linear
regime. To generalize the $V_i(\tau)$ solutions to the $V_i(\tau,\phi)$ form
would require to divide the $\phi$ domain up into $N_\phi$ orbiting grid points
and numerically integrate from the initial condition for each one. An interpolation
from the orbiting grid points to a fixed $\phi$ grid then yields the solution
$V_i(\tau,\phi)$ in numerical form. While this is technically possible, it is
not clear whether in this non-linear regime the local shearing box approach is
still justified in the first place. We will here, however, not consider this.

\section{From sloshing motion to internal torque}
\label{sec-from-sloshing-to-torque}
The internal torque (written with the symbol ${\boldsymbol {\cal G}}$ in
\citet{2013MNRAS.433.2403O}) is the torque that one annulus of the disc (at
radius $r_0$) exerts on its adjacent annulus just outside of it (at radius
$r_0+dr$). In other words, it is the outward flow of angular momentum per unit
time, integrated over vertical height $z$ and azimuth $r_0\phi$. For convenience
we will, however, define ``internal torque'' to be not the integral over azimuth
$r_0\phi$, but the azimuthal mean. In other words, in our definition the
internal torque vector is ${\bf G}={\boldsymbol {\cal G}}/(2\pi r_0)$. To
compute it, we first have to compute the local torque, then integrate this
vertically, and finally compute the mean over all azimuths.

\subsection{Internal torque for the laminar solutions}
\label{sec-internal-torque-lam-sols}
We first compute the local internal torque, at a given annulus at $r=r_0$ and
height above the midplane $z'$, {largely following the same procedure as
in \citet{2013MNRAS.433.2403O}}. We have to start with the total stress tensor
\begin{equation}\label{eq-stress-tensor-total}
\ttot{}_{ij} = \rho u^{(t)}_i u^{(t)}_j + p \delta_{ij} - \tvisc{}_{ij}
\end{equation}
which is defined as the flux of $i$-momentum in $j$-direction, and where
$u^{(t)}_i$ is the velocity including the orbital velocity:
\begin{equation}
u^{(t)}_i = \Omega_0r_0\delta_{iy} + u_i
\end{equation}
where $\delta_{iy}=1$ only for $i=y$. The first term is the unperturbed orbital
motion part, while the second term is the perturbed velocity.

We need, however, the flux of {\em angular} momentum, for which we need the
lever arm
\begin{equation}\label{eq-lever-arm-warped}
{\bf r}= r_0{\bf e}_x + z{\bf e}_z
\end{equation}
where we make the assumption that we will constrain our analysis to the
vertical column defined by $x=0$, $y=0$. The angular
momentum flux tensor $\ggloctens{}_{ij}$ in the $(x,y,z)$ shearing box system is
\begin{equation}\label{eq-angmom-flux-tensor}
\ggloctens{}_{ij} = \varepsilon_{ikl} r_{k}\ttot{}_{lj} = \varepsilon_{ikl} [r_{0}\delta_{kx}+z \delta_{kz}]\ttot{}_{lj}
\end{equation}
with $\varepsilon_{ikl}$ the Levi-Civita pseudo-tensor, defined as being
$\varepsilon_{xyz}=+1$ and switching sign for each permutation of $x$, $y$ and
$z$, and zero otherwise. Einstein's summation convention is used for
indices $k$ and $l$. The tensor $\ggloctens{}_{ij}$ represents the transport of the $i$
component of angular momentum into $j$ direction. Note that $\ggloctens{}_{ij}$ is not a
symmetric tensor (in contrast to $\ttot{}_{ij}$).

To get the local internal torque, we need to compute the $x-$ (outward)
component of this tensor:
\begin{eqnarray}
  \ggloccomp{}_x &=& \ggloctens{}_{xx} = -z\ttot{}_{xy} \\
  \ggloccomp{}_y &=& \ggloctens{}_{yx} = -r_0\ttot{}_{xz} + z\ttot{}_{xx} \\
  \ggloccomp{}_z &=& \ggloctens{}_{zx} = r_0\ttot{}_{xy}
\end{eqnarray}
(see Appendix \ref{app-local-internal-torque}).
We then vertically integrate these over $z'$
\begin{eqnarray}
  \Ggvertcomp_x &\equiv& \int_{-\infty}^{+\infty}\ggloccomp_x dz'\\
  \Ggvertcomp_y &\equiv& \int_{-\infty}^{+\infty}\ggloccomp_y dz'\\
  \Ggvertcomp_z &\equiv& \int_{-\infty}^{+\infty}\ggloccomp_z dz'
\end{eqnarray}
(see Appendix \ref{app-vertint-internal-torque}), apply a
rotation into the global $X$ and $Y$ directions,
\begin{eqnarray}
  \Ggvertcomp_X &=& \cos(\phi)\Ggvertcomp_x - \sin(\phi)\Ggvertcomp_y\label{eq-from-x-to-X}\\
  \Ggvertcomp_Y &=& \sin(\phi)\Ggvertcomp_x + \cos(\phi)\Ggvertcomp_y\label{eq-from-y-to-Y}\\
  \Ggvertcomp_Z &=& \Ggvertcomp_z
\end{eqnarray}
(see Appendix \ref{app-azimuthal-mean-internal-torque})
and finally compute the azimuthal mean:
\begin{eqnarray}
  \GGbigcomp_X &=& \frac{1}{2\pi}\int_{0}^{2\pi}\Ggvertcomp_X\,d\phi\label{eq-ggbigx-integral}\\
  \GGbigcomp_Y &=& \frac{1}{2\pi}\int_{0}^{2\pi}\Ggvertcomp_Y\,d\phi\label{eq-ggbigy-integral}\\
  \GGbigcomp_Z &=& \frac{1}{2\pi}\int_{0}^{2\pi}\Ggvertcomp_Z\,d\phi
\end{eqnarray}
(see Appendix \ref{app-azimuthal-mean-internal-torque}). The
result is:
\begin{eqnarray}
  2\GGbigcomp_X/\Xii &=& -\VVx(\tau) - i\alpha_t\VVx(\tau) - \alpha_t\psi \label{eq-ggbig-cmplx-x}\\
  2\GGbigcomp_Y/\Xii &=& -i\VVx(\tau) + \alpha_t\VVx(\tau) - i\alpha_t\psi \label{eq-ggbig-cmplx-y}\\
   \GGbigcomp_Z/\Xii &=&  q\alpha_t\label{eq-ggbig-cmplx-z}
\end{eqnarray}
where $\GGbigcomp_i$ are complex variables, and $\Xii$ is defined as
\begin{equation}\label{eq-xii-def}
\Xii \equiv \Omega_0^2r_0\Sigma h_p^2
\end{equation}
Note that $\GGbigcomp_Y=i\GGbigcomp_X$.
Since, from a physics perspective, we are only interested in the real values,
the result is:
\begin{eqnarray}
  2\GGbigcomp^{\mathrm{re}}_X/\Xii &=& -\VVx^{\mathrm{re}}(\tau) + \alpha_t\VVx^{\mathrm{im}}(\tau) - \alpha_t\psi \label{eq-ggbig-re-x}\\
  2\GGbigcomp^{\mathrm{re}}_Y/\Xii &=& \VVx^{\mathrm{im}}(\tau) + \alpha_t\VVx^{\mathrm{re}}(\tau)\label{eq-ggbig-re-y}\\
   \GGbigcomp^{\mathrm{re}}_Z/\Xii &=&  q\alpha_t\label{eq-ggbig-re-z}
\end{eqnarray}
where the superscripts $^{\mathrm{re}}$ and $^{\mathrm{im}}$ denote the real and
imaginary part, respectively. These expressions are the ones that can be
directly evaluated for a known solution $\VVx(\tau)$ consisting of a particular
solution and a homogeneous solution
$\VVx(\tau)=V_{xp0}+V_{xh0}e^{i\omega_0\tau}$, with $V_{xp0}$ given by
Eq.~(\ref{eq-part-sol-vx}), $V_{xh0}=\VVx(\tau=0)-V_{xp0}$, and
$\omega_0=\kappa-1+i\alpha_t$ (cf.~Eq.~\ref{eq-omega0-dispersion-definition}).

The real and imaginary parts of the steady-state particular solution
(Eq.~\ref{eq-part-sol-vx}) are:
\begin{eqnarray}
V_{xp0}^{\mathrm{re}} &\simeq& \alpha_t\,\frac{2+3\epsilon+5\alpha_t^2-\epsilon(\epsilon+\alpha_t^2)}{(\epsilon+\alpha_t^2)^2+4\alpha_t^2}\,\psi \label{eq-particular-sol-for-Vx-real}\\
V_{xp0}^{\mathrm{im}} &\simeq& \frac{\epsilon-5\alpha_t^2+\alpha_t^2(3\epsilon+\alpha_t^2)}{(\epsilon+\alpha_t^2)^2+4\alpha_t^2}\,\psi\label{eq-particular-sol-for-Vx-imag}
\end{eqnarray}
where we defined $\epsilon=\kappa^2-1$. Inserting these into
Eqs.~(\ref{eq-ggbig-re-x}-\ref{eq-ggbig-re-y}) gives the ``steady-state particular solution''
of the internal torque $\GGbigcomp^{\mathrm{re}}_{ip0}$ (with $i=X,Y,Z$).

It is at this point where the dominance of the torque by the sloshing motion
over the viscous torque becomes clear, at least for the steady-state particular
solution. Looking at Eq.~(\ref{eq-ggbig-re-x}) the $-\VVx^{\mathrm{re}}(\tau)$
term, when inserting Eq.~(\ref{eq-particular-sol-for-Vx-real}), can easily
become of order unity, while the viscous torque term, $\alpha_t\psi$, will
clearly be much less than unity.

To make the comparison to the literature easier, we will write this ``steady-state particular
solution'' of the internal torque as the $Q$ coefficients introduced by
\citet{1999MNRAS.304..557O} and \citet{2013MNRAS.433.2403O}. Consistent with
these papers, we define $Q_1$, $Q_2$ and $Q_3$ as
\begin{eqnarray}
  \GGbigcomp^{\mathrm{re}}_{Zp0} &=& -\Xii\; Q_1 \\
  \GGbigcomp^{\mathrm{re}}_{Xp0} &=& -\Xii\psi\; Q_2\label{eq-Q2-afo-GXp} \\
  \GGbigcomp^{\mathrm{re}}_{Yp0} &=& -\Xii\psi\; Q_3\label{eq-Q3-afo-GYp}
\end{eqnarray}
We then arrive at
\begin{eqnarray}
  Q_1 &=& -q\alpha_t\label{eq-Q1-fullsol}\\
  Q_2 &=& \frac{1+7\alpha_t^2+\epsilon(1-\alpha_t^2)}{(\epsilon+\alpha_t^2)^2+4\alpha_t^2}\alpha_t\label{eq-Q2-fullsol}\\
  Q_3 &=& -\frac{1}{2}\frac{\epsilon-3\alpha_t^2+(6-\epsilon)(\alpha_t^2+\epsilon)\alpha_t^2}{(\epsilon+\alpha_t^2)^2+4\alpha_t^2}\label{eq-Q3-fullsol}
\end{eqnarray}
with, we recall, $\epsilon=\kappa^2-1$. These expressions are consistent
with Eq.~(A39) of \citet{1999MNRAS.304..557O} and Eqs.~(9,10) of
\citet{2019MNRAS.487.4965Z}\footnote{There is a typo in Eq.~(10) of
  \citet{2019MNRAS.487.4965Z}, where the first occurrence of
  $+2\tilde\kappa$ in the numerator should be $-2\tilde\kappa$}.

The physical interpretation of the three $Q$s is that $Q_1$ causes the viscous
evolution of the disc (the radial flow of mass $\Sigma(r,t)$), $Q_2$ damps the
warp, and $Q_3$ rotates the warp vector and may thus produce a
twist\footnote{We define the word ``twist'' to denote the case when
  the direction of the warp vector $\warpvector/|\warpvector|$ varies with $r$.} in the
disc. If the torque vector points into the negative $X$-direction ($Q_2>0$ and
$Q_3=0$), the warp damps. If the torque vector points into the negative or
positive $Y$-direction ($Q_2=0$ and $Q_3\neq 0$) the warp twists (precession of
the annulli). In general ($Q_2>0$ and $Q_3\neq 0$) both a twist and a damping of
the warp occurs.

The magnitudes of $Q_2$ and $Q_3$ (or equivalently of
$\GGbigcomp^{\mathrm{re}}_{Xp0}$ and $\GGbigcomp^{\mathrm{re}}_{Yp0}$) are
intimitely connected to each other, because they are both determined primarily
by the amplitude and phase of the sloshing motion. \citet{2013MNRAS.433.2403O}
symbolize this by defining the complex variable $Q_4\equiv Q_2+iQ_3$, which we
shall write here as $Q_4=|Q_4|e^{i\phi_0}$, with $|Q_4|=\sqrt{Q_2^2+Q_3^2}$ and
\begin{equation}\label{eq-phi0-from-q3-q2}
\phi_0=\mathrm{atan}(Q_3/Q_2)
\end{equation}
The horizontal components of $\GGbig^{\mathrm{re}}_{p0}$ are therefore
\begin{eqnarray}
  \GGbigcomp^{\mathrm{re}}_{Xp0} &=& -\Xii\psi |Q_4| \cos(\phi_0)\\
  \GGbigcomp^{\mathrm{re}}_{Yp0} &=& -\Xii\psi |Q_4| \sin(\phi_0)
\end{eqnarray}
The physical interpretation of $\phi_0$ is the phase of the torque.  For
$\phi_0=0$ we have only damping of the warp, while for $\phi_0\neq 0$ we also
have a twisting component of the torque. In the extreme cases of
$\phi_0=(+/-)\pi/2$ we have only twisting, no damping, with the internal torque pointing in
(negative/positive) $Y$-direction. For reasons of energy conservation $-\pi/2\le
\phi_0\le \pi/2$ (no anti-damping). For $\alpha_t\ll 1$, the complex-valued
torque vector components $\GGbigcomp_{Xp0}$ and $\GGbigcomp_{Yp0}$ are nearly
proportional to the complex-valued sloshing velocity $-V_{xp0}$
(cf.~Eqs.~\ref{eq-ggbig-re-x}, \ref{eq-ggbig-re-y} with the terms proportional
to $\alpha_t$ neglected), and the phase $\phi_0$ of $Q_4$ corresponds to the
phase $V_{xp0}=|V_{xp0}|e^{-i\phi_0}$ of the sloshing motion.

\begin{figure*}
  \centerline{\includegraphics[width=0.47\textwidth]{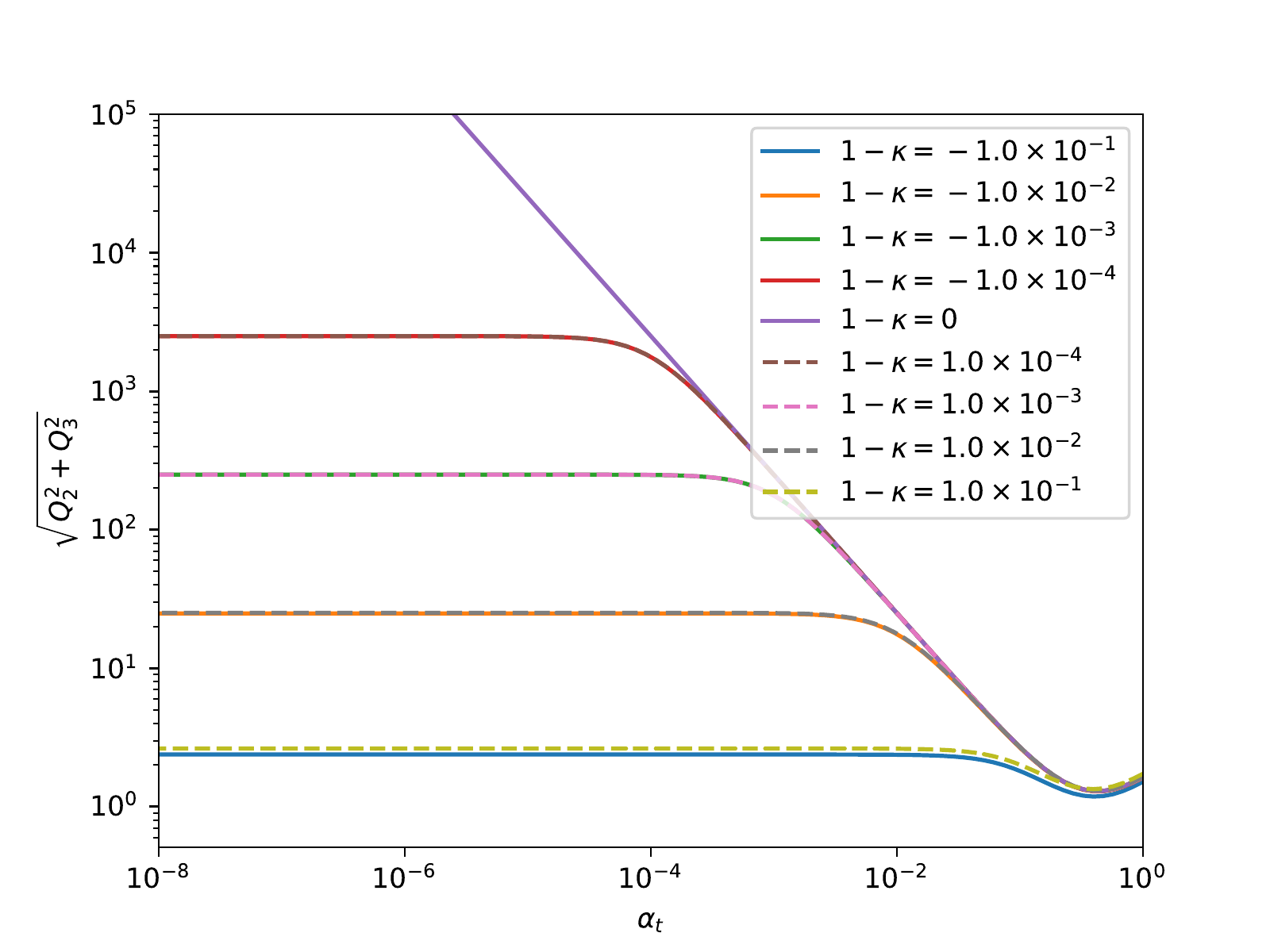}
    \includegraphics[width=0.47\textwidth]{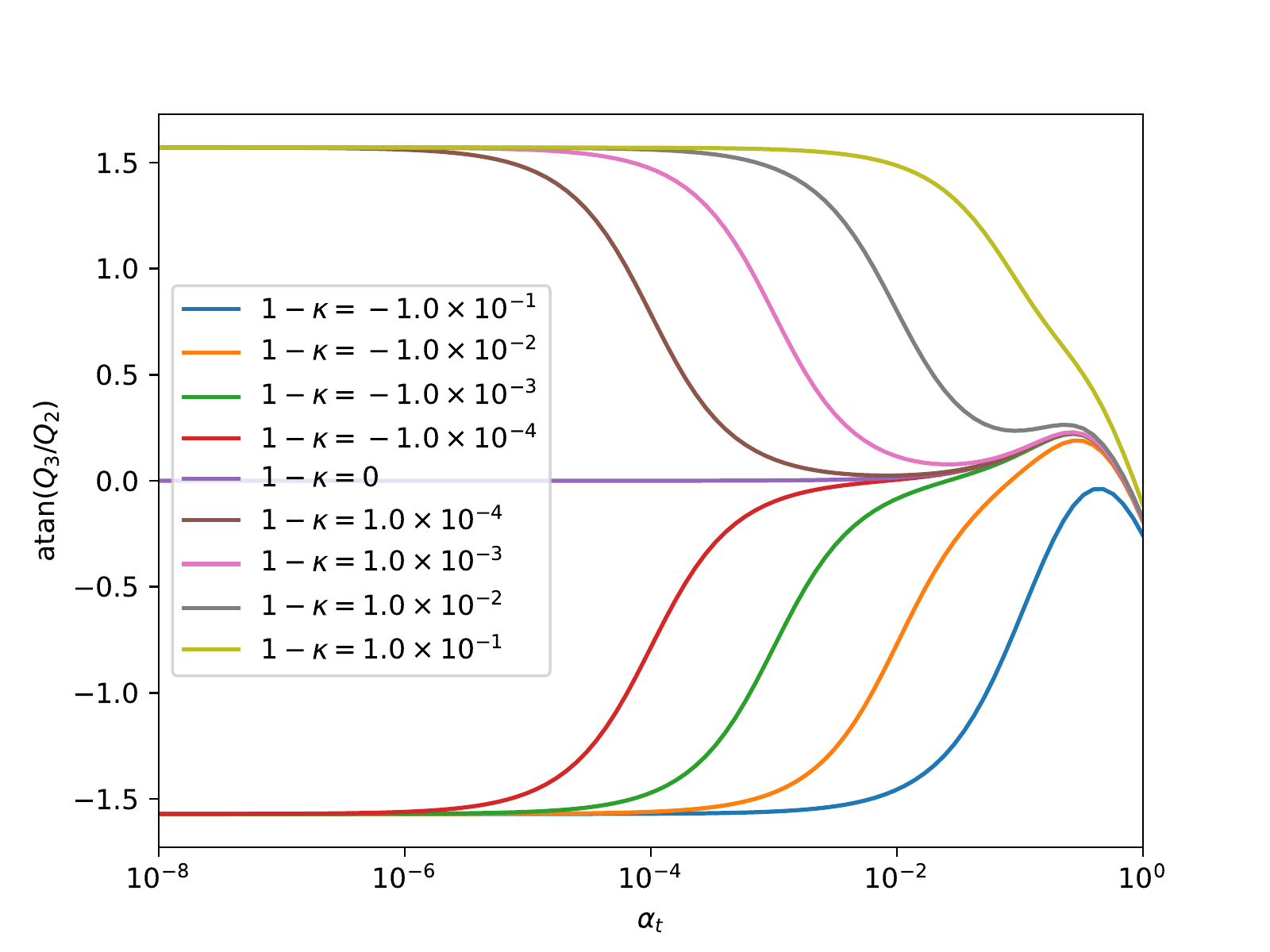}}
  \caption{\label{fig-Q-values-afo-alphat-kappa}The horizontal torque components
    for the steady-state particular solution, expressed in terms of Ogilvie's
    $Q_4= Q_2+iQ_3$, where $Q_2$ and $Q_3$ are defined by
    Eqs.~(\ref{eq-Q2-afo-GXp}, \ref{eq-Q3-afo-GYp}). Left:
    $|Q_4|=\sqrt{Q_2^2+Q_3^2}$. Right:
    $\phi_0=\mathrm{arg}(Q_4)=\mathrm{atan}(Q_3/Q_2)$.}
\end{figure*}

In Fig.~\ref{fig-Q-values-afo-alphat-kappa} the $|Q_4|$ and $\phi_0$ are shown.
This figure also makes clear the resonant behavior of the torque: For
$\kappa\rightarrow 1$ and for $\alpha_t\rightarrow 0$ the amplitude tends to
infinity. Physically the warp is like a oscillator with natural dimensionless
frequency $\kappa$, driven by the warp at dimensionless frequency $1$. When
$\kappa$ approaches unity, the oscillator approaches the resonance, and only
viscous damping can prevent it from becoming infinite.

For the special case $\kappa=1$ (i.e.~$\epsilon=0$) we find:
\begin{eqnarray}
  Q_1 &=& -q\alpha_t\\
  Q_2 &=& \left(\frac{1+7\alpha_t^2}{4+\alpha_t^2}\right)\,\frac{1}{\alpha_t}\\
  Q_3 &=& \left(\frac{\tfrac{3}{2}-3\alpha_t^2}{4+\alpha_t^2}\right)
\end{eqnarray}
which are identical to Eqs.~(95, 97, 98) of \citet{2013MNRAS.433.2403O}. This was
to be expected, since we use the same framework. For the special case that both
$0<\alpha_t\ll 1$ and $0<|\kappa-1|\ll 1$ we have:
\begin{eqnarray}
  Q_1 &=& -q\alpha_t\label{eq-Q1-small-alpha-kappa}\\
  Q_2 &\simeq& \frac{\alpha_t}{(\kappa^2-1)^2+4\alpha_t^2}\label{eq-Q2-small-alpha-kappa}\\
  Q_3 &\simeq& -\frac{\tfrac{1}{2}(\kappa^2-1)-\tfrac{3}{2}\alpha_t^2}{(\kappa^2-1)^2+4\alpha_t^2}\label{eq-Q3-small-alpha-kappa}
\end{eqnarray}
This limit is of particular interest to protoplanetary discs, where it is
thought that the viscosity is low, and the deviations from Keplerian rotation
($\kappa\neq 1$, $\epsilon\neq 0$) are caused by the radial pressure gradient.

\subsection{The role of the homogeneous solution of the sloshing motion}
\label{sec-role-hom-sol}
Let us now insert the full (particular+homogeneous) solution of $\VVx(\tau)$
into the internal torque components:
\begin{eqnarray}
  2\GGbigcomp_X/\Xii &=& - (V_{xp0} + i\alpha_tV_{xp0} + \alpha_t\psi ) \nonumber\\
  && -(V_{xh0} + i\alpha_tV_{xh0})e^{i\omega_0\tau} \\
  2\GGbigcomp_Y/\Xii &=& - ( iV_{xp0} - \alpha_tV_{xp0} + i\alpha_t\psi ) \nonumber\\
  && -(iV_{xh0} - \alpha_tV_{xh0})e^{i\omega_0\tau} \\
   \GGbigcomp_Z/\Xii &=&  q\alpha_t
\end{eqnarray}
with $V_{xh0}=V_{x0}-V_{xp0}$ where $V_{x0}$ is the initial condition of the
sloshing motion (a complex number), and $\omega_0=\kappa-1+i\alpha_t$
(cf.~Eq.~\ref{eq-omega0-dispersion-definition}). This can be re-written as
\begin{eqnarray}
  \GGbigcomp_X &=& \GGbigcomp_{Xp0} + (\GGbigcomp_{X0}-\GGbigcomp_{Xp0})e^{i\omega_0\tau} \label{GGbig-X-part-hom}\\
  \GGbigcomp_Y &=& \GGbigcomp_{Yp0} + (\GGbigcomp_{Y0}-\GGbigcomp_{Yp0})e^{i\omega_0\tau} \label{GGbig-Y-part-hom}\\
  \GGbigcomp_Z &=& \Xii q\alpha_t \label{GGbig-Z-part-hom}
\end{eqnarray}
where $\GGbigcomp_{X0}=\GGbigcomp_{X}(\tau=0)$ is the initial condition for
$\GGbigcomp_{X}(\tau)$, and likewise for the $Y$ component. The ``steady-state
particular solutions'' for the internal torque $\GGbigcomp_{Xp0}$ and
$\GGbigcomp_{Yp0}$ are defined as in
Eqs.~(\ref{eq-ggbig-cmplx-x},\ref{eq-ggbig-cmplx-y}) with $\VVx(\tau)$ set to the
steady-state particular solution for the sloshing motion $V_{xp0}$.  So now we
have expressed $\GGbigcomp_X(\tau)$ and $\GGbigcomp_Y(\tau)$ in a similar way as
for $\VVx(\tau)$: as a sum of a steady-state particular solution and a transient
homogeneous solution. The dimensionless time scale of the decay of the transient
is $1/\alpha_t$, and the dimensionless time scale for the rotation of the transient
around the steady-state particular solution is $1/|\kappa-1|$. 

The interpretation of this result is that the $Q_1$, $Q_2$ and $Q_3$ values
determine the internal torque vector if the sloshing motion has reached its
steady-state oscillation, i.e.\ if the homogeneous solution has decayed to zero
and only the steady-state particular solution remains. For large $\alpha_t$ this
is reached in a few orbits, on a time scale shorter than the time scale at which
the warp changes. For small $\alpha_t$ this steady-state oscillation may never
be reached, as the warp may change faster than that. The distinction between
the diffusive regime and the wavelike regime of warped discs is then simply
whether or not the transient part of $\GGbig{}$ decays faster or slower than
the change of the warp $\psi$.

\section{Ordinary differential equation for $\GGbig{}$}
\label{sec-ode-for-ggbig}
With the results of Section \ref{sec-role-hom-sol} we now have a complete
description of how $\GGbig{}(\tau)$ behaves as a function of the warp amplitude
$\psi$, dimensionless time $\tau=\Omega_0 t$ and the initial condition
$\GGbig{}(0)$. However, our solution
(Eqs.~\ref{GGbig-X-part-hom}-\ref{GGbig-Z-part-hom}) is only valid for a warp
$d\llunit{}(r)/d\ln r$ that is constant in time. As a result of the torque
$\GGbig{}$, however, the orientations $\llunit{}(r)$ of the annuli change, and
thereby the warp vector $\warpvector=d\llunit{}(r)/d\ln r$ changes as well: both
its amplitude $\psi=|\warpvector|$ and its direction
$\psidir{}=\warpvector/\psi$. This introduces two problems:
\begin{enumerate}
\item\label{enum-problem-1} If $\psi$ changes with time, the simple expressions for $\GGbig(\tau)$,
  Eqs.~(\ref{GGbig-X-part-hom}-\ref{GGbig-Z-part-hom}), no longer apply.
\item\label{enum-problem-2} The orbital plane of the annulus changes, meaning that at a later
  time we need to perform a coordinate transformation from $(X,Y,Z)$ to
  the new coordinates $(X',Y',Z')$ where the new orbital plane lies again
  in the $(X',Y')$-plane.
\end{enumerate}
Problem \ref{enum-problem-1} can be relatively easily solved by formulating an ordinary
differential equation (ODE) for $\GGbig$ that has
Eqs.~(\ref{GGbig-X-part-hom}-\ref{GGbig-Y-part-hom}) as a solution, as we will
describe in Section \ref{sec-ode-for-GxGy}. Problem \ref{enum-problem-2} is harder to solve. We
will propose a solution to this problem in Section \ref{sec-rotating-G}.

Both solutions rely on the splitting of $\GGbig$ into a dynamic sloshing torque
$\GGbig^{(s)}$ and a non-dynamic viscous torque $\GGbig^{(v)}$:
\begin{equation}\label{eq-split-G-into-s-and-v}
\GGbig = \GGbig^{(s)} + \GGbig^{(v)}
\end{equation}
which, in the usual $(X,Y,Z)$ coordinate system, have the following
components (see Eqs.~\ref{eq-ggbig-re-x},~\ref{eq-ggbig-re-z}):
\begin{equation}\label{eq-Gsplit-Gslosh}
\GGbig^{(s)}(\tau) = \frac{\Xii}{2}\left(\begin{matrix} 
-\VVx^{\mathrm{re}}(\tau) + \alpha_t\VVx^{\mathrm{im}}(\tau)\\
\VVx^{\mathrm{im}}(\tau) + \alpha_t\VVx^{\mathrm{re}}(\tau)\\
0
\end{matrix}\right)
\end{equation}
and
\begin{equation}\label{eq-Gsplit-Gvisc}
\GGbig^{(v)} = \Xii\left(\begin{matrix} 
-\alpha_t\psi\\
0\\
q\alpha_t
\end{matrix}\right)
\end{equation}
The latter can be expressed in coordinate-free form as follows:
\begin{equation}\label{eq-G-visc-coordfree}
\GGbig^{(v)} = \Xii q\alpha_t\llunit - \Xii\alpha_t \warpvector
\end{equation}
For the dynamic $\GGbig^{(s)}(\tau)$ we need to set up an ODE, which
we do next.

\subsection{Solution to problem \ref{enum-problem-1}: ODE for the sloshing torque}
\label{sec-ode-for-GxGy}
The most straightforward ODE for the sloshing torque components
$\GGbigcomp^{(s)}_X(\tau)$ and $\GGbigcomp^{(s)}_X(\tau)$ that has
Eqs.~(\ref{GGbig-X-part-hom}-\ref{GGbig-Y-part-hom}) as a solution is:
\begin{eqnarray}
\partial_\tau \GGbigcomp^{(s)}_X(\tau) &=& i\omega_0 \big(\GGbigcomp^{(s)}_X(\tau) - \GGbigcomp^{(s)}_{Xp0}\big)\\
\partial_\tau \GGbigcomp^{(s)}_Y(\tau) &=& i\omega_0 \big(\GGbigcomp^{(s)}_Y(\tau) - \GGbigcomp^{(s)}_{Yp0}\big)
\end{eqnarray}
If $\psi$ changes, then the solution automatically adapts to the changing value of $\psi$.
Using $\omega_0=\kappa-1+i\alpha_t$ and $\GGbigcomp^{(s)}_Y=i\GGbigcomp^{(s)}_X$, and
taking the real parts of both sides of the equations we obtain:
\begin{eqnarray}
\partial_\tau \GGbigcomp^{(s)\,\mathrm{re}}_X(\tau) &=& (\kappa-1) \big(\GGbigcomp^{(s)\,\mathrm{re}}_Y(\tau) - \GGbigcomp^{(s)\,\mathrm{re}}_{Yp0}\big) \nonumber\\&& - \alpha_t \big(\GGbigcomp^{(s)\,\mathrm{re}}_X(\tau) - \GGbigcomp^{(s)\,\mathrm{re}}_{Xp0}\big)\label{eq-GGbig-hompart-X}\\
\partial_\tau \GGbigcomp^{(s)\,\mathrm{re}}_Y(\tau) &=& -(\kappa-1) \big(\GGbigcomp^{(s)\,\mathrm{re}}_X(\tau) - \GGbigcomp^{(s)\,\mathrm{re}}_{Xp0}\big) \nonumber\\&& - \alpha_t \big(\GGbigcomp^{(s)\,\mathrm{re}}_Y(\tau) - \GGbigcomp^{(s)\,\mathrm{re}}_{Yp0}\big)\label{eq-GGbig-hompart-Y}
\end{eqnarray}
where it is important to note the different sign of the first term of the
right-hand-side in both equations. This suggests (but see Section \ref{sec-rotating-G} for an important
addition) the
following vectorial form of these equations:
\begin{equation}\label{eq-vect-form-of-G-hom-to-part}
  \frac{\partial \GGbig^{(s)}}{\partial \tau} = -(\kappa-1){\bf l}\times (\GGbig^{(s)}-\GGbig^{(s)}_{p0})
  -\alpha_t(\GGbig^{(s)}-\GGbig^{(s)}_{p0})
\end{equation}
where here the $\GGbig^{(s)}$ is considered to be a real-valued vector, and $\GGbig^{(s)}_{p0}$ is given by
\begin{equation}\label{eq-Gp-afo-Q123-repeat}
\GGbig^{(s)}_{p0} = -\Xii \left(Q_2\frac{d{\bf l}}{d\ln r}+Q_3{\bf l}\times \frac{d{\bf l}}{d\ln r}\right)
\end{equation}
with $Q_2$ and $Q_3$ given by Eqs.~(\ref{eq-Q2-fullsol}-\ref{eq-Q3-fullsol}), and
$\Xii\equiv \Omega_0^2r_0\Sigma h_p^2$ (cf.\ Eq.~\ref{eq-xii-def}).

\begin{figure*}
  \centerline{\includegraphics[width=0.47\textwidth]{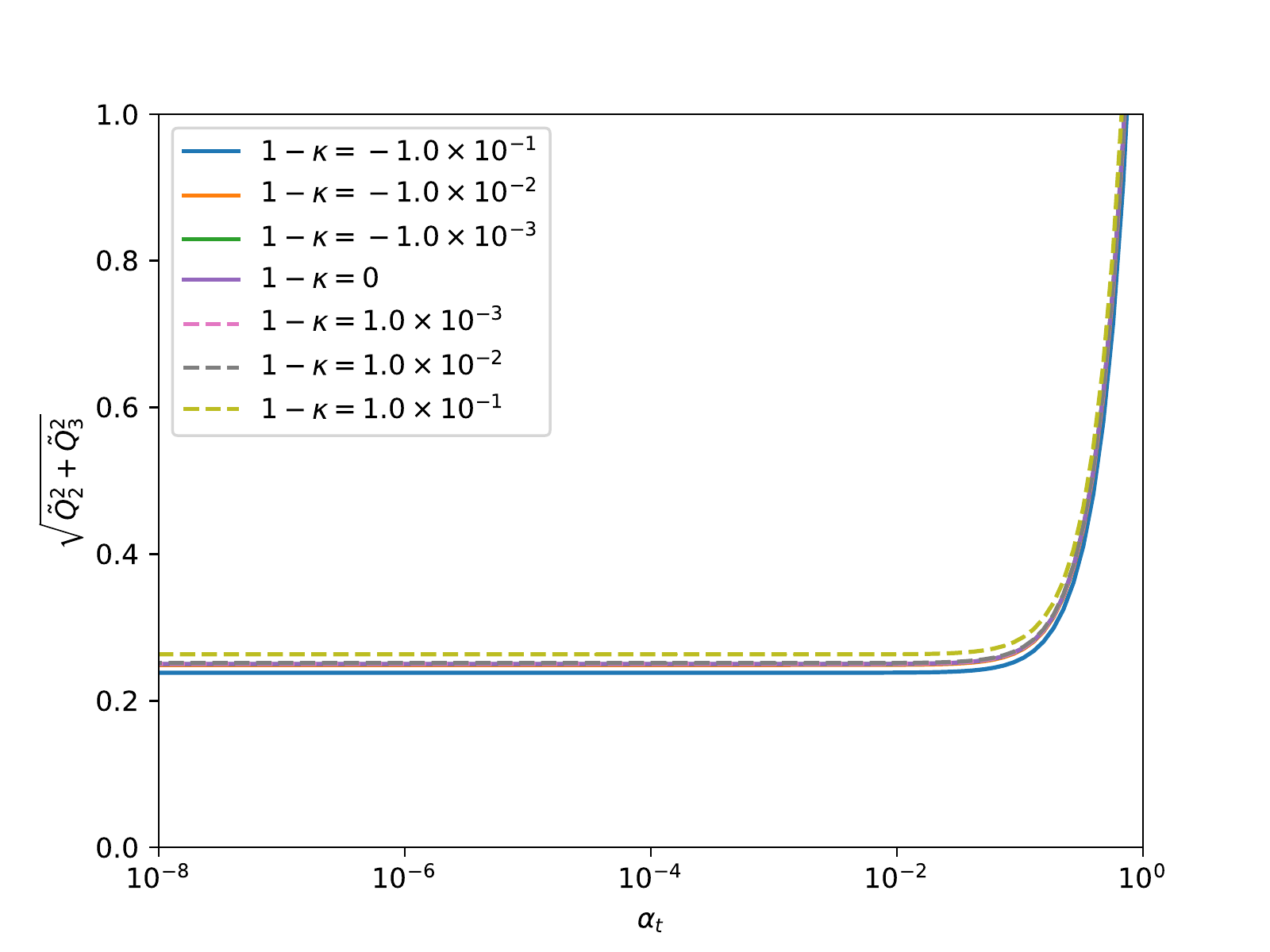}
    \includegraphics[width=0.47\textwidth]{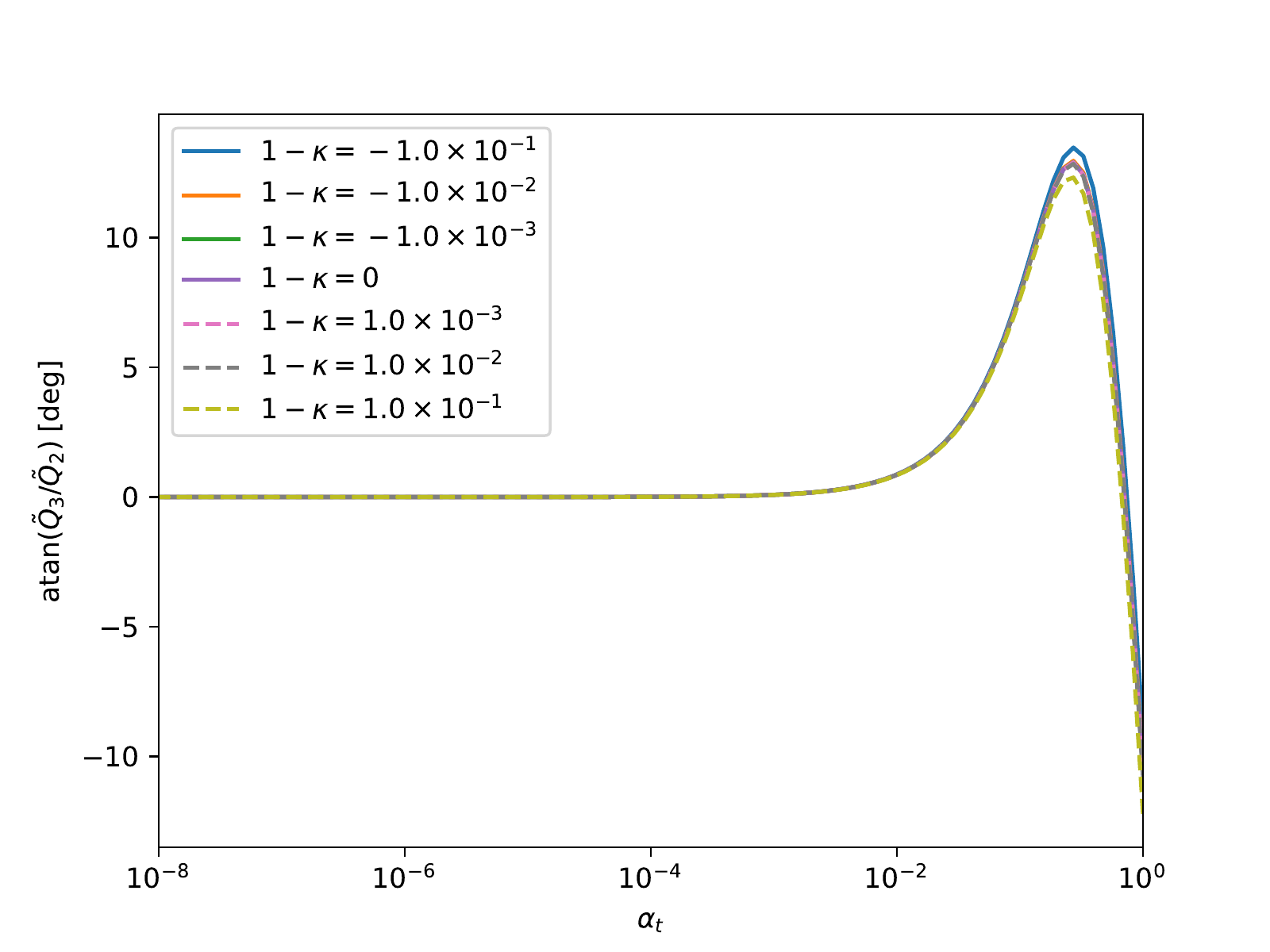}}
  \caption{\label{fig-Qtilde-values-afo-alphat-kappa}As Fig.~\ref{fig-Q-values-afo-alphat-kappa},
    but for $\tilde Q_2$ and $\tilde Q_3$, which are the coefficients entering the time-dependent
    equation for the torques, Eq.~(\ref{eq-vect-form-of-G-hom-to-part-alt-Qtilde}), where
    $\tilde Q_2$ and $\tilde Q_3$ are defined by
    Eqs.~(\ref{eq-qtilde-2-def}, \ref{eq-qtilde-3-def}). Left:
    $|\tilde Q_4|=\sqrt{\tilde Q_2^2+\tilde Q_3^2}$. Right:
    $\tilde\phi_0=\mathrm{arg}(\tilde Q_4)=\mathrm{atan}(\tilde Q_3/\tilde Q_2)$.}
\end{figure*}

To make it easier to compare our dynamic equation for $\GGbig^{(s)}$
(Eq.~\ref{eq-vect-form-of-G-hom-to-part}) to the literature, we can
rewrite it by bringing the $\GGbig^{(s)}$ terms to the left-hand-side, and keeping the
$\GGbig^{(s)}_{p0}$ terms on the right-hand-side:
\begin{equation}\label{eq-vect-form-of-G-hom-to-part-alt}
  \frac{\partial \GGbig^{(s)}}{\partial \tau} + (\kappa-1)\llunit{}\times \GGbig^{(s)} +
  \alpha_t\GGbig^{(s)} = (\kappa-1)\llunit{}\times \GGbig^{(s)}_{p0} +
  \alpha_t\GGbig^{(s)}_{p0}
\end{equation}
By inserting the expression for $\GGbig^{(s)}_{p0}$ (Eq.~\ref{eq-Gp-afo-Q123-repeat})
we obtain
\begin{equation}\label{eq-vect-form-of-G-hom-to-part-alt-Q}
  \begin{split}
    \frac{\partial \GGbig^{(s)}}{\partial \tau}  + & (\kappa-1)\llunit{}\times \GGbig^{(s)} +
    \alpha_t\GGbig^{(s)} \\=& -(\kappa-1)\Xii \left(Q_2\llunit{}\times\frac{d\llunit{}}{d\ln r}-Q_3\frac{d\llunit{}}{d\ln r}\right)
    \\&-\alpha_t\Xii \left(Q_2\frac{d\llunit{}}{d\ln r}+Q_3\llunit{}\times \frac{d\llunit{}}{d\ln r}\right)
  \end{split}
\end{equation}
This equation is exactly the same as Eq.~(\ref{eq-vect-form-of-G-hom-to-part}),
just written out more explicitly. For notational convenience let us regroup the
terms on the right-hand-side:
\begin{equation}\label{eq-vect-form-of-G-hom-to-part-alt-Qtilde}
  \begin{split}
    \frac{\partial \GGbig^{(s)}}{\partial \tau}  + & (\kappa-1)\llunit{}\times \GGbig^{(s)} +
    \alpha_t\GGbig^{(s)} \\=& 
    -\Xii \left(\tilde Q_2\frac{d\llunit{}}{d\ln r}+\tilde Q_3\llunit{}\times \frac{d\llunit{}}{d\ln r}\right)
  \end{split}
\end{equation}
with
\begin{eqnarray}
\tilde Q_2 &=& \alpha_t Q_2 -(\kappa-1)Q_3\label{eq-qtilde-2-def}\\
\tilde Q_3 &=& (\kappa-1)Q_2 + \alpha_tQ_3\label{eq-qtilde-3-def}
\end{eqnarray}
Fig.~\ref{fig-Qtilde-values-afo-alphat-kappa} shows the dependency of $\tilde Q_2$
and $\tilde Q_3$ on $\alpha_t$ and $\kappa-1$.
In the limit of $\alpha_t\ll 1$ and $|\kappa-1|\ll 1$ we find that
\begin{equation}
\tilde Q_2\rightarrow \frac{1}{4},\qquad \frac{\tilde Q_3}{\tilde Q_2}\rightarrow 0
\end{equation}
so that, in this limit, our equation simplifies to
\begin{equation}\label{eq-vect-form-of-G-hom-to-part-alt-Qtilde-smalllimit}
    \frac{\partial \GGbig^{(s)}}{\partial \tau}  +  (\kappa-1)\llunit{}\times \GGbig^{(s)} +
    \alpha_t\GGbig^{(s)} \simeq -\frac{\Xii}{4}\frac{d\llunit{}}{d\ln r}
\end{equation}
This equation is identical to Eq.~(13) of \citet{2000ApJ...538..326L} if we set
the total torque to be the sloshing torque $\GGbig=\GGbig^{(s)}$ (neglecting the
viscous torque for reasons that $\alpha_t\ll 1$), and define
\begin{equation}\label{eq-def-omega-a}
\omega_a \equiv \frac{\kappa^2-1}{2} \simeq \kappa-1
\end{equation}
where the last step is approximately valid for $|\kappa-1|\ll 1$.

\subsection{Solution to problem \ref{enum-problem-2}: Rotating $\GGbig^{(s)}$}
\label{sec-rotating-G}
\begin{figure}
  \centerline{\includegraphics[width=0.25\textwidth]{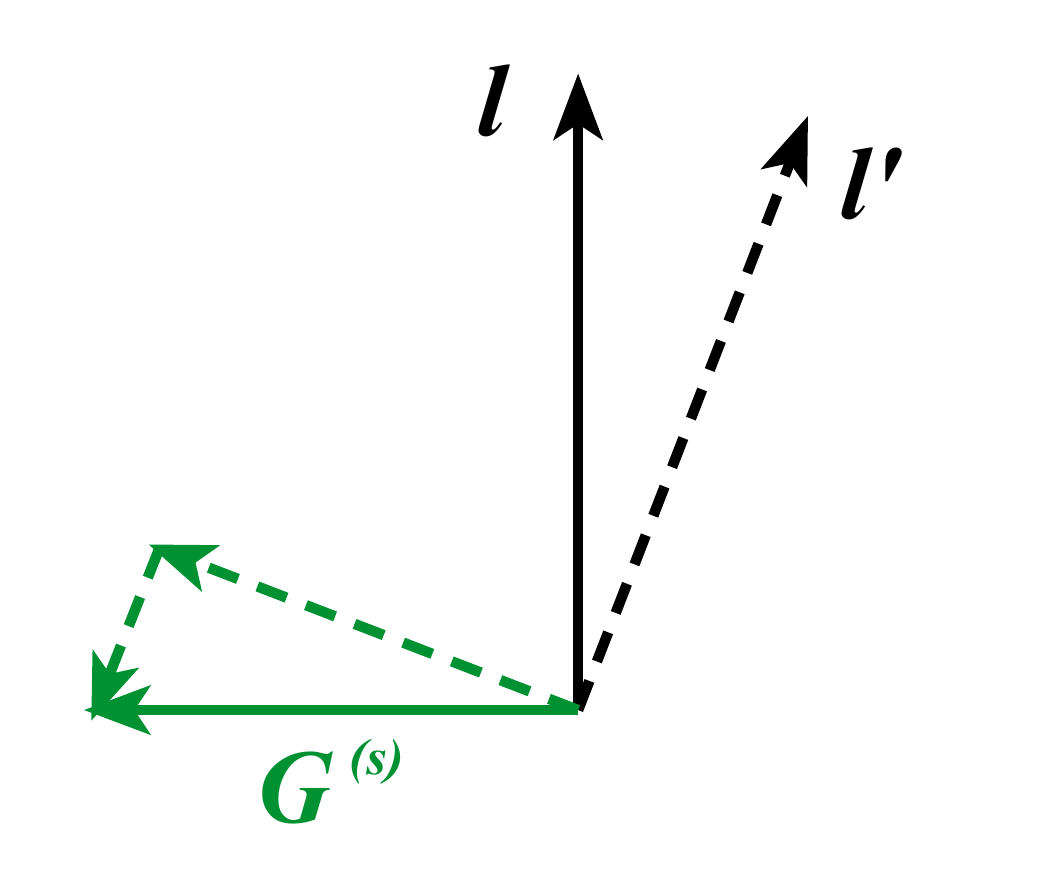}}
  \caption{\label{fig-cartoon-leakage}Cartoon of how the in-plane sloshing
    torque acquires a perpendicular component upon a change of the orientation
    vector of the orbital plane $\llunit\rightarrow \llunit'$. Solid lines:
    before the change, dashed lines: after the change, at which point
    $\GGbig^{(s)}$ has acquired a $\GGbig^{(s)}\cdot\llunit'\neq 0$
    component. This ``leakage'' of the sloshing torque into the perpendicular
    viscous torque can quickly dominate over the true perpendicular viscous
    torque $\GGbig^{(v)}$, and can unphysically affect the viscous evolution of
    $\Sigma(r,t)$.  }
\end{figure}

Problem \ref{enum-problem-2} is harder to solve, because it requires knowledge about if and how
$\GGbig^{(s)}$ changes when the orbital plane changes, i.e.\ when $\llunit$
changes. If we look again at the individual components of $\GGbig^{(s)}$ given
in Eq.~(\ref{eq-Gsplit-Gslosh}), we see that the nature of $\GGbig^{(s)}$ is
fundamentally fixed to the orientation of the orbital plane. The sloshing
motion, and thereby the sloshing torque, is strictly in the plane of the disc
(the $X$-$Y$-plane), or in other words: $\GGbig^{(s)}\cdot\llunit=0$. If the
orientation of the orbital plane changes, i.e.~$d\llunit/d\tau\neq 0$, and if we
leave the 3-D orientation of $\GGbig^{(s)}$ unchanged, then in the new
coordinates $(X',Y',Z')$, which describe the new orbital plane, we will see
$\GGbigcomp^{(s)}_X$ and $\GGbigcomp^{(s)}_Y$ rotationally mix into
$\GGbigcomp^{(s)}_{Z'}$.  This means that the latter becomes non-zero, and the
condition $\GGbig^{(s)}\cdot\llunit=0$ is broken. This is illustrated in
Fig.~\ref{fig-cartoon-leakage}.

Given that the sloshing torque is usually orders of magnitude larger than the
viscous torque ($|\GGbigcomp^{(s)}|\gg|\GGbigcomp^{(v)}|$), even a comparatively
small non-zero $\GGbigcomp^{(s)}_{Z'}$ can easily dominate the much smaller
perpendicular viscous torque $\GGbigcomp^{(v)}_{Z'}=\Xii q\alpha_t$.
So this ``leakage'' of the sloshing torque into the perpendicular torque
component is not harmless: It can strongly affect the perpendicular component of
$\GGbig=\GGbig^{(s)}+\GGbig^{(v)}$ (i.e.~the value of $\GGbig\cdot\llunit$),
which is responsible for the radial accretion of the disc material. This can
therefore lead to strange effects in the evolution of $\Sigma(r,t)$, as is
demonstrated in \citet{2019ApJ...875....5M}, their Fig.~(1, upper right panel),
and will be further discussed in Section \ref{sec-numerical-study}.

The question is now: Is this ``pollution'' of the $\GGbig\cdot\llunit$ component
by the sloshing torque physical or not? If it is physical, then it should be
possible to find a sloshing motion that produces a non-zero component
of $\GGbig^{(s)}$ in the direction of $\llunit$. At least in the linear regime, and
assuming that $\vi(z')=V_{i}\Omega_0 z'$
(cf.\ Eqs.~\ref{eq-vx-in-VOmz}-\ref{eq-vz-in-VOmz}), no such motion can be
found, because under these conditions the general form of the torque is given by
Eqs.~(\ref{eq-ggbig-cmplx-x}-\ref{eq-ggbig-cmplx-z}), which does not have a
component of $V$ in the expression for $\GGbig_Z$.

Assuming that this holds also in the non-linear regime, we can say that when
$\llunit$ changes, the sloshing torque $\GGbig^{(s)}$ changes as well, such that
$\GGbig^{(s)}\cdot\llunit$ stays zero at all times.

We conjecture at this point that the sloshing torque $\GGbig^{(s)}$ is simply
co-rotated with the change (i.e.\ rotation) of $\llunit$. The
rotation-per-unit-dimensionless-time is defined by two vectors, $\llunit$ and
$d\llunit/d\tau$. Together they define a rotation axis ${\bf R}\equiv
\llunit\times d\llunit/d\tau$. Using Rodrigues' formula in the limit of
infinitesimal rotations we find that the rotation of $\GGbig^{(s)}$ is
given by
\begin{equation}\label{eq-rotation-of-G-slosh}
\frac{\partial\GGbig^{(s)}}{\partial\tau} = \left(\llunit\times\frac{d\llunit}{d\tau}\right)\times \GGbig^{(s)}
\end{equation}

Putting it all together, we add the rotation term of
Eq.~(\ref{eq-rotation-of-G-slosh}) to the evolution equation for $\GGbig^{(s)}$
(Eq.~\ref{eq-vect-form-of-G-hom-to-part-alt-Q}, or
Eq.~\ref{eq-vect-form-of-G-hom-to-part-alt-Qtilde-smalllimit} in the limit of
small $\alpha_t$ and $|\kappa-1|$), and we obtain the full evolution
equation for $\GGbig^{(s)}$ which avoids the unphysical ``leakage'' of the
sloshing torque into the perpendicular torque component. The general result is:
\begin{equation}\label{eq-vect-form-of-Gslosh-with-rot}
  \begin{split}
    \frac{\partial \GGbig^{(s)}}{\partial \tau}  + & (\kappa-1)\llunit{}\times \GGbig^{(s)} +
    \alpha_t\GGbig^{(s)} \\=& 
    -\Xii \left(\tilde Q_2\frac{d\llunit{}}{d\ln r}+\tilde Q_3\llunit{}\times \frac{d\llunit{}}{d\ln r}\right)\\
    &+\left(\llunit\times\frac{d\llunit}{d\tau}\right)\times \GGbig^{(s)}
  \end{split}
\end{equation}
In the limit of small $\alpha_t$ and $|\kappa-1|$ this simplifies to (cf.\ Eq.~\ref{eq-vect-form-of-G-hom-to-part-alt-Qtilde-smalllimit}):
\begin{equation}\label{eq-vect-form-of-Gslosh-with-rot-limit}
  \begin{split}
    \frac{\partial \GGbig^{(s)}}{\partial \tau}  &+  (\kappa-1)\llunit{}\times \GGbig^{(s)} +
    \alpha_t\GGbig^{(s)} \\
    &\simeq -\frac{\Xii}{4}\frac{d\llunit{}}{d\ln r} +\left(\llunit\times\frac{d\llunit}{d\tau}\right)\times \GGbig^{(s)}
  \end{split}
\end{equation}
The total torque, at each moment in time, is then
\begin{equation}\label{eq-G-split-worked-out}
  \GGbig = \GGbig^{(s)}+\GGbig^{(v)} = \GGbig^{(s)} +
  q\Xii\alpha_t\llunit - \Xii\alpha_t \warpvector
\end{equation}
where $\GGbig^{(s)}$ is the solution of
Eq.~(\ref{eq-vect-form-of-Gslosh-with-rot}) or
Eq.~(\ref{eq-vect-form-of-Gslosh-with-rot-limit}), and where we used
Eq.~(\ref{eq-G-visc-coordfree}) in the final step.

Together with the global mass and angular momentum equations (see Section
\ref{sec-global-equations}) the above equations for $\GGbig{}(\tau)$
(Eqs.~\ref{eq-vect-form-of-Gslosh-with-rot}/\ref{eq-vect-form-of-Gslosh-with-rot-limit},
\ref{eq-G-split-worked-out}) forms a full set of dynamic equations for the
evolution of a warped disc in the limit of small $\alpha_t$.
They unify the diffusive and wave-like regime.

It should be kept in mind, however, that the rotational behavior of
$\GGbig^{(s)}$ according to Eq.~(\ref{eq-rotation-of-G-slosh}) is a conjecture.
It makes intuitive sense, and it removes an unphysical property of the previous
equations, but in itself it remains unproven.

The extra rotational term in Eq.~(\ref{eq-vect-form-of-Gslosh-with-rot-limit}) is
not seen in the equations of \citet{1999MNRAS.304..557O,2000ApJ...538..326L}.
Given that these papers were not concerned with the co-evolution of $\Sigma(r,t)$,
they were not concerned with the $\GGbig^{(s)}\cdot\llunit$ component of their
torque. Since
\begin{equation}
\GGbig^{(s)}\cdot \left(\llunit\times\frac{d\llunit}{d\tau}\right)\times \GGbig^{(s)} = 0
\end{equation}
this term, at least in the limit of small warps, does not affect their results
for the wavelike propagation of the warp. But it is important when, in addition,
the viscous evolution of $\Sigma(r,t)$ is considered. 

When using Eq.~(\ref{eq-vect-form-of-Gslosh-with-rot}) or
Eq.~(\ref{eq-vect-form-of-Gslosh-with-rot-limit}) in a numerical integration
algorithm, it should, however, be kept in mind that even the slightest numerical
error in the rotation may still ``leak'' too much of the sloshing torque into
the perpendicular torque component. This is because, for small $\alpha_t$, the
magnitude of the in-plane sloshing torque is so much larger than the
perpendicular viscous torque. In practice it may therefore improve the stability
and reliability of the algorithm to ``reset'' the perpendicular component of the
sloshing torque to zero every time step:
\begin{equation}\label{eq-reset-Gz}
\GGbig^{(s)} := \GGbig^{(s)} - (\GGbig^{(s)}\cdot\llunit)\llunit
\end{equation}
In fact, numerical experimentation shows (see Section \ref{sec-numerical-study})
that this ``resetting trick'' is so effective for sufficiently small time steps,
that the rotational term is no longer strictly needed: the ``resetting'' does
the rotation automatically. This is because the rotational term is perpendicular
to the orbital plane, and thus perpendicular to the sloshing torque (as is true
for any infinitesimal rotation). The numerical errors in the orbital plane are
quadratic in the rotation angle, and thus become vanishingly small for
sufficiently small time steps.

\subsection{Alternative solution to problem \ref{enum-problem-2}: Martin's $\beta$ terms}
\label{sec-martin-beta-terms}
\citet{2019ApJ...875....5M} choose a different path to overcome the ``leakage''
problem. They do not distinguish between the sloshing torque $\GGbig^{(s)}$ and
the viscous torque $\GGbig^{(v)}$, but instead add a strong damping term to their
equation for $\partial \GGbig/\partial\tau$ that forces the $\GGbig\cdot\llunit$
component to converge to the correct value on a short time scale. This time scale
$\tau_{\mathrm{damp}}=1/\beta$, with $\beta$ the proportionality parameter of their
damping term, can be chosen at will. The shorter this damping time scale
(i.e.\ the larger $\beta$), the better the artificial ``leakage'' is suppressed,
but at the cost of the equations becoming more stiff and thus harder to solve
numerically.

Let us derive their equation starting from our
Eqs.~(\ref{eq-vect-form-of-G-hom-to-part}, \ref{eq-Gp-afo-Q123-repeat}) but with
$\GGbig^{(s)}$ replaced by $\GGbig$
\begin{equation}\label{eq-vect-form-of-G-hom-to-part-full}
  \frac{\partial \GGbig}{\partial \tau} = -(\kappa-1){\bf l}\times (\GGbig-\GGbig_{p0})
  -\alpha_t(\GGbig-\GGbig_{p0})
\end{equation}
and with $\GGbig_{p0}$ given by
\begin{equation}\label{eq-Gp-afo-Q123-repeat-full}
\GGbig_{p0} = -\Xii \left(Q_1\llunit + Q_2\frac{d{\bf l}}{d\ln r}+Q_3{\bf l}\times \frac{d{\bf l}}{d\ln r}\right)
\end{equation}
(instead of Eq.~\ref{eq-Gp-afo-Q123-repeat}). In other words, the
$\GGbig\cdot\llunit$ component is now included in the equation, and the
$Q_1\llunit$ term on the right-hand-side ensures that the $\GGbig\cdot\llunit$
dynamically converges to the correct value $\Xii q\alpha_t$ over a
time scale $1/\alpha_t$. Following the same path as before, we can derive the
equivalent of Eq.~(\ref{eq-vect-form-of-G-hom-to-part-alt-Qtilde})
\begin{equation}\label{eq-vect-form-of-G-hom-to-part-alt-Qtilde-full}
  \begin{split}
    \frac{\partial \GGbig}{\partial \tau}  + & (\kappa-1)\llunit{}\times \GGbig +
    \alpha_t\GGbig \\=& 
    -\Xii \left(\tilde Q_1\llunit + \tilde Q_2\frac{d\llunit{}}{d\ln r}+\tilde Q_3\llunit{}\times \frac{d\llunit{}}{d\ln r}\right)
  \end{split}
\end{equation}
where $\tilde Q_1=\alpha_t Q_1$, and the $\tilde Q_2$ and $\tilde Q_3$ are
defined by Eqs.~(\ref{eq-qtilde-2-def}, \ref{eq-qtilde-3-def}). Following the
same reasoning as before, we arrive, for small $\alpha_t$ and $|\kappa-1|$, at the equivalent of
Eq.~(\ref{eq-vect-form-of-G-hom-to-part-alt-Qtilde-smalllimit})
\begin{equation}\label{eq-vect-form-of-G-hom-to-part-alt-Qtilde-smalllimit-full}
    \frac{\partial \GGbig}{\partial \tau}  +  (\kappa-1)\llunit{}\times \GGbig +
    \alpha_t\GGbig \simeq \Xii q\alpha_t^2\; \llunit -\frac{\Xii}{4}\frac{d\llunit{}}{d\ln r}
\end{equation}
This is the equivalent of \citet{2019ApJ...875....5M}'s Equation (15), without
their $\beta$-terms. If $\llunit$ changes very slowly with time, then
Eq.~(\ref{eq-vect-form-of-G-hom-to-part-alt-Qtilde-smalllimit-full}) assures
that the $\GGbig\cdot \llunit$ component dynamically converges to the correct
value. In principle this convergence also automatically damps away any
``leaked'' sloshing torque, but the time scale of
$\tau_{\mathrm{damp}}=1/\alpha_t\gg 1$ is much too long to suppress this
leakage. To ameliorate this, \citet{2019ApJ...875....5M} add two terms to their
equation:
\begin{equation}\label{eq-martin-beta-terms}
  \begin{split}
    \frac{\partial \GGbig}{\partial \tau}  &+  (\kappa-1)\llunit{}\times \GGbig +
    \alpha_t\GGbig + \beta(\GGbig\cdot\llunit)\llunit \\
    & = \Xii q\alpha_t(\alpha_t+\beta)\; \llunit -\frac{\Xii}{4}\frac{d\llunit{}}{d\ln r}
  \end{split}
\end{equation}
For values of $\beta\gg |d\llunit/d\tau|^{-1}$ we can estimate the behavior of the
perpendicular torque component $\GGbig\cdot\llunit$ by taking the inner product of this
equation with $\llunit$:
\begin{equation}
  \frac{\partial(\GGbig\cdot\llunit)}{\partial\tau} + (\alpha_t+\beta)\GGbig\cdot\llunit
  = \Xii q\alpha_t(\alpha_t+\beta)
\end{equation}
where we neglected the $\GGbig\cdot d\llunit/d\tau$ term. The solution to this equation
is the usual exponential convergence to the value
\begin{equation}
\GGbig\cdot\llunit \rightarrow \Xii q\alpha_t
\end{equation}
and the time scale is $\tau_{\mathrm{damp}}=1/(\alpha_t+\beta)$. \citet{2019ApJ...875....5M}
report values of $\beta\gtrsim 10$ to lead to reasonably good results.

In the hypothetical limit of $\beta\rightarrow\infty$ this method always
guarantees that $\GGbig\cdot\llunit=\Xii q\alpha_t$, or equivalently that
$\GGbig^{(s)}\cdot\llunit=0$ at all times. In an explicit numerical integration
algorithm this can, alternatively, be achieved by applying the ``resetting
condition'' Eq.~(\ref{eq-reset-Gz}) at each time step. This avoids the numerical
stiffness problem for large $\beta$ values, and achieves the same effect as the
$\beta$-terms in Eq.~(\ref{eq-martin-beta-terms}), as will be discussed in
Section \ref{sec-numerical-study}.

\section{Discussion}

\subsection{Numerical analysis of the behavior of the equations}
\label{sec-numerical-study}
To investigate the behavior of our new equations, we perform the same test
calculation as presented by \citet{2019ApJ...875....5M} in their Section 4.
It is a dimensionless setup, with a disk ranging from $r_\mathrm{in}=1$ to
$r_\mathrm{out}=20$, with a surface density that has smooth cut-offs at both
the inner and outer radius, and a warp of 10 degrees, {given by the equation
\citep[see][]{2019ApJ...875....5M}
\begin{equation}
  \Sigma(r,t=0) = \Sigma_0 \left( \frac{r}{r_\mathrm{in}} \right)^{-p} \left[1-\left(\frac{r_\mathrm{in}}{r} \right)^{1/2} \right] \left[1 - e^{r-r_\mathrm{out}} \right].
\end{equation}
Here, $\Sigma_0$ is a scaling constant and can be chosen arbitrarily.
Initially, the disk is only warped in one direction (`untwisted') and remains
that way throughout the simulations. This is due to the absence of precession
terms in the equations we use. Therefore, the initial shape of the disk can be
determined by the inclination angle $i$ which is the angle between the local
orbital plane and the $x$-$y$--plane. We set the initial inclination profile to
\citep[see][]{2019ApJ...875....5M}
\begin{equation}
i(r,t=0) = 10° \left[\frac{1}{2} \tanh \left( \frac{r-r_\mathrm{warp}}{r_\mathrm{width}} \right) + \frac{1}{2} \right],
\end{equation}
with the location of the warp $r_\mathrm{warp}=10$ and the warp width $r_\mathrm{width}=2$.
}
At the start of the calculation, $\GGbig(r,t=0)$ is set to the
viscous torque $\GGbig^{(v)}(r)$, i.e.\ the sloshing torque is set to zero. The
viscosity is set to $\alpha_t=0.01$ and the disc aspect ratio is set
to $H_p/r=0.1$.

The procedure for the numerical integration method of our generalized warped
disc equation is described in Section \ref{sec-numerical-recipe}.

\begin{figure*}
  \centerline{\includegraphics[width=0.97\textwidth]{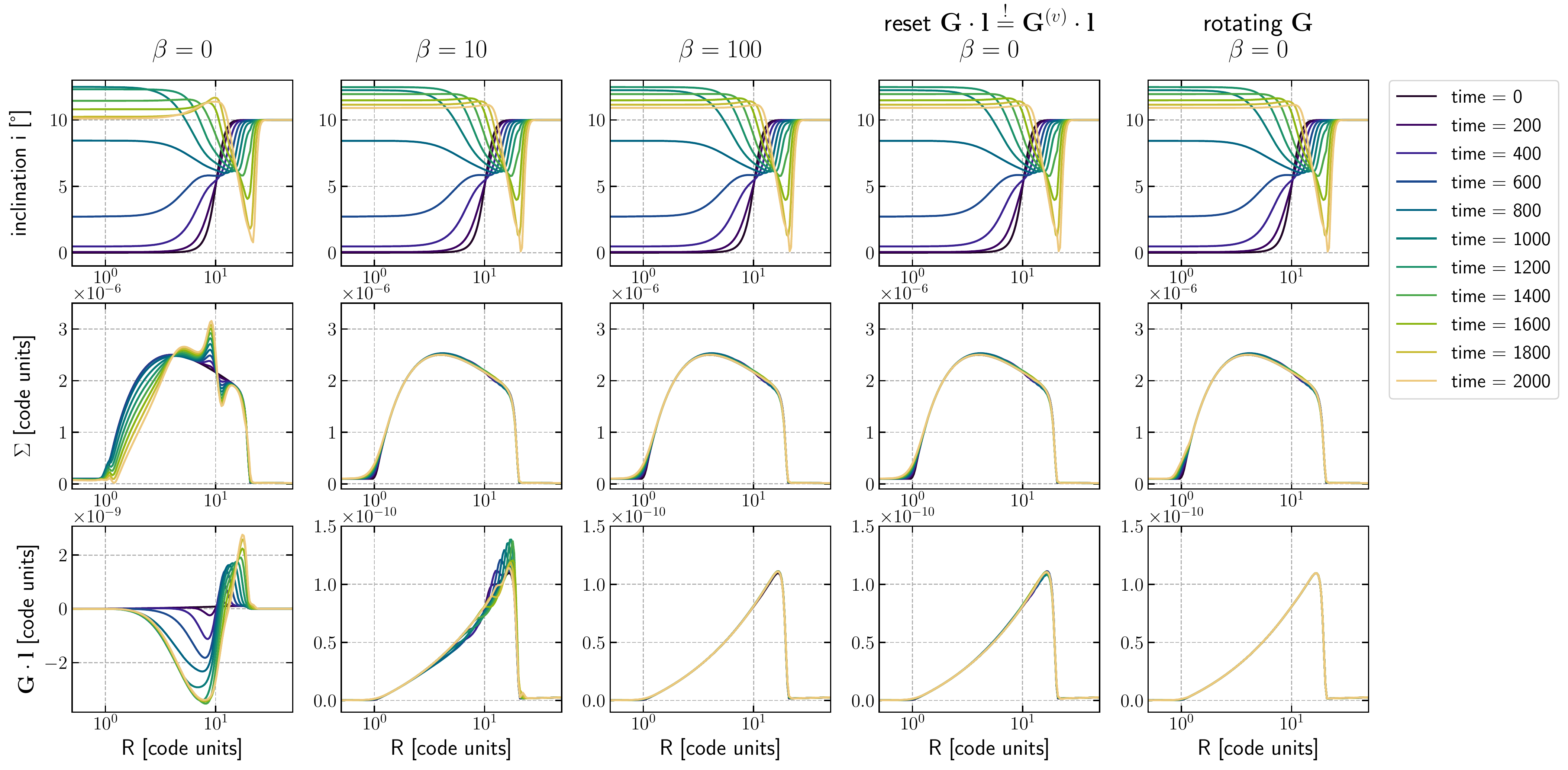}}
  \caption{\label{fig-beta-rotframe} Results of the test model of
    \citet{2019ApJ...875....5M} for five different methods of handling the
    $\GGbig\cdot\llunit$ component of the internal torque (from left to
    right). In the first four, the equations of \citet{2019ApJ...875....5M} are
    used. In the fourth panel, $\beta$ is again 0, but the $\GGbig\cdot\llunit$
    component is reset to the correct value $\GGbig^{(v)}\cdot\llunit$ after
    each macroscopic time step (i.e.\ not during the sub-timestepping done by
    the {\small\tt odeint} integrator of {\small\tt scipy}). In the final panel,
    the rotation term is added to the equation for $\GGbig$ (final term in
    Eq.~\ref{eq-vect-form-of-Gslosh-with-rot-limit}), which is the equation
    proposed in this paper. Top row: the inclination $i$ in degrees. Middle row:
    the surface density $\Sigma(r,t)$. Bottom row: the $\GGbig\cdot\llunit$.}
\end{figure*}

We integrate until dimensionless time 2000, which is roughly one wave-crossing
time for the warp. We carry out these calculations in five different ways, from
left to right in Fig.~\ref{fig-beta-rotframe}:
\begin{enumerate}
\item In the first calculation, we use use the equations of \citet{2019ApJ...875....5M},
  with $\beta=0$. This is equal to our equations without the rotation term (i.e.\ without
  the $(\llunit\times d\llunit/d\tau)\times\GGbig$ term). This calculation shows what
  goes wrong when the $\GGbig\cdot\llunit$ component of the internal torque is left
  ``untreated''. The surface density acquires a strong wiggle, first identified by
  \citet{2019ApJ...875....5M}, which is not seen in 3-D hydrodynamic simulations
  of warped discs \citep[e.g.][]{2015MNRAS.448.1526N}. We also see a strong effect on the
  inner edge of the disc, which is equally suspicious. In the bottom panel the cause
  of these phenomena can be seen. The $\GGbig\cdot\llunit$ component should, physically,
  be just the viscous torque. Yet, the ``leakage'' of the sloshing torques into this
  perpendicular component of $\GGbig$ completely overpowers the viscous torque,
  and even causes (inward of a radius of 10 to 20 code units) a {\em negative}
  torque. This de-facto acts as a negative viscosity, and thus violates the second
  law of thermodynamics. It must therefore be an unphysical effect.
\item The same, but now with $\beta=10$, to damp the $\GGbig\cdot\llunit$
  component back to what it should be:
  $\GGbig\cdot\llunit\rightarrow\GGbig^{(v)}\cdot\llunit$. Apart from
  small deviations this is indeed successful. The remaining ``triangle
  like'' curve is the expected curve for the viscous $\GGbig^{(v)}\cdot\llunit$.
  As a result, no spurious features appear in the $\Sigma(r,t)$. The remaining
  viscous evolution (very minor effects on these time scales) are physical.
\item The same, but now with $\beta=100$, i.e.\ even stronger damping.  The
  small remaining deviations from $\GGbig^{(v)}\cdot\llunit$ are nearly
  gone. Both this and the $\beta=10$ case demonstrate that the method of
  \citet{2019ApJ...875....5M} indeed works as advertised. However, a large
  $\beta$ makes the equations stiff and thus harder to solve numerically.
\item Instead of using the $\beta$ damping, we reset
  $\GGbig\cdot\llunit\rightarrow\GGbig^{(v)}\cdot\llunit$ after each time step.
  This is also succesful, and achieves the same result as for $\beta=100$.
\item Finally, including the rotational term to the equation (last term of
  Eq.~\ref{eq-vect-form-of-Gslosh-with-rot-limit}), but without $\beta$ and
  without resetting. One can see that $\GGbig\cdot\llunit$ does not have
  any ``leakage'' of the sloshing torque into the perpendicular torque, and
  it is even closer to the correct value, without any damping or resetting.
\end{enumerate}

We conclude that our equation (Eq.~\ref{eq-vect-form-of-Gslosh-with-rot-limit})
with the rotation term, possibly with an occasional ``reset'',
is the most physically consistent equation for $\GGbig$ for modeling the
evolution of a warped disc.

%
%
%

\subsection{Shearing box versus tilted slab interpretation}
\label{sec-shearingbox-tiltedslab}
\begin{figure*}
  \centerline{\includegraphics[width=0.80\textwidth]{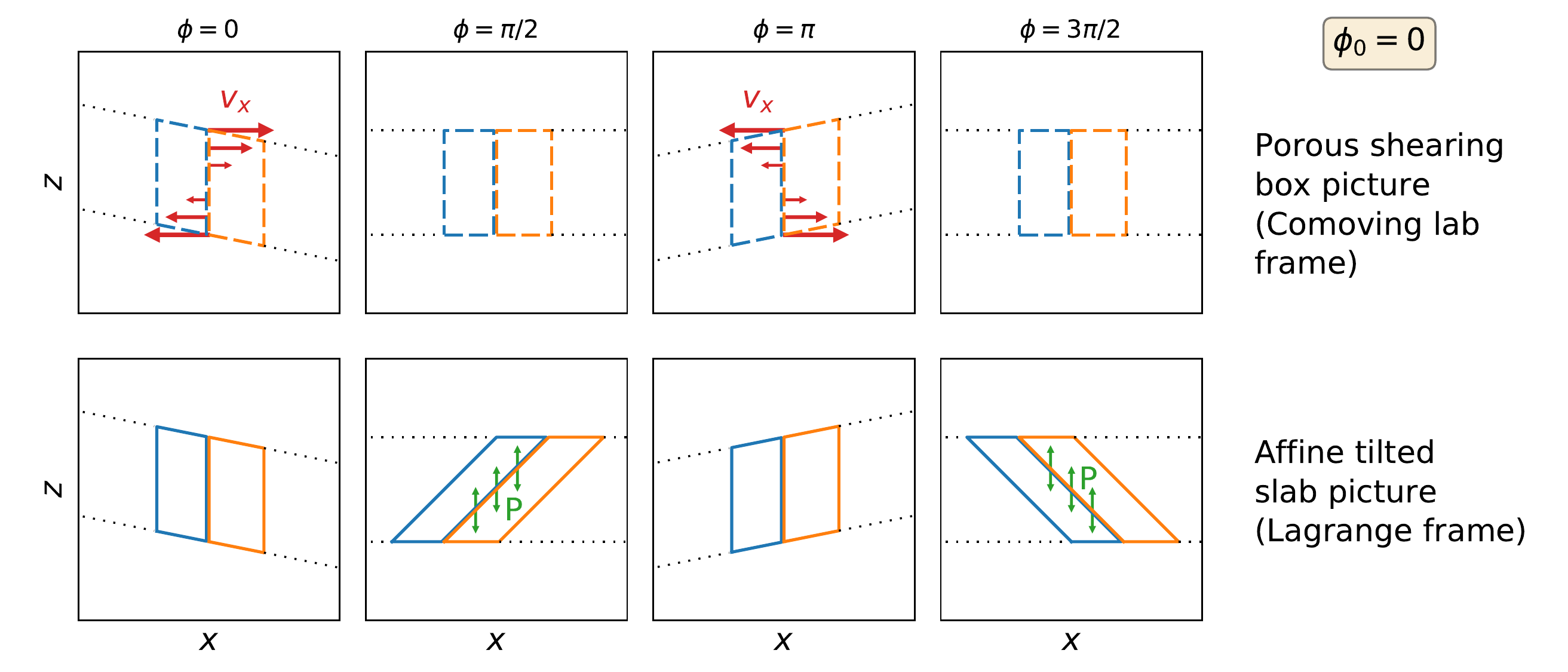}}
  \caption{\label{fig-cartoon-phi0-0}Cartoon of the sloshing motion for the case
    $\phi_0=0$ ($Q_2>0$, $Q_3=0$), i.e.\ pure warp-damping, without
    twisting. Two interpretations are shown. Top row: The comoving lab frame
    picture, which is the way by which all equations in this paper have been
    derived (using a porous shearing box). The red arrows show the radial
    velocity $\vx(z)$ of the gas, transporting angular momentum between the blue
    and the orange annuli. Bottom row: The Lagrange picture, in which the slabs
    (annuli) are tilted due to the horizontal sloshing motion (see the affine
    tilted slab model of \citet{2018MNRAS.477.1744O}). The green arrows show the
    vertical pressure force exchange, which lead to a torque that the blue and
    the orange annuli exert on each other. The cartoons are qualitative, and not
    to scale.}
\end{figure*}

The fact that the sloshing motion causes an internal torque is known since
\citet{1983MNRAS.202.1181P}. However, the physics behind this fact is not
intuitively obvious. In the top row of Fig.~\ref{fig-cartoon-phi0-0} the case of
pure warp-damping ($\phi_0=0$) without twisting is pictographically shown. In
this porous shearing box picture, the torque is entirely due to angular momentum
advected between adjacent annuli (here shown in blue and orange) by the
horizontal velocity $\vx(z')$. Gas pressure plays no role from this point of
view, because the box walls are vertical, so that the blue box cannot exert a
vertical force on the orange box and vice versa. The derivations in this paper
were done in this porous shearing box framework.

In the bottom row of Fig.~\ref{fig-cartoon-phi0-0} the exact same situation is
depicted from a Lagrange perspective \citep[see the affine model
  of][]{2018MNRAS.477.1744O}. Here, the annuli are not porous, but get tilted by
the horizontal velocity $\vx(z')$. The advective transport of angular momentum no
longer plays a role. Instead, the angular momentum exchange between the blue and
the orange annuli is now entirely mediated by the pressure, because due to the
tilt, the pressure now causes the two annuli to extert a vertical force on each
other.

To show that the two perspectives are equivalent, we can compute the tilt angle
$\theta$ as a function of $\tau$. {Comoving with the orbital motion,
it obeys the following ODE:}
\begin{equation}
\frac{d\tan\theta(\tau)}{d\tau} = V^{\mathrm{re}}_x(\tau)
\end{equation}
{
If $\VVx(\tau)$ is the $m=1$ periodic solution, then in the
$(\tau,\phi)$-picture (see Section \ref{sec-phi-tau}), the $\tau$-coordinate is
replaced by the $\phi$ coordinate. So the above differential equation in $\tau$
becomes the same differential equation in $\phi$. Given that the $m=1$ mode
goes as $e^{i\phi}$, the solution is:}
\begin{equation}
\tan\theta(\phi) = \mathrm{Re}[-i\VVx(\phi)] = V^{\mathrm{im}}_x(\phi)
\end{equation}
{where we set the integration constant to zero.}
Here $\theta=0$ means the slab is vertical and $\theta>0$ means the slab is
tilted clockwise, i.e.~$\tan\theta=x/z$ for a gas parcel belonging to the slab
going through $(x,z)=(0,0)$.  With a non-zero tilt, the gas pressure acquires
the possibility to transport $z$-momentum in $z$ direction from one slab to the
next. In other words, neighboring tilted slabs can exchange $y$-angular
momentum with each other through the pressure force. The internal torque is
defined as the transmission of angular momentum from one slab at $r_0$ to its
neighbor at $r_0+\Delta r$:
\begin{equation}
\begin{split}
  \Ggvertcomp_y &= r_0\int_{-\infty}^{\infty}p(z')\tan(\theta)dz'\\
  &= \Omega_0^2r_0\Sigma h_p^2\tan(\theta(\phi)) \equiv \Xii\tan(\theta(\phi))
  \equiv \Xii V^{\mathrm{im}}_x(\phi)
\end{split}
\end{equation}
In complex notation we thus have
\begin{equation}
  \Ggvertcomp_y(\phi) = -i\Xii \VVx(\phi)
\end{equation}
and $\Ggvertcomp_x=0$. Inserting these into Eqs.~(\ref{eq-from-x-to-X}, \ref{eq-from-y-to-Y})
and the results into Eqs.~(\ref{eq-ggbigx-integral}, \ref{eq-ggbigy-integral}),
{following the method outlined in Appendix \ref{app-azimuthal-mean-internal-torque}},
gives
\begin{eqnarray}
2\GGbigcomp_X/\Xii &=& - \VVx(\tau)\\
2\GGbigcomp_Y/\Xii &=& -i \VVx(\tau)
\end{eqnarray}
which reproduces the same results as for the comoving lab-frame (porous shearing
box) analysis (Eqs.~\ref{eq-ggbig-cmplx-x}, \ref{eq-ggbig-cmplx-y}) in the limit
$\alpha_t\ll 1$.

\begin{figure}
  \centerline{\includegraphics[width=0.45\textwidth]{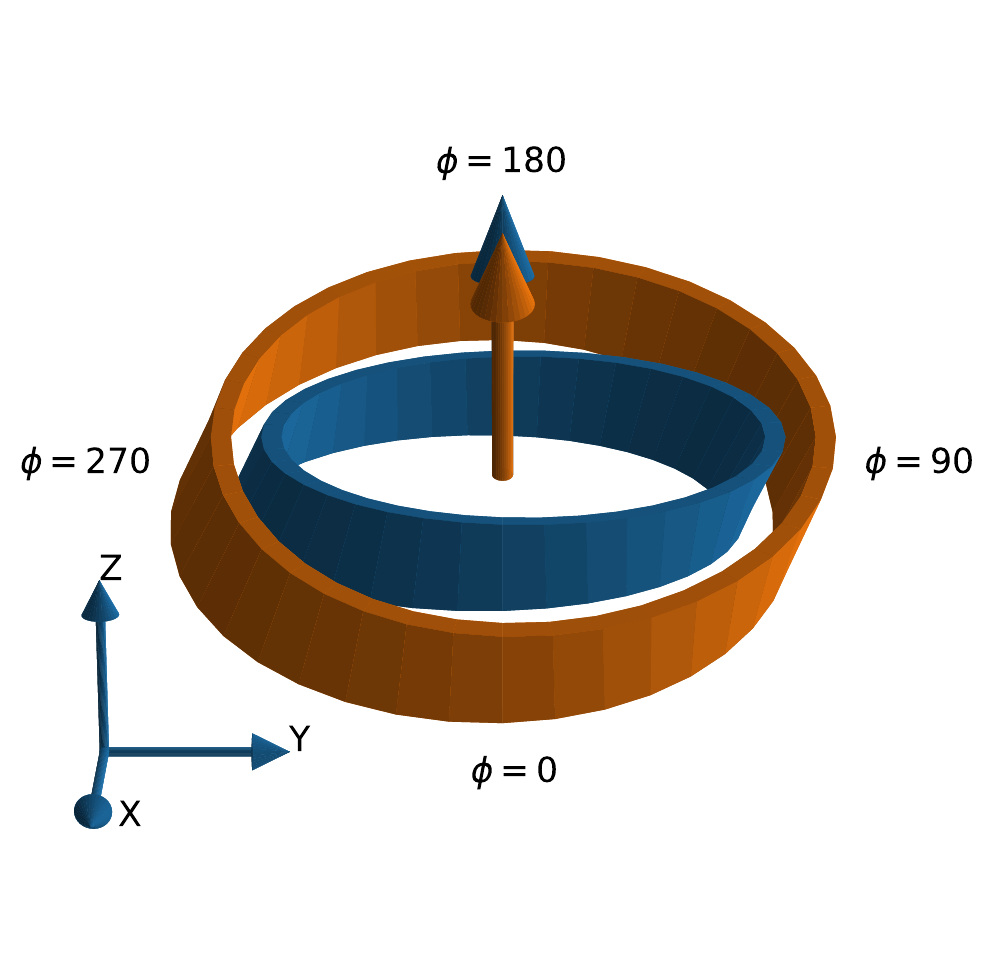}}
  \caption{\label{fig-two-tilted-slabs}A 3-D rendering of the sloshing / tilting
    motion shown in Fig.~\ref{fig-cartoon-phi0-0} (i.e.\ for $\phi_0=0$, a purely
    damping torque), shown for two adjacent slabs,
    following the ``affine tilted slab'' approach of Ogilvie (2018). The inner
    blue slab is at $x=0$ and is thus not inclined. The outer orange slab is at
    $x>0$ and is inclined with respect to the blue slab. The large arrows are
    the ${\bf l}$ unit vectors for the blue and orange annulus,
    respectively. The orbital motion is counter-clockwise when seen from above
    (right-hand rule). Model parameters are: $q=1.5$, $\psi=0.1$ and
    $\alpha_t=0.1$.  The maximum tilting of the slabs is at $\phi=\pi/2$ and
    $\phi=3\pi/2$, and therefore the torque damps the warp.}
\end{figure}

For perfectly keplerian discs ($q=1.5$) the amplitude of the tilting (for the
case of strong viscous damping) is greatest at $\phi\simeq \pi/2$ and
$\phi\simeq 3\pi/2$, as illustrated in Fig.~\ref{fig-cartoon-phi0-0}. A 3-D
version of this is shown in Fig.~\ref{fig-two-tilted-slabs}. This means that the
pressure-driven torque acts as an $X$-torque and thus damps the warp in this
case. 

Twisting torques ($Q_3\neq 0$) can appear if the sloshing motion is
phase-shifted with respect to the pure damping case ($\phi_0\neq 0$). This leads to
a shift in the {location $\phi$ along the annulus where the tilt is maximal},
and thus a rotation of the $\GGbig$ vector around the $\llunit$ vector {by an
amount $\phi_0$}. This
leads to non-zero $\GGbigcomp_Y$, which twists the disc.

\subsection{Caveats of the model}
\label{sec-caveats-shearing-box}
{
While the equations for warped disks discussed in this paper are easy to use and
numerically cheap to integrate, they have strong limitations. First, they assume
that all orbits are circular. Eccentricity can in principle be included
\citep[see e.g.][]{2021MNRAS.500.4110L}, but that requires also a treatment of
the disk variables along each orbit, making the model essentially 2-D.  There
are also other reasons why a 2-D treatment (in the coordinates $r$ and $\phi$)
may be necessary. For instance, any out-of-plane companion (planet or binary
companion) that induces the warp in the disk may also induce $m=2$ waves
(spirals) or higher modes in the disk that may be of equal importance as the
warp itself. A treatment similar to the 2-D affine model of
\citet{2018MNRAS.477.1744O} may be necessary to overcome these limitations
without directly resorting to 3-D models.

A further major caveat is that the equations of this paper are not capable of
treating the breaking of a warped disk. It has been found in many 3-D
simulations that under certain conditions the disk cannot maintain a smooth
continuous shape, but instead will break into two (or more) disconnected disks
\citep{1997MNRAS.285..288L, 2010MNRAS.405.1212L, 2010A&A...511A..77F,
  2012MNRAS.421.1201N, 2013MNRAS.434.1946N, 2013MNRAS.433.2142F,
  2016MNRAS.455L..62N, 2021arXiv210105824R}. This phenomenon likely requires a fully 3-D treatment.

On the other hand, for computational feasibility, fully global 3-D models of
warped disks necessarily have limited numerical resolution at the disk interior
scale, i.e.\ at spatial scales smaller than the pressure scale height. At low
viscosity, the strong sloshing motion induced by even small warps can induce
turbulence \citep[][]{1993MNRAS.260..323K, 2013MNRAS.433.2420O} that may be
stronger than the background turbulence \citep[][]{2000MNRAS.318...47T}.  A
local 2-D or 3-D shearing box calculation such as by \citet{2019MNRAS.483.3738P}
can, however, be used to update the local turbulent viscosity coefficient
$\alpha_t$, which might then be used in the 1-D warped disk equations again.
Whether a simple analytic ``fitting formula'' can be found for this self-induced
turbulent viscosity is not yet clear.

Finally there is the issue of irradiation. For disks around black holes or
neutron stars the radiation pressure of this radiation may be strong enough to
induce and enhance warping of a disk \citep{1996MNRAS.281..357P} and cause
precession \citep{2001MNRAS.320..485O}.  But even for
less strong irradiation: the warp causes the disk to be irradiated from one side
only, switching side each half orbit. This issue has been studied very recently
in the context of the chemistry \citep{2021arXiv210602660Y}, but to our
knowledge it has not yet been investigated if it
affects the internal dynamics of the disk and thus affects the
internal torque.}

\subsection{Applications of the model}
\label{sec-applications}
{One of the reasons why the two regimes of warped disk dynamics have
  traditionally been treated separately is because accretion disks around black
  holes and neutron stars are usually geometrically very thin ($h_p\ll r$) and
  hot ($T\gg 1000\,\mathrm{K}$) so that they are firmly in the diffusive regime
  ($\alpha_t>h_p/r$). This is because for sufficiently ionized disks the
  magnetohydrodynamics is nearly ideal, and the saturated magnetorotational
  turbulence reaches values of the order of $\alpha_t\sim 10^{-2}\cdots 10^{-1}$
\citep[e.g.][]{2015A&A...579A.117M}. Protoplanetary disks, on the other hand, are
  relatively cold ($T\lesssim 1500\,\mathrm{K}$ in the dusty outer parts) and
  geometrically not very thin ($h_p/r\sim 0.05$). They may thus be much less
  turbulent than hot disks. The precise value of $\alpha_t$ in those disks
  is still a matter of active debate, \citep[see][]{lesur:ppvii}, but
  values of the order of $\alpha_t\sim 10^{-4}$ are often suggested, which
  would put protoplanetary disks firmly in the wavelike regime.}

{However, at early times and close enough to the star, protoplanetary disks
  are likely to be much hotter, and fully saturated magnetorotational turbulence
  ($\alpha_t\sim 10^{-2}\cdots 10^{-1}$) should be expected, putting these
  regions of the protoplanetary disk right in between the two regimes, where the
  unified set of equations have to be used. Furthermore, as discussed in
  Section \ref{sec-caveats-shearing-box}, the vertical shear instability
  driven by the sloshing motion may induce strong turbulence, perhaps even stronger
  than the ``traditional'' vertical shear instability (VSI) driven by the
  baroclinic structure of flat disks \citep{2004A&A...426..755A,
    2013MNRAS.435.2610N, 2014A&A...572A..77S}. It is not yet clear if
  these instabilities may lead to $\alpha_t$ of the order of $h_p/r$, but
  if they do, then again, we would be in between the two regimes.}

{And even if the disk is in the wavelike regime, this does not mean that the
  viscous radial transport of mass within the disk is not important. The warp
  waves have a short timescale for travelling through the disk, given that
  their phase velocity is $c_s/2$ \citep{1999MNRAS.309..929N}. However, the life
  time of these disks is much longer, and during that time the viscous evolution
  of the surface density $\Sigma$ is likely to be non-negligible.  If the warp
  continues to be driven, for instance by a companion, then the warp dynamics
  and viscous radial mass redistribution have to be treated simultaneously. This
  required the use of the generalized warp equations.}

{When applying the equations to protoplanetary disks, it becomes important
  to be able to include external torques from companions (stars or planets) onto
  the disk. This can be done in the usual way by adding an external torque
  $\TTbig{}$ to the angular momentum conservation equation
  (Eq.~\ref{eq-angmom-cons}), for instance using the equations from
  \citet{2001ApJ...562.1012E}, \citep[see also e.g.][]{2015MNRAS.447..747L,
    2018MNRAS.473..603Z}. However, this torque will change the orbital
  orientation $\llunit$ of the annuli of the disk, leading to the same
  ``leakage'' problem as we discussed in Section \ref{sec-ode-for-ggbig}
  (``problem ii''). The solution is then the same as before: rotating
  $\GGbig^{(s)}$ such that it stays in the plane. In other words, the
  $d\llunit/d\tau$ in the rotational terms in
  Eq.~(\ref{eq-vect-form-of-Gslosh-with-rot} or
  \ref{eq-vect-form-of-Gslosh-with-rot-limit}) now should include the change of
  $\llunit$ due to the external torque $\TTbig{}$.}

\subsection{Numerical recipe for modeling warped discs}
\label{sec-numerical-recipe}
Given the complexity of the mathematical description of viscously evolving
warped discs, as described in this paper, we here summarize our recommended set
of equations and a method of solution for modeling the dynamics of warped discs.
\begin{enumerate}
\item The main conservation equation to solve is the equation of angular momentum
  conservation Eq.~(\ref{eq-angmom-cons}).
\item Conservation of mass is given by Eq.~(\ref{eq-mass-cons}), but this
  equation is not explicitly needed. The reason is because once we know the
  angular momentum density $\LLbig(r,t)$, we can use Eq.~(\ref{eq-def-llbig}) to
  compute the surface denisty $\Sigma(r,t)$ and the unit vector $\llunit(r,t)$,
  because the Kepler frequency $\Omega(r)$ is known. Deviations from Keplerian
  rotation are not considered here. Optionally one can solve the mass
  conservation equation in addition to the angular momentum
  conservation. This leads to a redundancy (see Eq.~\ref{eq-def-llbig}), which can then serve as an
  indicator of the magnitude of the numerical error made. 
\item The torque vector $\GGbig(r,t)$ is given by
  Eq.~(\ref{eq-G-split-worked-out}), where $\GGbig^{(s)}(r,t)$ obeys
  Eq.~(\ref{eq-vect-form-of-Gslosh-with-rot}).
\item The values of $\tilde Q_2$ and $\tilde Q_3$ are obtained from
  Eqs.~(\ref{eq-qtilde-2-def}, \ref{eq-qtilde-3-def}), which depend on $Q_2$ and
  $Q_3$, the values of which are obtained from Eqs.~(\ref{eq-Q2-fullsol},
  \ref{eq-Q3-fullsol}). For small values of $\alpha_t$ and $|\kappa-1|$ one can,
  instead, use the values $\tilde Q_2=1/4$ and $\tilde Q_3=0$.
\item The value for $\Xii$ is given in Eq.~(\ref{eq-xii-def}).
\item The symbol $q$ is defined in Eq.~(\ref{eq-definition-q}), and
  $\kappa$ is defined in Eq.~(\ref{eq-epicyclic-freq-dimless}).
\item The dimensionless time $\tau$, used in the above equations,
  is defined by Eq.~(\ref{eq-tau-t}).
\end{enumerate}

Numerically one can make a linearly or logarithmically spaced grid in $r$ with
$N_r$ grid points, and on each grid point one places 6 variables:
$\LLbigcomp_X$, $\LLbigcomp_Y$, $\LLbigcomp_Z$, $\GGbigcomp_X$, $\GGbigcomp_Y$,
$\GGbigcomp_Z$ (or 7 variables, if one opts to include $\Sigma$ as mentioned
above), where here the coordinate system $(X,Y,Z)$ can be freely chosen
(i.e.\ it is not necessarily aligned with the warp, as it was during the
derivation of the equations). This gives, at each time step, $6N_r$ variables.
Given that the equations for $\LLbigcomp_i$ are conservation equations, they are
best discretized in finite volume conserved form. The equations for
$\GGbigcomp_i$ are strictly local equations (no spatial derivatives
involved). They are best formulated at a staggered mesh (grid point in between
those of $\LLbigcomp_i$), although also a co-spatial gridding for $\LLbigcomp_i$
and $\GGbigcomp_i$ is possible. An explicit integration of these equations sets
the usual time step limit (Courant-Friedrichs-Lewy condition). Implicit and/or
higher-order integration can be done using e.g.\ the integration library of
SciPy, e.g.~the {\small\tt solve\_ivp()} or {\small\tt odeint()} routines of
SciPy, where the full $6N_r$ variables are given as a single vector to
{\small\tt solve\_ivp()} or {\small\tt odeint()} as initial condition, and
for each time point, the $6N_r$ variables of that time are returned. The
advantage of using such a library is that these routines have internal handling
of integration step size and use an automatic internal computation of the
Jacobian for handling stiffness.

Since the rotation term in Eq.~(\ref{eq-vect-form-of-Gslosh-with-rot}) involves
$d\llunit/d\tau$, which is a time-derivative itself, it is necessary to compute
the time derivatives of the 6 variables at each grid point in a particular
order: First compute the time derivatives of $\LLbigcomp_X$, $\LLbigcomp_Y$ and
$\LLbigcomp_Z$, from which $d\llunit/d\tau$ can be computed. This can then be
used to compute the time derivatives of $\GGbigcomp_X$, $\GGbigcomp_Y$ and
$\GGbigcomp_Z$.

Although mathematically the rotational term $(\llunit\times
d\llunit/d\tau)\times\GGbig^{(s)}$ in
Eq.~(\ref{eq-vect-form-of-Gslosh-with-rot}) should keep
$\GGbig^{(s)}\cdot\llunit=0$, it is recommended, for numerical stability, to
``reset'' the $\GGbig^{(s)}\cdot\llunit$ component to zero using
Eq.~(\ref{eq-reset-Gz}) once in a while.

\section{Conclusions}

\begin{enumerate}
\item The ``sloshing'' motion (the resonant epicyclic motion) induced by a warp
  obeys a time-dependent ODE, and approaches a steady-state oscillation that
  depends on the local warp vector $d\llunit{}/dr$. The time scale for reaching
  this steady-state oscillation is $\delta t\sim 1/(\alpha_t\Omega_0)$.
\item In the limit of small warps and weak ``sloshing'' motion, this
  time-dependence can be described as the homogenous solution of two coupled
  linear ODEs (Eqs.~\ref{eq-slosh-Vx-fullring}, \ref{eq-slosh-Vy-fullring}). The
  full solution is the sum of the steady-state particular solution and a
  homogeneous solution (Eqs.~\ref{eq-Vx-fullring-fullsol},
  \ref{eq-Vy-fullring-fullsol}). The amplitude of the homogeneous solution is
  set by the initial condition: the difference between the initial condition and
  the steady-state particular solution. The time dependence of the homogeneous
  solution is through a factor $e^{i\omega_0\tau}$, with $\tau=\Omega_0t$ and
  $\omega_0=\kappa-1+i\alpha_t$, where $\kappa$ is the dimensionless epicyclic
  frequency $\Omega_e/\Omega_0$ and $\alpha_t$ is the turbulence parameter. For
  $\alpha_t>0$ this homogeneous solution decays on a time scale $\delta t\sim
  1/(\alpha_t\Omega_0)$, so that the full solution approaches the steady-state
  one on that time scale.
\item The resulting torque vector $\GGbig$ consists of the sum of the sloshing
  torque $\GGbig^{(s)}$ plus the viscous torque $\GGbig^{(v)}$. The viscous
  torque is given by Eq.~(\ref{eq-G-visc-coordfree}). The sloshing torque is
  caused by the sloshing motion, and obeys a time-dependent local ordinary
  differential equation (Eq.~\ref{eq-vect-form-of-Gslosh-with-rot}). For a fixed
  disc orientation $\llunit$ and warp $\warpvector$, the solution to this
  equation is the sum of the steady-state particular solution
  $\GGbig^{(s)}_{p0}$ (Eq.~\ref{eq-Gp-afo-Q123-repeat}) and time-dependent
  decaying homogeneous part $\GGbig^{(s)}_{h}(\tau)$
  (Eqs.~\ref{eq-GGbig-hompart-X}, \ref{eq-GGbig-hompart-Y}). The decay time
  scale is $t_{\mathrm{decay}}=1/(\Omega_0\alpha_t)$. 
\item If the disc orientation $\llunit$ and warp $\warpvector$ change slowly
  compared to $1/\alpha_t$, the sloshing torque $\GGbig^{(s)}$ will be close to
  $\GGbig^{(s)}_{p0}$. Taking $\GGbig^{(s)}=\GGbig^{(s)}_{p0}$ will then be a
  good approximation, meaning that the torque vector will be uniquely
  determined by the local conditions given by $\llunit$ and $\warpvector$.
  This is the diffusive regime.
\item If the disc orientation $\llunit$ and warp $\warpvector$ change rapidly
  compared to $1/\alpha_t$, the sloshing torque $\GGbig^{(s)}$ will never
  converge to $\GGbig^{(s)}_{p0}$. Instead, $\GGbig^{(s)}$ will be a dynamic
  quantity, for which the ordinary differential equation
  Eq.~\ref{eq-vect-form-of-Gslosh-with-rot} has to be solved time-dependently.
  This is the wavelike regime.
\item Eq.~\ref{eq-vect-form-of-Gslosh-with-rot} thus unifies the well-known
  diffusive and wavelike regimes of warped discs. For small values of
  $\alpha_t$ and $|\kappa-1|$, appropriate for protoplanetary discs for instance,
  this equation simplifies to Eq.~\ref{eq-vect-form-of-Gslosh-with-rot-limit}.
\item The in-plane $-\Xii\alpha_t\warpvector$ component of the viscous torque
  $\GGbig^{(v)}$ (Eq.~\ref{eq-G-visc-coordfree}) is always much smaller
  than the sloshing torque, and is thus insignificant.
\item The perpendicular $\Xii q\alpha_t\llunit$ component of the viscous torque
  $\GGbig^{(v)}$ (Eq.~\ref{eq-G-visc-coordfree}) drives the viscous evolution of
  the surface density $\Sigma(r,t)$ of the disc. On the time scale of the
  typical wave crossing of a warp through the disc, this viscous evolution is
  typically relatively slow, because warp waves move with half the sound speed
  \citep[see e.g.][]{2016LNP...905...45N}, while the viscous disc time scales
  are of the order of $1/((h_p/r)\alpha_t)$ times longer. However, if a warp is
  continuously driven by an external body (through the external torque $\TTbig$
  in Eq.~\ref{eq-angmom-cons}), the warped disc equations have to be integrated
  over long enough time scales for the viscous evolution to matter. Under these
  conditions, the viscous torque $\GGbig^{(v)}$ cannot be neglected.
\item Compared to earlier work \citep{1999MNRAS.304..557O, 2000ApJ...538..326L},
  our Eq.~(\ref{eq-vect-form-of-Gslosh-with-rot}), or its simplified form
  Eq.~(\ref{eq-vect-form-of-Gslosh-with-rot-limit}), contains an extra term
  which rotates the sloshing torque $\GGbig^{(s)}$ along with the rotation of
  the orbital plane as $\llunit$ changes with time. This rotation avoids the
  emergence of an unphysical out-of-plane component of the sloshing torque.
  This is necessary to avoid unphysical effects on the evolution of
  $\Sigma(r,t)$. \citet{2019ApJ...875....5M} first discussed the emergence
  of such unphysical effects, and presented another approach to prevent
  them, by introducing the $\beta$-damping of this out-of-plane component.
\item The physical meaning of the sloshing torque $\GGbig^{(s)}$ can be
  elegantly described in terms of the affine tilted slab picture of
  \citet{2018MNRAS.477.1744O}, where the sloshing causes the vertical slab to
  tilt back and forth, allowing it to exert pressure in vertical direction on
  its neighbors, thereby exerting a torque on them. The azimuthal phase of the
  oscillating tilt, $\phi_0$, determines whether the torque is purely damping
  ($\phi_0=0$) or has a twisting effect too ($\phi_0\neq 0$).
\end{enumerate}

\section*{Acknowledgements}
We thank Gordon Ogilvie for insightful comments on an early version of this work
that helped us greatly to complete the picture we present in this paper. We also
thank Rebecca Nealon, Rebecca Martin for helpful discussions. For this paper we
used {\small\tt Python} with the {\small\tt NumPy}, {\small\tt SciPy} and
{\small\tt Matplotlib} libraries. Finally, we thank the anonymous referee
for helpful suggestions that allowed us to improve the manuscript. 

\section*{Data Availability}
This paper does not rely on external data.



\bibliographystyle{mnras}
\bibliography{paper} 




\appendix

\section{Alternative form of global angular momentum conservation equation}
\label{app-global-warp-eqs-alternative}
{
While the equations of Section \ref{sec-global-equations} are
complete and general, the angular momentum conservation equation has often
been formulated in different ways. For instance, in the diffusive regime,
\citet{1983MNRAS.202.1181P}
formulate it as (generalized to include the twisting term):
\begin{equation}\label{eq-angmom-cons-alt}
  \begin{split}
  \frac{\partial\LLbig}{\partial t} + \frac{1}{r}\frac{\partial}{\partial r}\big(
  r\LLbig{}v_r\big) =& -\frac{1}{r}\frac{\partial}{\partial r}
  \big(\nu_1Lq\;\llunit\big)
  +\frac{1}{r}\frac{\partial}{\partial r}
  \Big(\frac{1}{2}\nu_2L\;\warpvector\Big)\\
  &+\frac{1}{r}\frac{\partial}{\partial r}
  \big(\nu_3L\;\llunit\times\warpvector\big)
  +\TTbig{}
  \end{split}
\end{equation}
where we define $L\equiv|\LLbig|=\Sigma\Omega r^2$, and we
remind that $\warpvector=d\llunit/d\ln r$ is a vectorial
quantity (cf.~Eq.~\ref{eq-def-psivec-warp}). The three
viscosities are related to the $Q$s as follows:
\begin{eqnarray}
  \nu_1 &=& -\Omega h_p^2 q^{-1} Q_1 \\
  \nu_2 &=& 2\Omega h_p^2\; Q_2 \\
  \nu_3 &=& \Omega h_p^2\; Q_3
\end{eqnarray}
If these viscosity coefficients are formulated in the standard way
in terms of $\alpha$ values according to $\nu=\alpha \Omega h_p^2$
one finds:
\begin{eqnarray}
  \alpha_1 &=& -q^{-1} Q_1 = \alpha_t\\
  \alpha_2 &=& 2 Q_2 \\
  \alpha_3 &=& Q_3
\end{eqnarray}
where the $Q$-values are given in Eqs.~(\ref{eq-Q1-fullsol}-\ref{eq-Q3-fullsol}).
For $\alpha_t\ll 1$ and in the limiting case of $q=3/2$ (perfect Kepler), and thus 
$\epsilon=\kappa^2-1=0$ we get the familiar results that
$\alpha_2\simeq 1/(2\alpha_1)$ and $\alpha_3\simeq 3/8$. But it should be noted
that this formulation only holds for the diffusive regime, because it does
not allow a dynamical torque.
}

\section{Derivations of sub-steps}
\label{app-derivations}
\subsection{Equations of motion: From unwarped to warped frame}
\label{app-derivations-unwarped-warped}
Here we provide background information for the transformation of the equations
of motion from the unwarped $(x,y,z)$ coordinates
Eqs.~(\ref{eq-eom-standard-box-dt-x}-\ref{eq-eom-standard-box-dt-uz}) to the
warped $(x',y',z')$ coordinates with the modified velocity definitions
$(\vx,\vy,\vz)$
Eqs.~(\ref{eq-eom-warped-box-dt-x}-\ref{eq-eom-warped-box-dt-vz}).  The
coordinates $(x,y,z)$ and $(x',y',z')$ are related via
Eqs.~(\ref{eq-def-xprime}-\ref{eq-def-zprime})

The relation between the $(u_x,u_y,u_z)\equiv (D_tx,D_ty,D_tz)$ and $(u_x',u_y',u_z')\equiv (D_tx',D_ty',D_tz')$ velocities
is derived by taking the comoving time derivative $D_t$ of
Eqs.~(\ref{eq-def-xprime}-\ref{eq-def-zprime}):
\begin{eqnarray}
u_x &=& u_x'\label{eq-rel-u-uprime-x}\\
u_y &=& u_y'\label{eq-rel-u-uprime-y}\\
u_z &=& u_z' + \psi \Omega_0 x' \sin(\phi) - \psi\cos(\phi)u_x'\label{eq-rel-u-uprime-z}
\end{eqnarray}
where comoving with the orbital motion means $\phi=\Omega_0t$.
The relations between the $u$-velocities and the $v$-velocities
are given by Eqs.~(\ref{eq-def-vxprime}-\ref{eq-def-vzprime}),
which we repeat here for convenience:
\begin{eqnarray}
  u_x &=& \vx\label{eq-def-vxprime-repeat}\\
  u_y &=& \vy - q\Omega_0 x'\label{eq-def-vyprime-repeat}\\
  u_z &=& \vz + \psi \Omega_0 x' \sin(\phi)\label{eq-def-vzprime-repeat}
\end{eqnarray}
The comoving derivative in the $(x,y,z)$ coordinate system is:
\begin{equation}\label{eq-xyz-comoving-derivative}
D_t  = \partial_t + u_x\partial_x + u_y\partial_y + u_z\partial_z
\end{equation}
The partial derivatives in the two systems are related via\footnote{Note that
  the distinction between $t$ and $t'$ is only made for the partial time
  derivative ($\partial_t$ versus $\partial_{t'}$), in which the only difference
  is which other coordinates are kept constant while taking this derivative. In
  all other aspects $t\equiv t'$.}
\begin{eqnarray}
\partial_t &=& \partial_{t'} -\psi\Omega_0\sin(\phi)x'\partial_{z'}\label{eq-partial-t-in-prime}\\
\partial_x &=& \partial_{x'} + \psi\cos(\phi)\partial_{z'}\label{eq-partial-x-in-prime}\\
\partial_y &=& \partial_{y'} \label{eq-partial-y-in-prime}\\
\partial_z &=& \partial_{z'} \label{eq-partial-z-in-prime}
\end{eqnarray}
The divergence of ${\bf u}$ becomes:
\begin{equation}
  \label{eq-div-u-warped}
\begin{split}
  \nabla\cdot{\bf u}
  &= \partial_x u_x + \partial_y u_y + \partial_z u_z  \\
  &= (\partial_{x'} + \psi\cos(\phi)\partial_{z'})\vx
  + \partial_{y'}\vy + \partial_{z'}\vz
\end{split}
\end{equation}
The comoving time derivative $D_t$ (Eq.~\ref{eq-xyz-comoving-derivative}) then
becomes in the $(x',y',z')$ system:
\begin{equation}\label{eq-warped-comoving-derivative}
\begin{split}
D_t  =& 
\partial_{t'} -\psi\Omega_0\sin(\phi)x'\partial_{z'}
+ \vx(\partial_{x'} + \psi\cos(\phi)\partial_{z'})\\
&+ (\vy-q\Omega_0x')\partial_{y'} 
+ (\vz + \psi \Omega_0 x' \sin(\phi))\partial_{z'}\\
=&\partial_{t'} + \vx\partial_{x'} 
+ (\vy-q\Omega_0x')\partial_{y'} \\
&+ (\vz + \psi \vx \cos(\phi))\partial_{z'}
\end{split}
\end{equation}
The comoving derivatives of the velocities are:
\begin{eqnarray}
  D_t u_x &=& D_t \vx\\
  D_t u_y &=& D_t \vy - D_t(q\Omega_0 x')\nonumber\\
  &=& D_t \vy - q\Omega_0\vx\\
  D_t u_z &=& D_t \vz + D_t(\psi\Omega_0 x'\sin(\phi)) \nonumber\\
  &=& D_t \vz + \psi\Omega_0 \sin(\phi) \vx + \psi\Omega_0^2 x'\cos(\phi) 
\end{eqnarray}
Finally we note that as a result of $\phi=\Omega_0t$, when we follow
the motion of a fluid parcel, the comoving derivative of $\phi$
becomes
\begin{equation}
D_t\phi = \Omega_0
\end{equation}
which is the origin of the $\sin(\phi)$ terms in the above
equations. With these equations, it becomes straightforward to derive
Eqs.~(\ref{eq-eom-warped-box-dt-x}-\ref{eq-eom-warped-box-dt-vz}) from
Eqs.~(\ref{eq-eom-standard-box-dt-x}-\ref{eq-eom-standard-box-dt-uz}).

\subsection{Shear viscosity forces in the warped frame}
\label{app-f-shearvisc}
The viscous shear stress tensor $\tvisc_{ij}$, with $i,j=x,y,z$, is, in the
comoving lab frame coordinates $(x,y,z)$, given by
\begin{equation}
\tvisc_{ij} = \rho\nu \shear_{ij}
\end{equation}
with the shear tensor given by
\begin{equation}
\shear_{ij} = \partial_i u_j+\partial_j u_i-\tfrac{2}{3}\delta_{ij}\nabla\cdot {\bf u}
\end{equation}
where $\delta_{ij}$ is the Kronecker delta. In the warped frame $(x',y',z')$ with the
mixed-frame velocities $(\vx,\vy,\vz)$ the shear tensor becomes:
\begin{eqnarray}
\shear_{xx} \kern-0.5em &=& \kern-0.5em 2(\partial_{x'}+\psi\cos(\phi)\partial_{z'}) \vx  - \tfrac{2}{3}\nabla\cdot {\bf u}\label{eq-sheartensor-warped-xx}\\
\shear_{yy} \kern-0.5em &=& \kern-0.5em 2\partial_{y'} \vy  - \tfrac{2}{3}\nabla\cdot {\bf u}\\
\shear_{zz} \kern-0.5em &=& \kern-0.5em 2\partial_{z'} \vz  - \tfrac{2}{3}\nabla\cdot {\bf u}\\
\shear_{xy} = \shear_{yx} \kern-0.5em &=& \kern-0.5em (\partial_{x'}+\psi\cos(\phi)\partial_{z'}) \vy + \partial_{y'} \vx - q\Omega_0\label{eq-sheartensor-warped-yx}\\
\shear_{yz} = \shear_{zy} \kern-0.5em &=& \kern-0.5em \partial_{y'} \vz + \partial_{z'} \vy\\
\shear_{zx} = \shear_{xz} \kern-0.5em &=& \kern-0.5em \partial_{z'} \vx + (\partial_{x'}+\psi\cos(\phi)\partial_{z'}) \vz \nonumber\\
&& +\psi \Omega_0\sin(\phi)\label{eq-sheartensor-warped-zx}
\end{eqnarray}
It is important to keep in mind that the components of the shear tensor are
still in the original $(x,y,z)$ orthogonal directions; they are just formulated
with partial derivatives in the $(x',y',z')$ coordinates.
The viscous force density $f_i^v$ in Eq.~(\ref{eq-all-body-forces}) can now be
expressed as:
\begin{equation}
f_i^v = (1/\rho)\big(\partial_x \tvisc_{xi}+\partial_y \tvisc_{yi}+\partial_z \tvisc_{zi}\big)
\end{equation}
where the partial derivatives are taken in the non-warped coordinates $(x,y,z)$.
To cast them into partial derivatives in the coordinates $(x',y',z')$ we use
Eqs.~(\ref{eq-partial-x-in-prime}-\ref{eq-partial-z-in-prime}):
\begin{equation}
f_i^v = (1/\rho)\big((\partial_{x'}+\psi\cos(\phi)\partial_{z'}) \tvisc_{xi}+\partial_{y'} \tvisc_{yi}+\partial_{z'} \tvisc_{zi}\big)
\end{equation}
Again, $i$ here refers to the $(x,y,z)$ directions, not the warped ones.
So far the expressions for the shear viscous tensor are general, and
can be used for numerical 3-D warped shearing box modeling.

For the laminar solutions that are translationally symmetric in $x'$ and
$y'$, all instances of $\partial_{x'}$ and $\partial_{y'}$ become zero.
The divergence of the velocity is reduced to
\begin{equation}
\nabla\cdot{\bf u}=\psi\cos(\phi)\partial_{z'}\vx + \partial_{z'}\vz 
\end{equation}
The components of $\shear_{ij}$ then simplify to:
\begin{eqnarray}
\shear_{xx} \kern-0.5em &=& \kern-0.5em \tfrac{4}{3}\psi\cos(\phi)\partial_{z'}\vx  - \tfrac{2}{3}\partial_{z'}\vz\label{eq-sheartensor-warped-laminar-xx}\\
\shear_{yy} \kern-0.5em &=& \kern-0.5em - \tfrac{2}{3}\psi\cos(\phi)\partial_{z'}\vx - \tfrac{2}{3}\partial_{z'}\vz \\
\shear_{zz} \kern-0.5em &=& \kern-0.5em \tfrac{4}{3}\partial_{z'} \vz  - \tfrac{2}{3}\psi\cos(\phi)\partial_{z'}\vx\\
\shear_{xy} = \shear_{yx} \kern-0.5em &=& \kern-0.5em \psi\cos(\phi)\partial_{z'}\vy  - q\Omega_0\label{eq-sheartensor-warped-laminar-yx}\\
\shear_{yz} = \shear_{zy} \kern-0.5em &=& \kern-0.5em \partial_{z'} \vy\\
\shear_{zx} = \shear_{xz} \kern-0.5em &=& \kern-0.5em \partial_{z'} \vx + \psi\cos(\phi)\partial_{z'}\vz+\psi \Omega_0\sin(\phi)\label{eq-sheartensor-warped-laminar-zx}
\end{eqnarray}

Next we employ the Ansatz that the velocities are linear in $z'$ and zero at $z'=0$,
by using Eqs.~(\ref{eq-vx-in-VOmz}-\ref{eq-vz-in-VOmz}). The derivatives of the
velocities then become
\begin{equation}
\partial_{z'}v_i  = \Omega_0 V_i
\end{equation}
with, as usual, $i=x,y,z$. The components of $\shear_{ij}$ then simplify even more:
\begin{eqnarray}
\shbig_{xx} \kern-0.5em &=& \kern-0.5em \tfrac{4}{3}\psi\cos(\phi)\VVx  - \tfrac{2}{3}\VVz\\
\shbig_{yy} \kern-0.5em &=& \kern-0.5em - \tfrac{2}{3}\psi\cos(\phi)\VVx - \tfrac{2}{3}\VVz \\
\shbig_{zz} \kern-0.5em &=& \kern-0.5em \tfrac{4}{3} \VVz  - \tfrac{2}{3}\psi\cos(\phi)\VVx\\
\shbig_{xy} = \shbig_{yx} \kern-0.5em &=& \kern-0.5em \psi\cos(\phi)\VVy  - q\\
\shbig_{yz} = \shbig_{zy} \kern-0.5em &=& \kern-0.5em  \VVy\\
\shbig_{zx} = \shbig_{xz} \kern-0.5em &=& \kern-0.5em  \VVx + \psi\cos(\phi)\VVz+\psi\sin(\phi)
\end{eqnarray}
where we defined
\begin{equation}
\shbig_{ij}\equiv \Omega_0^{-1}\shear_{ij}
\end{equation}
for notational convenience. The shear viscosity force is then
\begin{equation}\label{eq-shear-viscosity-force}
f_i^v = \Omega_0(1/\rho)\big(\psi\cos(\phi)\shbig_{xi}+\shbig_{zi}\big)\partial_{z'}(\rho\nu)
\end{equation}
where the $\partial_{z'}$ now only acts on $\rho\nu$, because
$\partial_{z'}\shbig_{ij}=0$ due to the linear velocity Ansatz. The viscosity
coefficient $\nu$ is written in the classical way as
\begin{equation}\label{eq-nu-in-alpha-hp-om0}
\nu = \alpha_t \frac{c_s^2}{\Omega_0} = \alpha_t h_p^2\Omega_0
\end{equation}

For the assumption of vertical isothermality, $(1/\rho)\partial_{z'}(\rho\nu)$
can be worked out further as:
\begin{equation}
  (1/\rho)\partial_{z'}(\rho\nu) = \nu\partial_{z'}\ln\rho
  = - \alpha_t \Omega_0 z'
\end{equation}
where we used the Gaussian vertical structure of
Eq.~(\ref{eq-simple-gaussian-vertical-structure-warped}).
The shear viscosity force then becomes
\begin{equation}
f_i^v = - \alpha_t \Omega_0^2 z' \big(\psi\cos(\phi)\shbig_{xi}+\shbig_{zi}\big)
\end{equation}
Following the notation of Eq.~(\ref{eq-bigf-ve}) we define a scaled, and
$z'$-independent version of this as $F^{v}_i=(\Omega_0^2z')^{-1}f^v_{i}$,
which then reads
\begin{equation}\label{eq-bigF-visc-definition}
F_i^v = - \alpha_t \big(\psi\cos(\phi)\shbig_{xi}+\shbig_{zi}\big)
\end{equation}
Concretely, these become:
\begin{eqnarray}
F_x^v &=& - \alpha_t \big((\tfrac{4}{3}\psi^2\cos^2(\phi)+1)\VVx \nonumber\\ && + \tfrac{1}{3}\psi\cos(\phi)\VVz  +\psi\sin(\phi) \big)\label{eq-Fvisc-x} \\
F_y^v &=& - \alpha_t \big((\psi^2\cos^2(\phi)+1)\VVy  - q\psi\cos(\phi) \big) \label{eq-Fvisc-y} \\
F_z^v &=& - \alpha_t \big((\psi^2\cos^2(\phi)+\tfrac{4}{3})\VVz  \nonumber\\ &&  + \tfrac{1}{3}\psi\cos(\phi)\VVx\label{eq-Fvisc-z} 
+\psi^2\sin(\phi)\cos(\phi)\big) 
\end{eqnarray}
These expressions of the viscous force can then be used for the $F^{\mathrm{ve}}_i$
in the equations of motion
Eqs.~(\ref{eq-eom-warped-box-dtau-rho-laminar-withpres}-\ref{eq-eom-warped-box-dtau-vz-laminar-withpres})

\subsection{Local internal torque for the laminar solution}
\label{app-local-internal-torque}
The components of the local internal torque are
\begin{eqnarray}
  \ggloccomp{}_x &=& -z\ttot{}_{xy} \\
  \ggloccomp{}_y &=& -r_0\ttot{}_{xz} + z\ttot{}_{xx} \\
  \ggloccomp{}_z &=& r_0\ttot{}_{xy}
\end{eqnarray}
where we should not forget that $z=z'-\psi x'\cos(\phi)$. 
The stress tensor components appearing here are:
\begin{eqnarray}
  \ttot{}_{xx} &=& \rho u_x u_x + p - \rho\nu \shear_{xx}\nonumber\\
  &=&  \rho \vx \vx + p  - \rho\nu[\tfrac{4}{3}\psi\cos(\phi)\partial_{z'}\vx \nonumber\\
    && - \tfrac{2}{3}\partial_{z'}\vz]\\
  \ttot{}_{xy} &=& \Omega_0r_0\rho u_x + \rho u_x u_y - \rho\nu \shear_{xy}\nonumber\\
  &=& \Omega_0r_0\rho \vx + \rho \vx \vy - \rho\nu [\psi\cos(\phi)\partial_{z'}\vy\nonumber\\
    &&- q\Omega_0]\\
  \ttot{}_{xz} &=& \rho u_x u_z - \rho\nu \shear_{xz}\nonumber\\
  &=& \rho \vx\vz - \rho\nu [\partial_{z'} \vx + \psi\cos(\phi)\partial_{z'}\vz \nonumber\\
    &&+\psi \Omega_0\sin(\phi)]
\end{eqnarray}
where we used the expressions for $\shear_{xx}$, $\shear_{xy}$ and $\shear_{xz}$ from
Eqs.~(\ref{eq-sheartensor-warped-laminar-xx}, \ref{eq-sheartensor-warped-laminar-yx},
\ref{eq-sheartensor-warped-laminar-zx}), and the expressions for $u_x$, $u_y$
and $u_z$ from Eqs.~(\ref{eq-def-vxprime}, \ref{eq-def-vyprime}, \ref{eq-def-vzprime}),
respectively. Furthermore we have, in the last step, set $x'=x=0$ and assumed a laminar solution
which implies $\partial_{x'}=\partial_{y'}=0$. Inserting these into the
expressions for $\ggloc{}$ yields:
\begin{eqnarray}
  \ggloccomp_x &=& -z'\Omega_0r_0\rho \vx -z'\rho \vx\vy \nonumber \\
  && + z'\rho\nu [\psi\cos(\phi)\partial_{z'}\vy - q\Omega_0]\label{eq-g-laminar-x}\\
  \ggloccomp_y &=& -r_0\rho \vx\vz+z'\rho \vx\vx + z'p\nonumber\\
  && \quad + r_0\rho\nu[\partial_{z'} \vx + \psi\cos(\phi)\partial_{z'}\vz+\psi \Omega_0\sin(\phi)] \nonumber\\
  && \quad - z'\rho\nu[\tfrac{4}{3}\psi\cos(\phi)\partial_{z'}\vx - \tfrac{2}{3}\partial_{z'}\vz] \label{eq-g-laminar-y}\\
  \ggloccomp_z &=& \Omega_0r_0^2\rho \vx + r_0\rho \vx\vy \nonumber \\
  &&- r_0\rho\nu[\psi\cos(\phi)\partial_{z'}\vy - q\Omega_0]\label{eq-g-laminar-z}
\end{eqnarray}
where $\ggloccomp_{xyz}$, $\rho$, $p$ and $\vxyz$ are to be understood as functions of $(\tau,z',\phi)$.
Now we insert $\vi=\Omega_0 V_i z'$ (cf.~Eqs.~\ref{eq-vx-in-VOmz}-\ref{eq-vz-in-VOmz}), which
is the Ansatz for the laminar solutions, and we obtain:
\begin{eqnarray}
  \ggloccomp_x &=& -(z')^2\Omega_0^2r_0\rho \VVx -(z')^3\Omega_0^2\rho \VVx\VVy \nonumber\\
  &&+ z'\rho\nu [\psi\Omega_0\cos(\phi)\VVy - q\Omega_0]\label{eq-g-ansatz-x}\\
  \ggloccomp_y &=& -(z')^2\Omega_0^2r_0\rho \VVx\VVz+(z')^3\Omega_0^2\rho \VVx\VVx + z'p\nonumber\\
  && \quad + \Omega_0r_0\rho\nu[\VVx + \psi\cos(\phi)\VVz+\psi \sin(\phi)] \nonumber\\
  && \quad - z'\Omega_0\rho\nu[\tfrac{4}{3}\psi\cos(\phi)\VVx
    - \tfrac{2}{3}\VVz] \label{eq-g-ansatz-y}\\
  \ggloccomp_z &=& \Omega_0^2r_0^2(z')\rho \VVx + r_0(z')^2\Omega_0^2\rho \VVx\VVy \nonumber\\
  &&- r_0\Omega_0\rho\nu[\psi\cos(\phi)\VVy - q]\label{eq-g-ansatz-z}
\end{eqnarray}

\subsection{Vertically integrated local internal torque for the laminar solution}
\label{app-vertint-internal-torque}
By integrating Eqs.~(\ref{eq-g-ansatz-x}-\ref{eq-g-ansatz-z}) over $z'$
\begin{equation}
\Ggvertcomp_i \equiv \int_{-\infty}^{+\infty}\ggloccomp_i dz'
\end{equation}
all terms proportional to $z'$ and $(z')^3$ intergate to zero, because
the density $\rho(z')$ and pressure $p(z')$ are even functions in $z'$.
We obtain the following components for the vertically-integrated
internal torque vector $\Ggvert{}$:
\begin{eqnarray}
  \Ggvertcomp_x &=& -\Omega_0^2r_0\Sigma h_p^2 \VVx\label{eq-g-ansatz-x-intz}\\
  \Ggvertcomp_y &=& -\Omega_0^2r_0\Sigma h_p^2 \VVx\VVz + \nonumber\\
  &&\Omega_0r_0\Sigma\nu[\VVx + \psi\cos(\phi)\VVz+\psi \sin(\phi)] \label{eq-g-ansatz-y-intz}\\
  \Ggvertcomp_z &=& \Omega_0^2r_0\Sigma h_p^2 \VVx\VVy - \Omega_0r_0\Sigma\nu[\psi\cos(\phi)\VVy - q]\label{eq-g-ansatz-z-intz}
\end{eqnarray}
where 
\begin{equation}
\Sigma = \int_{-\infty}^{+\infty}\rho dz'
\end{equation}
and we assumed that $\nu$ is independent of $z'$. Furthermore, we used
\begin{equation}
  \begin{split}
  &\int_{-\infty}^{+\infty}\rho(z')(z')^2dz'\\&=\frac{\Sigma}{\sqrt{2\pi}h_p}
  \int_{-\infty}^{+\infty}\exp\left(-\frac{(z')^2}{2h_p^2}\right)(z')^2dz'=
  \Sigma h_p^2
  \end{split}
\end{equation}
Next we replace $\nu$ using Eq.~(\ref{eq-nu-in-alpha-hp-om0}) with
$\alpha_t h_p^2\Omega_0$. If we now define, for notational convenience
\begin{equation}
\Xii \equiv \Omega_0^2r_0\Sigma h_p^2
\end{equation}
then we can write
\begin{eqnarray}
\Ggvertcomp_x/\Xii \kern-0.5em &=& \kern-0.5em - \VVx\label{eq-g-ansatz-x-intz-Xi}\\
\Ggvertcomp_y/\Xii \kern-0.5em &=& \kern-0.5em - \VVx\VVz + \alpha_t[\VVx + \psi\cos(\phi)\VVz+\psi \sin(\phi)] \label{eq-g-ansatz-y-intz-Xi}\\
\Ggvertcomp_z/\Xii \kern-0.5em &=& \kern-0.5em \VVx\VVy - \alpha_t[\psi\cos(\phi)\VVy - q]\label{eq-g-ansatz-z-intz-Xi}
\end{eqnarray}
For non-linear (numerical) solutions of $V_i$ (see Subsection
\ref{sec-non-lin-sol-tau-phi}) this is the form of the torque that has to be
used. However, for sufficiently small $V_i$ we can ignore the $\VVx\VVz$, $\VVx\VVy$
and $\psi\cos(\phi)\VVz$ terms in the above equations. They then reduce to
\begin{eqnarray}
\Ggvertcomp_x/\Xii  &=&  - \VVx\label{eq-g-ansatz-x-intz-Xi-lin}\\
\Ggvertcomp_y/\Xii  &=&  \alpha_t[\VVx + \psi \sin(\phi)] \label{eq-g-ansatz-y-intz-Xi-lin}\\
\Ggvertcomp_z/\Xii  &=&  q\alpha_t\label{eq-g-ansatz-z-intz-Xi-lin}
\end{eqnarray}
The only velocity component that remains is $\VVx$. If we want to apply the
complex version of the linear solution for $\VVx$,
Eq.~(\ref{eq-Vx-linear-full-tau-phi}), then we should also replace
$\sin(\phi)$ with $-ie^{i\phi}$. 
In addition, we now use Eq.~(\ref{eq-V-tau-phi-with-expiphi})
to write $\VVx=\VVx(\tau)e^{i\phi}$, and obtain
\begin{eqnarray}
\Ggvertcomp_x/\Xii  &=&  - \VVx(\tau)e^{i\phi}\label{eq-g-ansatz-x-intz-Xi-cmplx-lin}\\
\Ggvertcomp_y/\Xii  &=&  \alpha_t[\VVx(\tau) - i\psi ]e^{i\phi} \label{eq-g-ansatz-y-intz-Xi-cmplx-lin}\\
\Ggvertcomp_z/\Xii  &=&  q\alpha_t\label{eq-g-ansatz-z-intz-Xi-cmplx-lin}
\end{eqnarray}
where $\VVx(\tau)$ is now a complex solution to Eqs.(\ref{eq-slosh-Vx-fullring},
\ref{eq-slosh-Vy-fullring}), i.e.~given by Eq.~(\ref{eq-Vx-fullring-fullsol}).

\subsection{Azimuthal mean internal torque for the laminar solution}
\label{app-azimuthal-mean-internal-torque}
The ultimate goal of the computation of the internal torque is to find the
azimuthal mean vertically integrated internal torque, because this is what is
needed for the evolution of the warp of a disc. Computing the mean requires
integration over azimuth $\phi$. However, we cannot simply integrate
Eqs.~(\ref{eq-g-ansatz-x-intz-Xi-lin}-\ref{eq-g-ansatz-z-intz-Xi-lin}) over
$\phi$ because the basis vectors of the local coordinate system $(x,y,z)$ rotate
with respect to the global coordinates $(X,Y,Z)$. \citet{2013MNRAS.433.2403O}
solve this by applying a rotation to $(\Ggvertcomp_x,\Ggvertcomp_y)$ to
obtain $(\Ggvertcomp_X,\Ggvertcomp_Y)$, where now the components point
in the global $X$ and $Y$ directions
\begin{eqnarray}
  \Ggvertcomp_X &=& \cos(\phi)\Ggvertcomp_x - \sin(\phi)\Ggvertcomp_y\label{eq-rotate-GX}\\
  \Ggvertcomp_Y &=& \sin(\phi)\Ggvertcomp_x + \cos(\phi)\Ggvertcomp_y\label{eq-rotate-GY}\\
  \Ggvertcomp_Z &=& \Ggvertcomp_z
\end{eqnarray}
We then integrate these over $\phi$ to obtain the azimuthal mean:
\begin{eqnarray}
  \GGbigcomp_X &=& \frac{1}{2\pi}\int_{0}^{2\pi}\Ggvertcomp_X\,d\phi\\
  \GGbigcomp_Y &=& \frac{1}{2\pi}\int_{0}^{2\pi}\Ggvertcomp_Y\,d\phi\\
  \GGbigcomp_Z &=& \frac{1}{2\pi}\int_{0}^{2\pi}\Ggvertcomp_Z\,d\phi
\end{eqnarray}
These integrals can be conveniently evaluated if we write
$\cos(\phi)=(e^{i\phi}+e^{-i\phi})/2$ and $\sin(\phi)=(e^{i\phi}-e^{-i\phi})/2i$
in Eqs.~(\ref{eq-rotate-GX}, \ref{eq-rotate-GY}). The $e^{i\phi}$ part
integrates to zero, while for the $e^{-i\phi}$ part only the terms in
Eqs.~(\ref{eq-g-ansatz-x-intz-Xi-cmplx-lin}, \ref{eq-g-ansatz-y-intz-Xi-cmplx-lin})
proportional to $e^{i\phi}$ survive. This leads to the following
expressions for the complex values of the internal torque:
\begin{eqnarray}
  2\GGbigcomp_X/\Xii &=& -\VVx(\tau) - i\alpha_t\VVx(\tau) - \alpha_t\psi \\
  2\GGbigcomp_Y/\Xii &=& -i\VVx(\tau) + \alpha_t\VVx(\tau) - i\alpha_t\psi \\
   \GGbigcomp_Z/\Xii &=&  q\alpha_t
\end{eqnarray}

\section{Symbols}
A list of symbols used in this paper, their meaning, the equation where
they are first used, and their relation to other papers literature, is
given in Table \ref{table-symbols}.

\begin{table*}
  \begin{tabular}{clcccc}
    Symbol & Meaning & Dimension & Eq/Sec & Ogilvie \& Latter & Martin et al. \\
    \hline
    $t$    & Time & $T$ & & $t$ & $t$ \\
    $\tau$ & Dimensionless time & & Eq.~\ref{eq-tau-t} & $\tau$ & \\
    $r_0$   & Radius of the annulus & $L$ & Sec.~\ref{sec-local-warped-shearing-box} & $r_0$ & $R$ \\
    $\phi$   & Azimuthal angle along annulus & & Sec.~\ref{sec-local-warped-shearing-box} & $\tau$ & \\
    $\llunit$   & Unit vector of orbit orientation of annulus & & Sec.~\ref{sec-local-warped-shearing-box} & $\llunit$ & $\llunit$ \\
    $\Omega_0$ & Orbital angular frequency at $r=r_0$ & $T^{-1}$ & Sec.~\ref{sec-local-warped-shearing-box} & $\Omega_0$ & $\Omega$ \\
    $q$         & $\Omega\propto r^{-q}$ & & Eq.~\ref{eq-definition-q} & $q$ & \\
    $\kappa$    & Dimensionless epicyclic freq. & & Eq.~\ref{eq-epicyclic-freq-dimless} &  & $\kappa/\Omega$ \\
    $\alpha_t$  & Turbulence parameter & & Eq.~\ref{eq-nu-in-alpha-hp-om0} & $\alpha$ & $\alpha$ \\
    $\warpvector$  & Warp vector & & Eq.~\ref{eq-def-psivec-warp} & $|\psi|{\bf m}$ & $d\llunit/d\ln R$ \\
    $\psi$  & Warp strength & & Eqs.~\ref{eq-def-psivec-warp}, \ref{eq-def-psi-warp} & $|\psi|$ & $|d\llunit/d\ln R|$ \\
    $X,Y,Z$   & Global coordinate system & $L$ & Sec.~\ref{sec-local-warped-shearing-box} & & \\
    $x,y,z$   & Local comoving coordinate system & $L$ & Sec.~\ref{sec-flat-shearing-box} & $x,y,z$& \\
    $x',y',z'$   & Local comoving warped coordinate system & $L$ & Sec.~\ref{sec-warped-shearing-box} & $x',(y'),z'$ & \\
    $u_x,u_y,u_z$ & Velocity in $(x,y,z)$ system & $L/T$ & Sec.~\ref{sec-flat-shearing-box} & $u_x,u_y,u_z$ & \\
    $\vx,\vy,\vz$ & Velocity in sheared/warped system & $L/T$ & Sec.~\ref{sec-warped-shearing-box} & $v_x,v_y,v_z$ & \\
    $\VVx,\VVy,\VVz$ & Dimensionless tilt velocity & & Eqs.~\ref{eq-vx-in-VOmz}-\ref{eq-vz-in-VOmz} & $u,v,w$ & \\
    $\omega_0$ & Dimensionless frequency of the hom.~solution &  & Eq.~\ref{eq-omega0-dispersion-definition} & & \\
    $H$ & Dimensionless pressure scale height & & Eq.~\ref{eq-def-dimless-H} & $H$ & \\
    $\rho$ & Gas density & $M/L^3$ & Sec.~\ref{sec-flat-shearing-box}, Eq.~\ref{eq-simple-gaussian-vertical-structure-warped} & $\rho$ & \\
    $p$ & Gas pressure & $M\,L^{-1}T^{-2}$ & Sec.~\ref{sec-flat-shearing-box}, Eq.~\ref{eq-pressure-force} & $p$ & \\
    $\Sigma$ & Surface density & $M/L^2$ & Eq.~\ref{eq-simple-gaussian-vertical-structure} & $\Sigma$ & $\Sigma$ \\
    $c_s$ & Isothermal sound speed & $L/T$ & & $c_s$ & \\
    $h_p$ & Pressure scale height of the disc & $L$ & $h_p=c_s/\Omega_0$ & $h_p$ & $H$ \\
    $v_r$ & Radial velocity of the gas in the global disc & $L/T$ & Eq.~\ref{eq-vr-formula} & & $v_R$ \\
    $f_i^p$ & Gas pressure acceleration & $L\,T^{-2}$ & Eqs.~\ref{eq-pressure-force} and \ref{eq-fp-x-prime}-\ref{eq-fp-z-prime}& & \\
    $f_i^v$ & Viscous acceleration & $L\,T^{-2}$ & Eq.~\ref{eq-shear-viscosity-force} & & \\
    $F_i^v$ & Dimensionless viscous acceleration of tilt &  & Eq.~\ref{eq-bigF-visc-definition} & & \\
    $D_t$ & Comoving time derivative & $L^{-1}$ & Sec.~\ref{sec-flat-shearing-box}, Eq.~\ref{eq-xyz-comoving-derivative} & $D$ & \\
    $D_\tau$ & Dimensionless comoving time deriv. & & Eq.~\ref{eq-Dtau-in-Dt} & & \\
    $\GGbig$ & Azimuthally averaged torque density & $M\,L\,T^{-2}$ & Eq.~\ref{eq-ggbig-cmplx-x}-\ref{eq-ggbig-cmplx-z} or \ref{eq-ggbig-re-x}-\ref{eq-ggbig-re-z} & ${\boldsymbol {\cal G}}/(2\pi r_0)$ & $-{\bf G}/R$ \\
    $\GGbig^{(s)}$ & The dynamic (``sloshing'') part of $\GGbig$ & $M\,L\,T^{-2}$ & Eq.~\ref{eq-Gsplit-Gslosh} & & \\
    $\GGbig^{(v)}$ & The static viscous part of $\GGbig$ & $M\,L\,T^{-2}$ & Eq.~\ref{eq-Gsplit-Gvisc} & & \\
    $\GGbig^{(s)}_{p0}$ or $\GGbig_{p0}$ & The steady-state particular solution of $\GGbig$ or $\GGbig^{(s)}$ & $M\,L\,T^{-2}$ & Eq.~\ref{eq-Gp-afo-Q123-repeat} or \ref{eq-Gp-afo-Q123-repeat-full} & & \\
    $\ggloc$ & Local torque density & $M\,T^{-2}$ & Eqs.~\ref{eq-g-laminar-x}-\ref{eq-g-laminar-z} & ${\bf g}$ & \\
    $\Xii$ & Dimensionality constant for $\GGbig$ & $M\,L\,T^{-2}$ & Eq.~\ref{eq-xii-def} & & \\
    $\beta$ & The damping coefficient of \citet{2019ApJ...875....5M} &  & Sec.~\ref{sec-martin-beta-terms} & & $\beta$ \\
    $Q_1, Q_2, Q_3$ & The $Q$ symbols of \citet{1999MNRAS.304..557O} &  & Eqs.~\ref{eq-Q1-fullsol}-\ref{eq-Q3-fullsol} & $Q_1, Q_2, Q_3$ & \\
    $\tilde Q_1, \tilde Q_2, \tilde Q_3$ & Alternative $Q$ symbols &  & $\tilde Q_1=\alpha_t Q_1$ and Eqs.~\ref{eq-qtilde-2-def}, \ref{eq-qtilde-3-def} & & \\
    $\phi_0$ & {Phase (orientation) of sloshing torque} & & Eq.~\ref{eq-phi0-from-q3-q2} & & \\
  \end{tabular}
  \caption{\label{table-symbols}Table of often-used symbols of this paper. The
    dimension column gives the dimension of the quantity, where $T$ is time, $L$
    is length, $M$ is mass. The equivalent $y'$ quantities in \citet{2013MNRAS.433.2403O} are
    in parentheses, because we define the $y'$-coordinate as non-winding-up, as opposed to
    \citet{2013MNRAS.433.2403O}. }
\end{table*}



\bsp	
\label{lastpage}
\end{document}